\documentclass[superscriptaddress,aps,prd,nofootinbib]{revtex4-2}

\usepackage{graphicx}
\usepackage{dcolumn}
\usepackage{hyperref}
\usepackage{ulem}
\usepackage{color}
\usepackage{bm}
\usepackage{siunitx}
\usepackage{braket}
\usepackage{amsfonts,amsmath,amssymb,bm,tensor}
\usepackage{mathtools}
\usepackage{comment}
\usepackage{ifpdf}
\usepackage{slashed}
\usepackage[mathscr]{eucal}
\usepackage[utf8]{inputenc}
\usepackage{cancel}
\begin{document}


\title{Detection of isotropic cosmic birefringence and its implications \texorpdfstring{\\}{}
for axion-like particles including dark energy}

\author{Tomohiro Fujita}
\affiliation{Institute for Cosmic Ray Research, The University of Tokyo, Kashiwa, 277-8582, Japan}

\author{Kai Murai}
\affiliation{Institute for Cosmic Ray Research, The University of Tokyo, Kashiwa, 277-8582, Japan}
\affiliation{Kavli Institute for the Physics and Mathematics of the Universe (WPI), The University of Tokyo, Kashiwa 277-8583, Japan}

\author{Hiromasa Nakatsuka}
\affiliation{Institute for Cosmic Ray Research, The University of Tokyo, Kashiwa, 277-8582, Japan}

\author{Shinji Tsujikawa}
\affiliation{Department of Physics, Waseda University, 3-4-1 Okubo, Shinjuku, Tokyo 169-8555, Japan}

\begin{abstract}
We investigate the possibility that axion-like particles (ALPs) with various potentials account for the isotropic birefringence recently reported by analyzing the Planck 2018 polarization data.
For the quadratic and cosine potentials, we obtain lower bounds on the mass, coupling constant to photon $g$, abundance and equation of state of the ALP to produce the observed birefringence. Especially when the ALP is responsible for dark energy, it is possible to probe the tiny deviation of dark energy equation of state from $-1$ through the cosmic birefringence. We also explore ALPs working as early dark energy (EDE), which alleviates the Hubble tension problem. Since the other parameters are limited by the EDE requirements, we narrow down the ALP-photon coupling to $10^{-19}\, {\rm GeV}^{-1}\lesssim g\lesssim 10^{-16}\, {\rm GeV}^{-1}$ for the decay constant $f=M_\mathrm{pl}$. Therefore, the Hubble tension and the isotropic birefringence imply that $g$ is typically the order of $f^{-1}$, which is a non-trivial coincidence.
\end{abstract}

\maketitle

\tableofcontents

\section{Introduction}
\label{sec: introduction}

The Cosmic Microwave Background (CMB) observation has played a crucial role in the development of the modern precision cosmology. The WMAP and Planck satellites determined the various cosmological parameters with high precision and established the standard cosmology
based on the $\Lambda$ Cold Dark Matter ($\Lambda$CDM) model \cite{Spergel:2003cb,weiland2011seven,Ade:2013zuv,Aghanim:2018eyx}.
Recently, however, the novel analysis on the Planck 2018 polarization data indicated a hint of a new ingredient beyond $\Lambda$CDM.
Ref.~\cite{MinamiKomatsu} reported a measurement of the isotropic cosmic birefringence, which excludes the null hypothesis at $99.2\,\%$
confidence level (CL).
Cosmic birefringence is the rotation of the photon polarization angle \cite{Carroll:1998zi,Lue:1998mq}, and the observation of isotropic cosmic birefringence may imply the presence of new physics.

One of the possible sources of cosmic birefringence is axion or axion-like particle (ALP) with a weak coupling to photon \cite{Pospelov:2008gg,Finelli:2008jv,Lee:2013mqa,Zhao:2014yna,Lee:2016jym,Liu:2016dcg}.
Axion is a hypothetical pseudo-Nambu-Goldstone boson originally introduced to solve the strong $CP$ problem~\cite{Peccei:1977hh,Kim:1979if,Shifman:1979if}, and ALPs have been introduced in many extensions of the standard model of particle physics \cite{Marsh:2015xka,Gong:2016zsb}.
Especially, those particles predicted in string theory have the broad range of mass and the couplings to gauge fields \cite{Svrcek:2006yi,Arvanitaki:2009fg}.

In the presence of ALPs coupled to photons through a Chern-Simons term, the difference of the ALP field along the light path induces cosmic birefringence \cite{Carroll:1998zi,Lue:1998mq,Feng:2004mq,Feng:2006dp,Liu:2006uh}.
While the ALP perturbation at the decoupling of photon induces anisotropic birefringence, the background motion of the same field induces isotropic birefringence.
Therefore, we are interested in the ALP field whose background value evolves in time after the decoupling of CMB photons.
If the ALP field rapidly oscillates during the photon decoupling epoch, its background value around the last scattering surface (LSS) is
averaged over the duration of thickness of the LSS and exponentially suppressed.
In this case, what is left over for
the contribution to polarization angle is the ALP field value today, and hence the isotropic birefringence is significantly suppressed. Therefore, we expect that the ALP field can induce a non-negligible birefringence if its mass is small enough not to oscillate until the last scattering epoch.

In particular, if the ALP mass $m$ is
as small as today's Hubble constant
$H_0$ and the field slowly rolls down the potential until now, the ALP may comprise all or part of dark energy~\cite{Fukugita:1994hq,Frieman:1995pm,Kim:1998kx,Kim:1999dc,Choi:1999xn,Nomura:2000yk,Kim:2002tq,Hall:2005xb,Kim:2009cp, Chatzistavrakidis:2012bb,Kim:2014tfa,Kang:2019vsk}.
In this case, the corresponding dark energy scenario is called thawing quintessence \cite{Caldwell:2005tm}
in which the field equation of state is initially close to $-1$ and it starts to
deviate from $-1$ only recently.
The joint likelihood analysis of
Planck 2018 combined with the data of supernovae type Ia and baryon acoustic oscillations showed that today's field equation of state is constrained to be
in the range $w_{\phi}<-0.95$ (95\,\%\,CL) \cite{Aghanim:2018eyx}
with the quintessence prior
$w_{\phi} \geq -1$
(see also Refs.~\cite{Chiba:2012cb,Tsujikawa:2013fta,Durrive:2018quo}).
Due to the observational degeneracy
of $w_{\phi}$ around $-1$, it is generally difficult to distinguish thawing quintessence from the cosmological constant. In other words, the time variation of the ALP field is hard to be detected by the standard distance measurements alone.
On the other hand, the cosmic birefringence provides an independent probe for constraining the background
dynamics of the ALP field.
Even if $w_{\phi}$ is very close to $-1$,
we will show that the existence of the ALP-photon coupling $g$
can explain the observed rotation angle
of cosmic birefringence, while satisfying
other experimental bounds of $g$.

In the $\Lambda$CDM model, there is
an observational
tension of the $H_0$ value between
CMB \cite{Aghanim:2018eyx} and local astrophysical measurements at low redshifts (cosmic distance ladder) \cite{Riess:2011yx,Riess:2016jrr,Bonvin:2016crt,Riess:2018byc,Birrer:2018vtm,Riess:2019cxk}, with the significance of $4.4\sigma$~\cite{Riess:2019cxk}.
Among various solutions to alleviate this problem \cite{Wyman:2013lza,DiValentino:2016hlg,Zhao:2017cud,DiValentino:2017zyq,DiValentino:2017iww,DiValentino:2017rcr,Khosravi:2017hfi,Mortsell:2018mfj,Poulin:2018zxs,Pandey:2019plg,Vattis:2019efj,Alexander:2019rsc,Vagnozzi:2019ezj,Knox:2019rjx,Sekiguchi:2020teg}, the modifications of the cosmological dynamics prior to the CMB decoupling epoch dubbed
early dark energy (EDE) \cite{Karwal:2016vyq,Poulin:2018cxd,Agrawal:2019lmo,Lin:2019qug,Smith:2019ihp,Niedermann:2019olb,Berghaus:2019cls,Sakstein:2019fmf,Ye:2020btb,Niedermann:2020dwg,Ye:2020oix} have been
in active study.
In these scenarios, the energy density of EDE behaves like a cosmological constant
at early times and increases the Hubble
expansion rate before the last scattering epoch, and then it dilutes away like or faster than radiation.
As a result, the presence of EDE reduces the sound horizon at the last scattering and increases $H_0$ inferred from the
observed temperature anisotropies in CMB.
This unique evolution of EDE can be realized if the bottom of the potential is equal to or higher order than the quartic potential.
With such a non-linear potential, the ALP oscillation gets slower as the amplitude decreases.
Consequently, the aforementioned suppression of cosmic birefringence due to the ALP oscillation during the last scattering epoch is not as severe as the quadratic potential, which drastically changes the prediction of birefringence compared to the conventional ALPs.

In this paper, we investigate the possibility that the ALPs account for the
recently reported isotropic cosmic birefringence with various potentials.
First, we examine the case where the ALP comprises a part of the energy component of our universe including dark energy and dark matter by extending
the previous work~\cite{Fujita:2020aqt}
which some of the authors conducted when only the upper bound on the isotropic cosmic birefringence angle was available.
We adopt the quadratic potential and cosine potential and discuss the relations among the ALP mass, the ALP energy fraction, initial conditions, and the inferred value of the coupling $g$.
In addition, the equation of state is related to the inferred value of $g$.
From these relations, we put lower bounds on $g$, the ALP mass, the current ALP abundance, and the equation of state.

We also investigate the isotropic birefringence produced in the proposed EDE scenarios for the first time, while the earlier work studied the anisotropic birefringence in an EDE model \cite{Capparelli:2019rtn}.
We study two typical models of EDE with  higher-order periodic potentials and
power-law potentials.
Since the ALP mass and initial conditions are restricted by the requirement that the ALP works as EDE, we can infer the ALP parameters with less free parameters than the former case.
In particular, for higher-order periodic potentials, the inferred $g$ value and the decay constant $f$ satisfy the non-trivial relation $g f = \mathcal{O} (1)$ for $f = M_{\mathrm{pl}}$.

While we mainly focus on the background
ALP field inducing the isotropic birefringence in this paper, the ALP fluctuation
at the observer also contributes to it.
We also discuss the latter effect.

This paper is organized as follows.
In Sec.~\ref{sec: CB}, we briefly review the cosmic birefringence induced by the ALPs.
On using the recent observed angle of
cosmic birefringence, we identify the inferred ALP parameters for the simple ALP models in Sec.~\ref{sec: DE ALP} and for the EDE models in Sec.~\ref{sec: EDE}.
We discuss other possible contributions to the isotropic birefringence in Sec.~\ref{sec: discussion}. We conclude in
Sec.~\ref{sec: conclusion}.

\section{Cosmic Birefringence by ALP}
\label{sec: CB}

In this paper, we consider an ALP field $\phi$ coupled to photons with the electromagnetic tensor $F_{\mu\nu}$. The gravitational sector is described by general relativity
with the Ricci scalar $R$. The action in such theories is given by
\begin{equation}
    \mathcal{S}
    =\int {\rm d}^4 x \sqrt{-\tilde{g}} \left[ \frac{M_{\rm pl}^2}{2}R
    - \frac{1}{2} g^{\mu \nu} \partial_{\mu} \phi \partial_{\mu} \phi - V(\phi)
    - \frac{1}{4} F_{\mu\nu} F^{\mu\nu} + \frac{1}{4}g \phi F_{\mu\nu} \tilde{F}^{\mu\nu} \right],
\end{equation}
where $\tilde{g}$ is the determinant of
metric tensor $g_{\mu \nu}$, $M_{\rm pl}=2.435 \times 10^{18}$~GeV is the reduced Planck mass, $V(\phi)$ is the ALP potential, $g$ is the ALP-photon coupling constant, and $\tilde{F}^{\mu\nu}$ is the dual
of electromagnetic tensor.
It is known that a linearly polarized photon propagating under the influence of ALP field rotates its polarization plane, because of the parity-violating nature of the ALP and its coupling to photon~\cite{Carroll:1998zi,Lue:1998mq,Feng:2004mq,Feng:2006dp,Liu:2006uh}.
To observe this phenomenon, the CMB photon is an ideal target.
When we observe CMB photons emitted at the last scattering surface (LSS), the rotation angle $\alpha$ of
their polarization plane
depends on the difference of ALP field values between the observer (``obs'')  and the LSS.
Since this rotation is caused by the ALP-photon coupling, $\alpha$ is also proportional to the coupling constant $g$.
The rotation angle $\alpha$ of the CMB photon coming from a direction on the sky (denoted as a unit vector
$\hat{\bm{n}}$)
is given by \cite{Harari:1992ea}
\begin{equation}
    \alpha(\hat{\bm{n}})
    =
    \frac{g}{2} \Big[ \phi(t_0,\bm{0} ) - \phi(t_\mathrm{LSS},d_\mathrm{LSS}\hat{\bm n})\Big],
\end{equation}
where $\bm{0}$ is the position of the observer, $t_0$ is the present time, $t_\mathrm{LSS}$ is the last scattering time, and $d_\mathrm{LSS}$ is a distance to the LSS.
In the following, we use the subscripts
``0'' and ``LSS'' as the values today and
at the LSS, respectively.
As $\phi$ depends on both time and space, we decompose it into the background and perturbed parts, as,
\begin{align}
    \phi (t_\mathrm{LSS},d_\mathrm{LSS}\hat{\bm n})
    &=
    \bar{\phi}_\mathrm{LSS} +  \delta\phi_\mathrm{LSS}\,,
\\
    \phi (t_0,\bm{0} )
    &=
    \bar{\phi}_\mathrm{obs} +  \delta\phi_\mathrm{obs}\,,
\end{align}
where a bar represents background values.
The rotation angle can be also separated into the isotropic and anisotropic terms, as
\begin{equation}
    \alpha(\hat{\bm{n}})
    =
    \bar{\alpha} + \delta \alpha(\hat{\bm n})
    =
    \frac{g}{2} \left(\bar{\phi}_{\mathrm{obs}} - \bar{\phi}_{\mathrm{LSS}} + \delta \phi_{\mathrm{obs}}\right) -     \frac{g}{2} \delta \phi_{\mathrm{LSS}}(\hat{\bm n}).
    \label{alpha full expression}
\end{equation}

Recently, re-analyzing the observational data of Planck satellite, Ref.~\cite{MinamiKomatsu} reported the detection of the isotropic birefringence, as
\begin{equation}
    \bar{\alpha}=0.35\pm 0.14\,{\rm deg}\,.
    \label{alphabar value}
\end{equation}
Since the null result is excluded at 99.2\,\% CL,
it provides a fascinating hint of new phenomena beyond the standard
$\Lambda$CDM cosmology.
On the other hand, the anisotropic birefringence has not yet been detected and only some constraints were derived~\cite{Bianchini:2020osu,Namikawa:2020ffr}.
Inspired by these latest
observations on $\alpha$, we focus on the isotropic birefringence and consider models
in which the background ALP field explains the observed value of $\bar\alpha$ through the background dynamics,
$\bar{\alpha} = \frac{g}{2}\left(\bar{\phi}_{\mathrm{obs}} - \bar{\phi}_{\mathrm{LSS}}\right).$
It should be noted that $\delta\phi_\mathrm{obs}$ also contributes to $\bar\alpha$ and may be able to account for
the observed value by itself.
However, since the ALP fluctuations, $\delta\phi_\mathrm{LSS}$ and $\delta\phi_\mathrm{obs}$, have the same origin and are tightly connected,
one should be careful not to violate the observational constraint on $\delta\alpha$, when seeking the possibility of $\bar\alpha\simeq g\delta\phi_\mathrm{obs}/2$.
In fact, we will see that its contribution is constrained as $|\bar\alpha| \le 0.13$\,deg in the simplest case.
We will further discuss this issue in Sec.~\ref{sec: discussion}.

In order to extract constraints on $g$ from the observed
value of $\bar{\alpha}$ in Eq.~(\ref{alphabar value}), we need to solve the time evolution of $\bar{\phi}(t)$ together
with the Friedmann equation.
We study the background cosmological dynamics
on the spatially flat Friedmann-Lema{\^\i}tre-Robertson-Walker
space-time with the line element,
\begin{equation}
\mathrm{d}s^2 = -\mathrm{d}t^2 +a^2(t)\delta_{ij}
{\rm d}x^i {\rm d}x^j\,,
\end{equation}
where $a(t)$ is the time-dependent scale factor.
Then, the scalar field obeys
\begin{equation}
\ddot{\bar{\phi}} + 3H\dot{\bar{\phi}} + V_{,\bar{\phi}}(\bar{\phi}) = 0\,,
\label{phieq}
\end{equation}
where $V_{,\bar{\phi}}={\rm d}V/{\rm d}\bar{\phi}$, a dot represents the derivative with respect to $t$, and $H \equiv \dot{a}/a$ is the Hubble expansion rate.
The energy density and pressure of the ALP field are given, respectively, by
$\rho_{\phi}=\dot{\phi}^2/2+V(\phi)$ and
$P_{\phi}=\dot{\phi}^2/2-V(\phi)$.
The associated density parameter and
equation of state are
\begin{equation}
\tilde{\Omega}_{\phi}=
\frac{\rho_{\phi}}{3M_{\rm pl}^2 H^2}\,,
\qquad
\tilde{w}_{\phi}=
\frac{P_{\phi}}{\rho_{\phi}}\,,
\end{equation}
where we denote their today's values as
$\Omega_{\phi}$ and $w_{\phi}$, respectively.

To solve Eq.~(\ref{phieq}) for $\bar{\phi}$, we need to know the evolution of $H$ from the late radiation era to today. For this purpose, we take the energy densities of radiation, nonrelativistic matter, and cosmological constant into account, which are denoted as $\rho_r$, $\rho_M$, and $\rho_{\Lambda}$ respectively.
They obey the continuity equations,
\begin{equation}
\rho_I+3H\left(1+w_I \right)\rho_I=0\,, \qquad
(I=r,M,\Lambda)\,,
\label{coneq}
\end{equation}
where $w_r=1/3$, $w_M=0$, and $w_{\Lambda}=-1$.
The Friedmann equation is given by
\begin{equation}
3M_{\rm pl}^2 H^2=\rho_r+\rho_M+\rho_{\Lambda}\,.
\label{Fri}
\end{equation}
When the ALP field is nearly frozen by the Hubble friction, $\tilde{w}_{\phi}$
is close to $-1$.
After $\bar{\phi}$ starts to oscillate around the minimum of a quadratic potential, it behaves as a dust with the averaged equation of state
$\tilde{w}_{\phi} \simeq 0$.
Then, the energy density of ALP field can be incorporated into either $\rho_{\Lambda}$
or $\rho_M$ in Eq.~(\ref{Fri}).

To compute an effective field value
$\langle \bar{\phi} \rangle_{\mathrm{LSS}}$ at the LSS, we also take into account
the effect of finite thickness of the LSS.
If the time variation of $\bar{\phi}$ is significant around the
decoupling epoch of photons, it is not trivial to identify the value of $\langle \bar{\phi} \rangle_{\mathrm{LSS}}$.
Indeed, the ALP field with a mass larger than $H$ at the LSS
starts to oscillate before the decoupling epoch.
A natural way to define the effective field value for the cosmic birefringence in CMB is to take a time average of $\bar\phi(t)$ as \cite{Capparelli:2019rtn,Fujita:2020aqt}
\begin{equation}
	\langle \bar{\phi} \rangle_{\mathrm{LSS}} = \int \mathrm{d}T\, {\cal V}(T)\,\bar{\phi}\left(t(T)\right)\,.
	\label{eq_LSSwashout}
\end{equation}
In Eq.~(\ref{eq_LSSwashout}) the ALP field is weighed by
a visibility function ${\cal V}(T)$.
This function describes the probability density that a CMB photon, now observed, is scattered at the temperature $T$.
We approximate ${\cal V}(T)$ by a Gaussian function,
\begin{equation}
	{\cal V}(T) \simeq \frac{1}{\sqrt{2\pi}\sigma_T}\exp\left[ -\frac{(T-T_L)^2}{2\sigma_T^2} \right],
	\label{VT}
\end{equation}
where $T_L = 2941\,\si{K}$ and $\sigma_T = 248\,\si{K}$ are
the numerical fitting parameters \cite{Weinberg:2008zzc}.
If $\bar{\phi}$ is nearly frozen around the decoupling epoch, then $\langle \bar{\phi} \rangle_{\mathrm{LSS}}$ is almost
equivalent to $\bar{\phi}$ at $T=T_L$.

If the field exhibits rapid oscillations
between positive and negative values around the LSS, however, $\langle \bar{\phi} \rangle_{\mathrm{LSS}}$
practically approaches 0 by taking the time average
(\ref{eq_LSSwashout}).
This is especially the case for the ALP
mass $m$ much larger than the Hubble
expansion rate $H$ at the LSS.
In such a case, even though the amplitude of $\bar{\phi}$ today
is smaller than that at the LSS,
$|\langle \bar{\phi} \rangle_{\mathrm{LSS}}|$ is suppressed
relative to $|\bar{\phi}_\mathrm{obs}|$
and hence the dominant contribution to $\bar{\alpha}$ comes from
$\bar{\phi}_\mathrm{obs}$.
In Sec.~\ref{sec: DE ALP}, we will discuss the mass range
of ALP field in which $|\langle \bar{\phi} \rangle_{\mathrm{LSS}}|$ becomes smaller
than $|\bar{\phi}_\mathrm{obs}|$.

In summary, we will address the possibility to account for the reported value of isotropic birefringence, Eq.~\eqref{alphabar value},
by considering several ALP potentials.
The background value of $\alpha$ is computed according to
\begin{equation}
    \bar{\alpha} = \frac{g}{2} \Delta \bar{\phi}\equiv \frac{g}{2}\left( \bar{\phi}_{\mathrm{obs}} - \langle \bar{\phi} \rangle_{\mathrm{LSS}}\right).
\label{master eq}
\end{equation}
On using Eq.~(\ref{coneq}), the Friedmann Eq.~(\ref{Fri})
can be expressed as
\begin{equation}
H(a)=H_0 \sqrt{\Omega_M a^{-4}
(a+a_{\rm eq})+\Omega_{\Lambda}}\,,
\label{Ha}
\end{equation}
where $H_0$ is today's Hubble constant, which is given by
\begin{equation}
H_0=2.1331 \times 10^{-33}\,h~{\rm eV}\,.
\end{equation}
We take the dimensionless constant $h$
to be $0.677$.
In Eq.~(\ref{Ha}), $\Omega_M$ and $\Omega_{\Lambda}$
are today's density parameters of nonrelativistic matter
and cosmological constant, respectively.
We adopt the values $\Omega_M=0.31$ and
$\Omega_{\Lambda}=0.69$ in our numerical simulation.
We also choose the value $a_{\rm eq}=1/3400$ for the scale factor at radiation-matter equality, where $a=1$ today.
The central temperature $T_L=2941$\,K
in Eq.~(\ref{VT}) corresponds to the
scale factor $a_L=T_0/T_L=9.266 \times
10^{-4}$, where we used today's temperature $T_0=2.725$\,K.
{}From Eq.~(\ref{Ha}), the Hubble parameter $H_{\rm LSS}$ at the peak of Gaussian
distribution (\ref{VT}) can be
estimated as
\begin{equation}
H_{\rm LSS} \simeq 3.3 \times
10^{-29}~{\rm eV}\,.
\end{equation}

When we study the background dynamics of the ALP field, there is a typical
constant mass
scale $m$ whose energy scale is related to
the second derivative $V_{,\bar{\phi}\bar{\phi}}$.
For the numerical purpose,
we write the field Eq.~(\ref{phieq})
in the form,
\begin{equation}
    \bar{\phi}''
    + 3\frac{H_0}{m} \sqrt{\Omega_M a^{-4}(a+a_{\rm eq})+\Omega_{\Lambda}}\,\bar{\phi}'
    + \frac{V_{,\bar{\phi}}}{m^2}
    =0\,,
\label{fieldeq1}
\end{equation}
where a prime represents a derivative with respect to
the dimensionless variable $\tau \equiv mt$.
Since the temperature has the dependence $T \propto a^{-1}$,
the integral (\ref{eq_LSSwashout}) can be expressed as
$\langle \bar{\phi} \rangle_{\mathrm{LSS}}=\int_{0}^{\tau_0}
{\rm d}\tau \, H T {\cal V} \bar{\phi}/m$. Then, from the past to today, we
need to integrate the following differential equation,
\begin{equation}
    \langle \bar{\phi} \rangle_{\mathrm{LSS}}'=
    \frac{H_0}{m} \frac{T_0}{a}
    \sqrt{\Omega_M a^{-4} (a+a_{\rm eq}) + \Omega_{\Lambda}}\,
    {\cal V}\left( \frac{T_0}{a} \right) \bar{\phi}\,.
    \label{fieldeq2}
\end{equation}
For a given potential $V$ and initial conditions,
we will integrate Eqs.~\eqref{Ha}, \eqref{fieldeq1}, and \eqref{fieldeq2}
to obtain the values of $\bar{\phi}_{\rm obs}$ and
$\langle \bar{\phi} \rangle_{\mathrm{LSS}}$.
We will choose one of the initial conditions of the ALP field to be $\bar{\phi}'(\tau = 0)=0$.

\section{Simple ALP models}
\label{sec: DE ALP}

In this section, we compute how much the isotropic birefringence is generated for
two different ALP potentials:
(A) $V_\text{mass}(\phi)=m^2 \phi^2/2$ and
(B) $V_\text{cos}(\phi)=m^2 f^2 [1-\cos(\phi/f)]$.
We explore the mass region, $10^{-42}\,\mathrm{eV}\lesssim m \lesssim 10^{-25.5}\,\mathrm{eV}$,
in which the ALP can act as either dark
energy or a sub-dominant component of dark matter, depending on its mass.
Respecting the constraint on the ALP energy fraction not to ruin the success of the $\Lambda$CDM cosmology, we will obtain the ALP-photon coupling constant $g$ consistent with
the observed isotropic birefringence
\eqref{alphabar value}.
We also derive lower bounds on the ALP mass and its energy fraction to generate the observed value of $\bar\alpha$ in these models.

\subsection{Quadratic potential}
\label{sec: quadratic potential}

We first consider a quadratic potential
given by
\begin{equation}
V_{\mathrm{mass}}(\phi) = \frac{1}{2} m^2\phi^2.
\label{eq_V_quadratic}
\end{equation}
In Ref.~\cite{Fujita:2020aqt}, some of the authors have studied the ALP with the same potential and computed the isotropic and anisotropic birefringence induced by its background and perturbation parts, respectively.
In the following, we
explain the background calculation in more detail than Ref.~\cite{Fujita:2020aqt}
and derive the new limits on some parameters by using the observed value of $\bar{\alpha}$.

For the quadratic potential,
the ALP field begins to oscillate when $m \simeq H$.
If $m \gg H_{\mathrm{LSS}}$, then the ALP field
exhibits rapid oscillations around the
CMB decoupling epoch. In this case,
the LSS value $\langle \bar{\phi} \rangle_{\mathrm{LSS}}$ of Eq.~(\ref{eq_LSSwashout})
is suppressed to be
smaller than $\bar{\phi}_{\rm obs}$ due to
the time averaging of fast oscillations
of $\bar{\phi}$.
If $ m \ll H_0$, on the other hand,
the Hubble friction does not allow the field
to roll down the potential by today,
and hence $\Delta\bar\phi$ is suppressed.
Therefore the cosmic birefringence is most sensitive
to the intermediate mass region, $H_0 \lesssim m \lesssim H_{\mathrm{LSS}}$.

With a given mass $m$, the different
choices of initial conditions of
the ALP field $\bar{\phi}_{\mathrm{init}}$
not only
affects the quantitative estimate of $\Delta \bar{\phi}$ but also today's
energy fraction $\Omega_{\phi}$.

For $m \lesssim H_0$, the energy density
of ALP field can be the source for all of dark energy. As $m$ increases, $\bar{\phi}$
starts to roll down the potential
around the redshift $z \lesssim 1$.
This leads to the deviation of today's
ALP equation of state $w_{\phi}$ from $-1$.
The likelihood analysis
based on the Planck 2018 data \cite{Aghanim:2018eyx} with the
prior $\tilde{w}_{\phi} \geq -1$ puts
the bound $w_{\phi}<-0.95$ at 95\,\%\,CL.
Applying this constraint to the potential
(\ref{eq_V_quadratic}), we find that  $\Omega_{\phi}$ can be as large as the density parameter of dark energy $\Omega_{\Lambda} = 0.69$ for
\begin{equation}
m \leq 8.5 \times 10^{-34}\,\mathrm{eV}\,.
\label{darkm}
\end{equation}
If $m$ is larger than this upper bound,
$w_{\phi}$ is larger than $-0.95$ and
hence the energy density of ALP field is not the main source for dark energy.

For $m \geq 10^{-32}\,\mathrm{eV}$, the
ALP field begins to oscillate in the past,
and $\tilde{w}_{\phi}$ shows a transition from $-1$ to $0$ by today.
In the mass range
\begin{equation}
10^{-32}~\mathrm{eV} \leq m \leq 10^{-25.5}~\mathrm{eV}\,,
\label{oscim}
\end{equation}
the observations of CMB and large-scale structures put constraints on the ALP with such a transition of $\tilde{w}_\phi$, as $\Omega_{\phi}h^2 \leq 0.006$  \cite{Hlozek:2014lca}.

From these constraints, we obtain the upper limit of $\Omega_{\phi}$, as
\begin{align}
	\Omega_{\phi,\mathrm{max}}
	=
	\begin{cases}
		0.69
		&(m \leq 8.5\times 10^{-34} \,\si{\eV}),
		\\
		0.006h^{-2}
		&(10^{-32}\,\si{\eV} \leq m \leq 10^{-25.5}\,\si{\eV}).
	\end{cases}
	\label{eq: Omega phi constraint}
\end{align}
In the intermediate mass region
$8.5 \times 10^{-34}\,\mathrm{eV} < m < 10^{-32}\,\mathrm{eV}$, we linearly connect these upper limits of $\Omega_{\phi}$ in the $\log m$-$\log \Omega_{\phi}$ plane.
Note that $\Omega_{\phi,\mathrm{max}}$ increases for the heavier mass region, $m > 10^{-25.5}\,\si{\eV}$.
The ALP behavior becomes more similar to that of normal dark matter there,
and several complications
such as the growth of perturbations due to clustering and the shorter oscillation period
may not be negligible. To make a conservative argument, we restrict ourselves
to the lighter mass region, $m \le 10^{-25.5}\,\si{\eV}$.
In Sec.~\ref{sec: discussion}, we will further discuss these phenomenological aspects of the higher mass ALP.

\begin{figure}[t]
	\centering
	\includegraphics[width=.60\textwidth ]{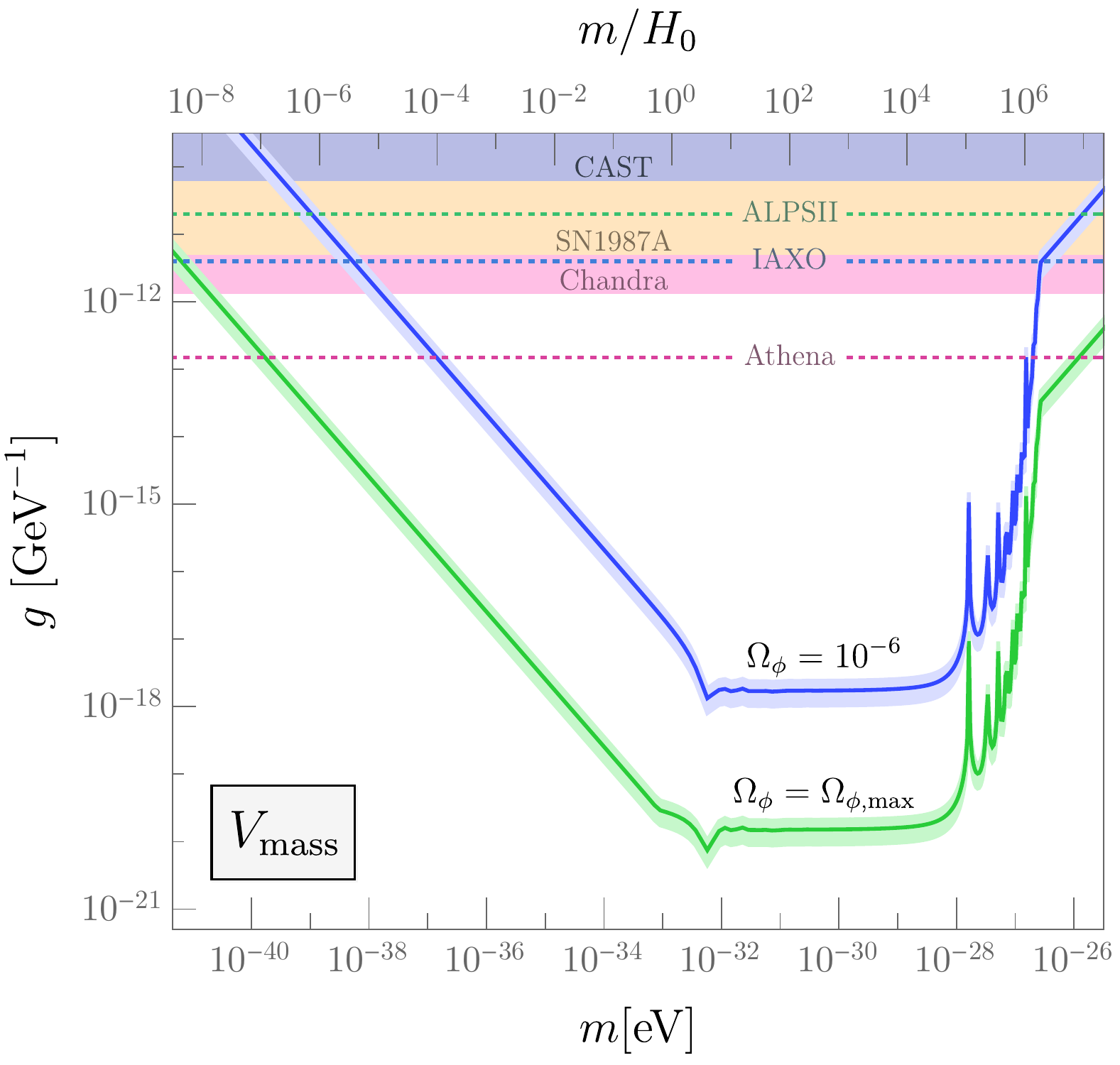}
	\caption{
	The ALP-photon coupling constant $g$ inferred by the isotropic birefringence $\bar{\alpha} = 0.35 \pm 0.14 \,~\mathrm{deg}$ versus the ALP mass $m$ for the quadratic potential
	$V_{\mathrm{mass}}(\phi)=m^2\phi^2/2$.
	The green line corresponds to the maximum energy fraction $\Omega_\phi=\Omega_{\phi,\mathrm{max}}$ given in
	Eq.~\eqref{eq: Omega phi constraint},
	whereas the blue line shows the case with $\Omega_{\phi} = 10^{-6}$.
	The shaded regions are excluded by the
	measurements of CAST~\cite{Anastassopoulos:2017ftl} (blue), SN1987A~\cite{Payez:2014xsa} (orange), and Chandra~\cite{Berg:2016ese} (pink).
	We also plot the projected sensitivities of the future experiments, ALPSII~\cite{Bahre:2013ywa}, IAXO~\cite{Armengaud:2014gea,Irastorza:2018dyq}, and Athena~\cite{Conlon:2017ofb}, from top to bottom as dotted lines.
	}
	\label{fig: mass}
\end{figure}

With the energy constraint $\Omega_{\phi} \le \Omega_{\phi,\mathrm{max}}$,
we compute the field excursion $\Delta\bar\phi$ by numerically solving
Eqs.~(\ref{fieldeq1}) and (\ref{fieldeq2}).
In Fig.~\ref{fig: mass}, we plot the ALP-photon coupling $g$ which generates
the observed isotropic birefringence $\bar{\alpha} = 0.35 \pm 0.14$~deg
for a given $m$.
The green line, which corresponds to the maximum ALP energy fraction $\Omega_{\phi}=\Omega_{\phi,\mathrm{max}}$,
can be interpreted as the lower bound on $g$, while the blue line corresponds to $\Omega_\phi=10^{-6}$.
In addition, we also show the current constraints on $g$ with shaded regions and the future sensitivities with dotted lines.
CAST~\cite{Anastassopoulos:2017ftl} and IAXO~\cite{Armengaud:2014gea,Irastorza:2018dyq} are axion helioscope experiments, Chandra~\cite{Berg:2016ese} and Athena~\cite{Conlon:2017ofb} are X-ray observatories, and
ALPSII~\cite{Bahre:2013ywa} is a light shining through a wall style experiment.
The ALP-photon coupling is also constrained by the conversion to photons in galactic magnetic fields of axions emitted by SN1987A~\cite{Payez:2014xsa}.

As for the behavior of $g$ in Fig.~\ref{fig: mass}, there are three
distinct regions depending on the mass $m$.
In the following, we will discuss each
of them in turn.

For $m \lesssim H_0$, the ALP field is nearly frozen until recently, in which
case $3H \dot{\bar{\phi}} \simeq -m^2 \bar{\phi}$. The field excursion $\Delta \bar{\phi}$
is approximately proportional to
$m^2 \bar{\phi}_{\rm obs}$, where
$\Omega_{\phi} \simeq m^2
\bar{\phi}_{\rm obs}^2/(6M_{\rm pl}^2 H_0^2)$.
This means that the rotation angle
(\ref{master eq}) has the dependence,
\begin{equation}
\left| \bar{\alpha} \right|
\propto g m^2
\left| \bar{\phi}_{\rm obs} \right|
\propto gm \sqrt{\Omega_{\phi}}\,,\qquad
(m \lesssim H_0)\,,
\label{ales1}
\end{equation}
where the constant of proportionality
can be determined numerically.

In the intermediate mass region
$H_0 \lesssim m \lesssim H_{\mathrm{LSS}}$,
$\bar\phi$ starts to oscillate at $m \simeq H_{\mathrm{osc}}$, where
$H_{\mathrm{osc}}<H_{\mathrm{LSS}}$.
The field value at the onset of oscillations (denoted as the scale
factor $a_{\rm osc}$) is practically
identical to $\langle \bar{\phi} \rangle_{\mathrm{LSS}}$, after which
the amplitude of $\bar{\phi}$ decreases
as $\Phi \propto a^{-3/2}$.
In this case $\bar\phi_\mathrm{obs}$ is negligible relative to $\langle \bar{\phi} \rangle_{\mathrm{LSS}}$, so that
$\bar{\alpha} \simeq -g\langle \bar{\phi} \rangle_{\mathrm{LSS}}/2$.
On using the matter-dominated
approximation ($a \propto H^{-2/3}$) to relate $\langle \bar{\phi} \rangle_{\mathrm{LSS}}$ with today's field
amplitude $\Phi_0$, we obtain
\begin{equation}
\langle \bar{\phi} \rangle_{\mathrm{LSS}} \simeq \left( \frac{a_{\mathrm{osc}}}{a_0} \right)^{-3/2} \Phi_0 \simeq
\frac{H_{\rm osc}}{H_0} \Phi_0
\simeq \frac{m}{H_0}\Phi_0\,.
\label{phire}
\end{equation}
Since $\sqrt{\Omega_{\phi}} \propto m \Phi_0$, it follows that
\begin{equation}
\left| \bar{\alpha} \right| \propto
g m \Phi_0
\propto g \sqrt{\Omega_{\phi}}\,,\qquad
(H_0 \lesssim m \lesssim H_{\mathrm{LSS}})\,,
\label{ales2}
\end{equation}
which means that $\left| \bar{\alpha} \right|$ does not depend on $m$.

For $m \gtrsim H_{\rm LSS}\simeq 3.3
\times 10^{-29}$\,eV, $\bar{\phi}$
starts to oscillate before the last
scattering epoch.
Up to the mass range
$m \lesssim 2.7 \times 10^{-27}~{\rm eV}
\simeq 80H_{\rm LSS}$,
$|\langle \bar{\phi} \rangle_{\mathrm{LSS}}|$ is still larger than
$|\bar{\phi}_{\rm obs}|$,
even though $\langle \bar{\phi} \rangle_{\mathrm{LSS}}$ is suppressed due to the oscillations of $\bar{\phi}$ around the LSS. As $m$ increases, the exponential
suppression of $\langle \bar{\phi} \rangle_{\mathrm{LSS}}$ tends to be
more significant.

For $m \gtrsim 2.7 \times
10^{-27}~{\rm eV} \simeq 80H_{\rm LSS}$,
$|\langle \bar{\phi} \rangle_{\mathrm{LSS}}|$ becomes smaller than $|\bar{\phi}_{\rm obs}|$.
In this regime, the rotational angle
has the dependence,
\begin{equation}
\left| \bar{\alpha} \right| \propto
g \left| \bar{\phi}_{\rm obs} \right|
\propto g m^{-1} \sqrt{\Omega_{\phi}}\,,\qquad
(m \gtrsim 2.7 \times
10^{-27}~{\rm eV})\,.
\label{ales3}
\end{equation}

Numerically, we obtain the constants of proportionality in Eqs.~(\ref{ales1}),
(\ref{ales2}), and (\ref{ales3}).
The resulting approximate expressions in three different regimes are given,
respectively, by
\begin{align}
    g =
    \begin{dcases}
        1.8 \times 10^{-18}\,\mathrm{GeV}^{-1}
            \left( \frac{|\bar \alpha|}{ 0.35 \,\mathrm{deg} } \right)
        \left( \frac{\Omega_{\phi}}{\Omega_{\Lambda}} \right)^{-1/2}
        \left( \frac{m/H_0}{10^{-2}} \right)^{-1},
        & (m\lesssim H_0),
\\
        1.5 \times 10^{-20}\,\mathrm{GeV}^{-1}
        \left( \frac{|\bar \alpha|}{ 0.35 \,\mathrm{deg} } \right)
        \left( \frac{\Omega_{\phi} h^2}{0.006} \right)^{-1/2},
        & (H_0 \lesssim m \lesssim H_{\mathrm{LSS}}),
\\
        1.8 \times 10^{-12}\,\mathrm{GeV}^{-1}
        \left( \frac{|\bar \alpha|}{ 0.35 \,\mathrm{deg} } \right)
        \left( \frac{\Omega_{\phi} h^2}{0.006} \right)^{-1/2}
        \left( \frac{m/H_0}{10^8} \right),
        & (m \gtrsim 2.7 \times 10^{-27} \, \mathrm{eV}).
    \end{dcases}
    \label{eq: mass m-g dependence}
\end{align}
For the mass range $H_{\rm LSS}<m \lesssim 2.7 \times 10^{-27}~{\rm eV}$,
$g$ exponentially increases with $m$,
while the dependence of $g$ on $|\bar{\alpha}|$ and $\Omega_{\phi}$ are the same as the third of
Eq.~\eqref{eq: mass m-g dependence}.
As we observe in Fig.~\ref{fig: mass},
the coupling $g$ generating the value
$\bar{\alpha} = 0.35$\,deg has
the mass dependence
$g \propto m^{-1}$ for $m \lesssim H_0$,
$g \propto m^0$ for
$H_0 \lesssim m \lesssim
H_{\mathrm{LSS}}$,
and $g \propto m$ for
$m \gtrsim 2.7 \times 10^{-27} \, \mathrm{eV}$.

As $\Omega_\phi$ decreases from
$\Omega_{\phi,{\rm max}}$,
the green line in Fig.~\ref{fig: mass}
moves upwards, i.e., to the region of
larger values of $g$.
Since $g$ is bounded from above
by Chandra measurements, the mass region which can explain the observed value of
$\bar\alpha$ is limited.
Combining the first of Eq.~(\ref{eq: mass m-g dependence})
with the observational bound by Chandra,
$g < 1.4 \times 10^{-12}\,\mathrm{GeV}^{-1}$, we obtain the constraint on the ALP mass, as
\begin{equation}
m> 1.8 \times 10^{-41}~\mathrm{eV}
\left( \frac{0.69}
{\Omega_{\phi}} \right)^{1/2}
\left( \frac{ \bar \alpha }{ 0.35 \,\mathrm{deg} } \right)\,.
\label{mbound}
\end{equation}
This result is, to our best knowledge, the first lower mass bound on a dark energy model.

In the literature, the time evolution of $\tilde{w}_{\phi}$ is often used to constrain the (effective) mass
of a quintessence field.
However, the observational allowed range
of $\tilde{w}_{\phi}$ is close to
the value $-1$ as we already mentioned
above, so it is difficult to distinguish
between quintessence and cosmological
constant from the observations of
supernovae type Ia and the
distant measurements of CMB and baryon acoustic oscillations.
In comparison to them, the lower ALP
mass bound (\ref{mbound}) accomplishes
the prominent sensitivity to the time variation of quintessence.
In a similar way, one can also derive the upper bound on the ALP mass, which is expected to be around
$10^{-25}$-$10^{-23}$ eV. Since it is beyond the applicable limit of Eq.~\eqref{eq: Omega phi constraint},
we leave its detailed calculation for future work.

Furthermore, as we see in Fig.~\ref{fig: mass}, the minimum value of $g$ is taken
in the intermediate mass region
$H_0 \lesssim m \lesssim H_{\mathrm{LSS}}$.
Applying the Chandra constraint on $g$
to the numerically calculated minimum $g$ for a fixed $\Omega_{\phi}$,
we obtain the lower bound on $\Omega_{\phi}$, as
\begin{equation}
    \Omega_{\phi} > 9.1 \times 10^{-19}\left( \frac{ \bar \alpha }{ 0.35 \,\mathrm{deg} } \right)^2\,,
    \label{Omega lowbound m}
\end{equation}
where the numerical coefficient is a bit
different from the one derived by the
second line of \eqref{eq: mass m-g dependence} because the actual minimum $g$ for a fixed $\Omega_{\phi}$ is slightly smaller.
It is particularly remarkable that
the observation of cosmic birefringence
gives rise to an extremely small lower bound of $\Omega_{\phi}$.
At the same time, it should be noted that Eq.~\eqref{Omega lowbound m} is derived under the assumption that
$\Delta\bar\phi$ generates the
observed value of $\bar\alpha$. The perturbation $\delta\phi_\mathrm{obs}$ sourced by an adiabatic mode
might provide a significant contribution to $\bar\alpha$ through Eq.~\eqref{alpha full expression}, and then the bound can be
subject to change. We will discuss such possibilities in Sec.~\ref{sec: discussion}.

\subsection{Axion potential}
\label{sec: axion potential}

In this section, we consider the ALP field with a cosine potential,
\begin{equation}
    V_{\cos}(\phi) = m^2f^2 \left[
    1 - \cos \left(\frac{\phi}{f}\right)\right]\,,
    \label{eq_V_cos}
\end{equation}
where $m$ and $f$ are constants having
a dimension of mass.
This potential is often used in the context of the QCD axion and ALP~\cite{Marsh:2015xka}.
Due to the periodicity of the
potential, we will consider the case in which the field initial value
$\bar{\phi}_{\mathrm{init}}$
is in the range $0 \leq |\bar{\phi}_{\mathrm{init}}|
\leq \pi f$.
We also choose $f=M_{\mathrm{pl}}$, but the similar calculation can be performed for
arbitrary values of $f$.

The difference from the quadratic potential (\ref{eq_V_quadratic}) is that
$V_{\rm cos}(\phi)$ is bounded from above,
as $V_{\rm cos}(\phi) \le 2m^2 f^2$.
Moreover, the potential (\ref{eq_V_cos}) has a plateau at $\phi=\pi f$, with
the inflection point at
$\phi=\pi f/2$.
In the region $\phi \ll f$, the potential
approximately reduces to the quadratic
one, i.e.,
$V_{\rm cos}(\phi) \simeq m^2 \phi^2/2$.
The background ALP field obeys
\begin{equation}
\ddot{\bar{\phi}} + 3H\dot{\bar{\phi}}
+ m^2 f \sin \left( \frac{\bar{\phi}}{f} \right)=0\,,
\label{phiaxi}
\end{equation}
where $H$ is given by Eq.~(\ref{Ha}).

\begin{figure}[t]
	\centering
	\includegraphics[width=.55\textwidth ]{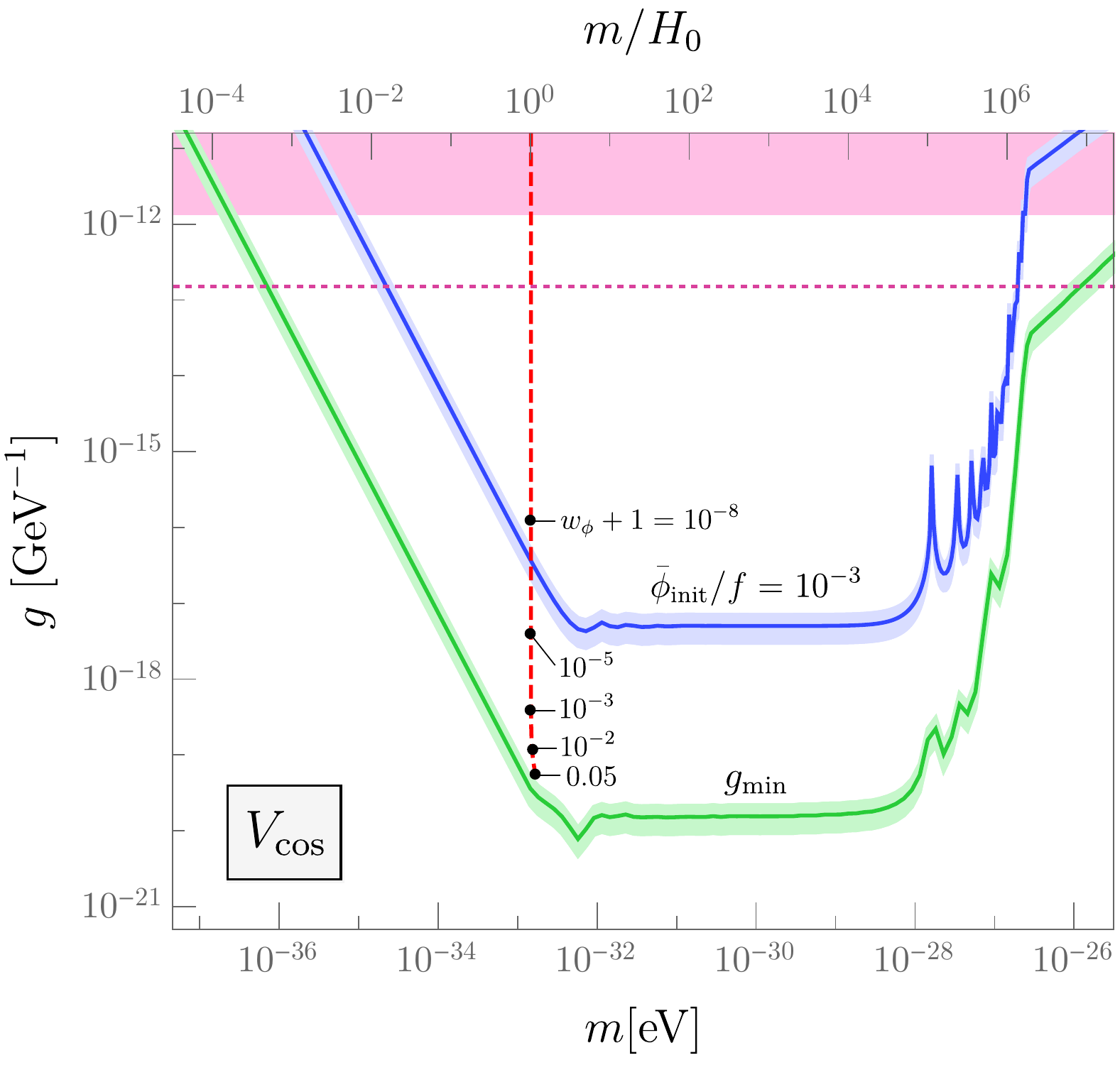}
	\caption{
	The ALP-photon coupling constant $g$ inferred by the isotropic birefringence $\bar{\alpha} = 0.35 \pm 0.14 \,\mathrm{deg}$ versus the ALP mass $m$ for the cosine potential
	$V_{\cos}(\phi)=m^2f^2[1-\cos(\phi/f)]$ with $f=M_{\rm pl}$.
	The green line corresponds to the case of minimum values of $g$ satisfying $\Omega_\phi \leq \Omega_{\phi,\mathrm{max}}$.
	The blue line adopts the initial condition $\bar\phi_{\mathrm{init}}/f
	= 10^{-3}$ and the parameter region above this line requires a fine tuning near $\bar\phi_{\mathrm{init}}\sim 0$.
	The ALP that accounts for all of dark energy by staying on the hilltop of the potential lies on the red dashed
	line. In this case, we show five different values of $w_{\phi}+1$ ($10^{-8}$, $10^{-5}$, $10^{-3}$, $10^{-2}$, 0.05) as the black dots, whose initial conditions $\bar\phi_\mathrm{init}$ can be found in Fig.~\ref{fig: axion m-phi contour}. Each point on and in the left side of the red dashed line has two corresponding values of $\bar\phi_\mathrm{init}$ in the regions $0<\bar{\phi}_{\rm init}/f<\pi/2$
	and $\pi/2<\bar{\phi}_{\rm init}/f<\pi$, with the degeneracy
	at the inflection point $\bar{\phi}_{\rm init}/f=\pi/2$
	on the green line.
	}
	\label{fig: axionic}
\end{figure}

For given values of $\bar{\alpha}$ and $m$, we expect that a larger $\bar{\phi}_{\mathrm{init}}$ leads to a
greater $\Delta \bar{\phi}$ or, equivalently, a smaller $g$.
However, the choice of a large $\bar{\phi}_{\mathrm{init}}$ can
give rise to a large $\Omega_{\phi}$
exceeding the maximum
$\Omega_{\phi,{\rm max}}$ given in Eq.~\eqref{eq: Omega phi constraint}.
Due to the non-linear property of the cosine potential, it is difficult to analytically relate $\Omega_{\phi}$ to $\bar{\phi}_{\mathrm{init}}$.
Therefore, for each $m$, we numerically scan over $\bar{\phi}_{\mathrm{init}}$ to find  minimum values of $g$ satisfying the condition $\Omega_\phi\le\Omega_{\phi,\mathrm{max}}$. The green line in Fig.~\ref{fig: axionic} corresponds to the minimum coupling $g_{\rm min}$ which accounts for the observed value
$\bar{\alpha}=0.35 \pm 0.14\,\mathrm{deg}$.

\begin{figure}[t]
	\centering
	\includegraphics[width=.45\textwidth ]{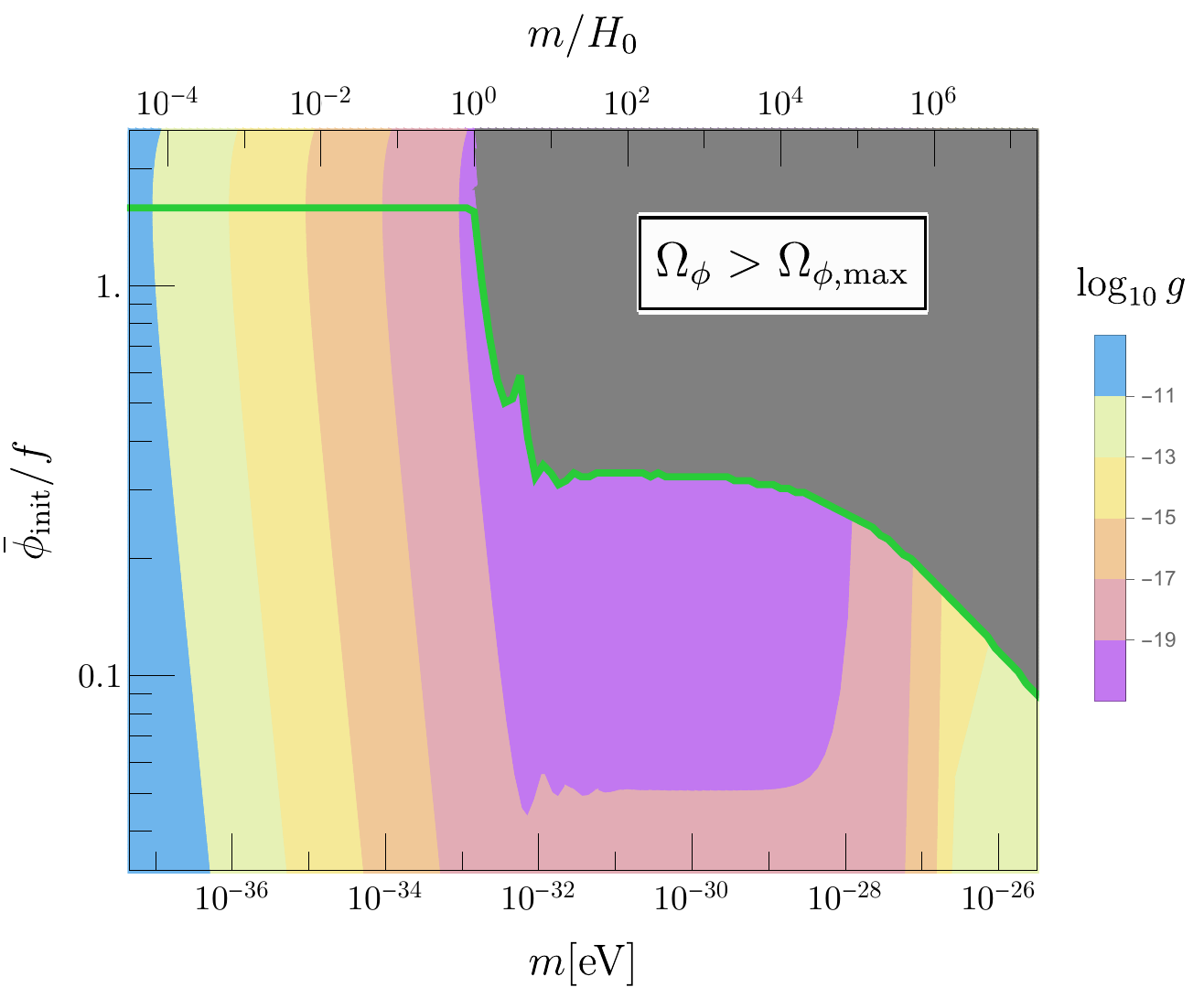}
    \hspace{8mm}
	\includegraphics[width=.45\textwidth ]{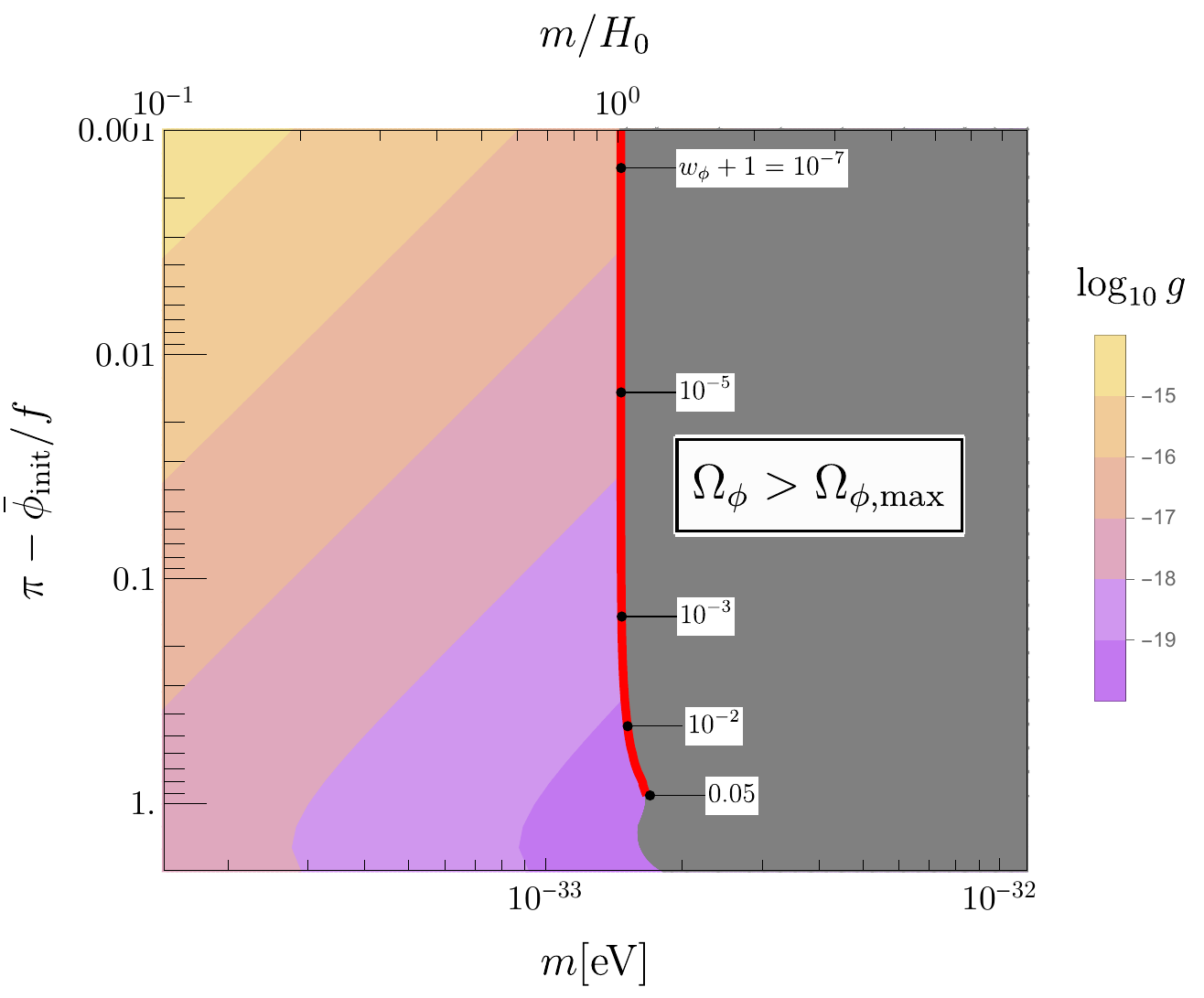}
	\caption{
	(Left panel)
	The contour of the ALP-photon coupling constant $g$ inferred by the isotropic birefringence
	$\bar{\alpha} = 0.35~\mathrm{deg}$
	in the ($m$, $\bar
	{\phi}_{\mathrm{init}}/f)$ plane for the cosine potential (\ref{eq_V_cos}) with $f=M_\text{pl}$.
	The gray region is excluded by the upper limit of $\Omega_{\phi}$.
	The green line represents $\bar{\phi}_{\mathrm{init}}$ that gives the minimum value of $g$ for a given mass,
	and it corresponds to the initial conditions of the green line in Fig.~\ref{fig: axionic}.
	(Right panel)
	The plot scheme is the same as the left panel except for the vertical axis that changes to focus on the hill-top initial condition, $\bar{\phi}_{\mathrm{init}} \sim
	\pi f$.
	The red line represents the ALP parameters accounting for all of dark energy and it corresponds to the
	red dashed line in Fig.~\ref{fig: axionic}.
	The black dots show five different values of $w_{\phi}+1$.}
	\label{fig: axion m-phi contour}
\end{figure}

In the left panel of Fig.~\ref{fig: axion m-phi contour}, we also show the theoretical line corresponding to the
coupling $g_{\rm min}$ in the ($m,\bar{\phi}_{\mathrm{init}}$) plane as the green line, over the contour of $g$ generating $\bar{\alpha} = 0.35~\mathrm{deg}$.
The gray region is excluded by
the violation of the condition $\Omega_{\phi} \le \Omega_{\phi,\mathrm{max}}$.
For $m \lesssim H_0$, the field value
that generates $g_{\rm min}$ is
$\bar{\phi}_{g_{\mathrm{min}}}
\simeq \pi f/2$.
This is because $\bar{\phi}$ slowly rolls down the potential until now in this mass region and the gradient of the potential is maximum at the inflection point,
$\bar{\phi} = \pi f/2$.
For $m \gtrsim H_0$, $\bar{\phi}_{g_{\mathrm{min}}}$ corresponds to $\Omega_{\phi} = \Omega_{\phi,\mathrm{max}}$ because $\bar{\phi}$ begins to oscillate by today, and the larger $\bar{\phi}_{\mathrm{init}}$ is, the greater $\Delta \bar{\phi}$ is.

The observational bound (\ref{alphabar value}) of $\bar{\alpha}$ infers
that, for smaller
$\bar{\phi}_{\mathrm{init}}$,
the coupling $g$ tends to be larger,
so that the allowed region is more
severely constrained by the Chandra bound.
Moreover, in the standard scenario, the homogeneous field $\bar{\phi}$ should
be generated by the misalignment mechanism, and $\bar{\phi}_{\mathrm{init}}/f$ is naturally expected to be of order unity.
Thus, the initial field displacement in the range $\bar{\phi}_{\mathrm{init}}/f \ll 1$ requires a fine tuning.
As a reference, we plot the theoretical
line corresponding to the initial condition $\bar{\phi}_{\mathrm{init}}/f = 10^{-3}$  as a blue line in Fig.~\ref{fig: axionic}.

The properties of theoretical lines in
Fig.~\ref{fig: axionic} look similar to those for the quadratic potential plotted in Fig.~\ref{fig: mass},
but there is the difference in the light
mass range $m \lesssim H_0$.
The third term on the left hand side of Eq.~(\ref{phiaxi}) approaches 0 as $\bar{\phi}$ increases toward $\pi f$,
so the field excursion $\Delta \bar{\phi}$
does not possess linear dependence in
$\bar{\phi}_{\mathrm{init}}$.
In comparison to the quadratic potential,
the coupling $g$ grows more rapidly
with the decrease of $m$.
The minimum coupling $g$ has
the following relation,
\begin{equation}
    g_{\rm min} =  3.6 \times 10^{-16}\,\mathrm{GeV}^{-1}
        \left( \frac{ \bar \alpha }{ 0.35 \,\mathrm{deg} } \right)
        \left( \frac{m/H_0}{10^{-2}} \right)^{-2},
        \quad (m\lesssim H_0)\,.
    \label{eq: cos m-g dependence}
\end{equation}
We note that the axion-photon coupling is typically given by $g = c_\phi \alpha_\text{EM}/f$, where $\alpha_\text{EM}\simeq 1/137$ is the fine structure constant and $c_\phi=\mathcal O(1)$ is a dimensionless constant.
Therefore, the tiny coupling constant
of order $g \sim 10^{-20}\,\mathrm{GeV}^{-1}$, which
corresponds to $m \sim H_0$ in
Eq.~(\ref{eq: cos m-g dependence}), can be naturally expected for the ALP with
$f\sim M_{\mathrm{pl}}$.
From Eq.~(\ref{eq: cos m-g dependence}),
the Chandra experiment gives the mass constraint,
\begin{equation}
m > 1.8 \times 10^{-37}\,\mathrm{eV}
\left( \frac{ \bar \alpha }{ 0.35 \,\mathrm{deg} } \right)^{1/2}\,,
\end{equation}
which is tighter than the bound
(\ref{mbound}).

For $m \gtrsim H_0$, the initial field value
corresponding to the coupling $g_{\rm min}$
is mostly in the range $\bar{\phi}_{\mathrm{init}}/f \lesssim 1$,
in which regime the potential approximately
reduces to $V_{\rm cos}(\phi) \simeq m^2 \phi^2/2$. Hence the field excursion from
the LSS to today is not different from
that for the quadratic potential studied
in Sec.~\ref{sec: quadratic potential}.
As we observe in Figs.~\ref{fig: mass}
and
\ref{fig: axionic}, the qualitative shapes of the minimum $g$ line in the two ALP potentials are similar to each
other for $m \gtrsim H_0$.
From the theoretical line
of $g_{\rm min}$
in the mass range $H_0<m<H_{\rm LSS}$, we obtain the constraint,
\begin{equation}
    \Omega_{\phi} > 1.2 \times 10^{-18}\left( \frac{ \bar \alpha }{ 0.35 \,\mathrm{deg} } \right)^2,
\end{equation}
which is close to the bound
(\ref{Omega lowbound m}) derived for the quadratic potential.

In the following, we study the
case in which the ALP accounts for
all of dark energy, i.e., $\Omega_{\phi}=\Omega_{\Lambda}=0.69$.
If the ALP field is near the top of
the potential during the epoch of cosmic
acceleration, the potential energy is
given by $V_{\cos}(\bar{\phi} \simeq \pi f) \simeq 2 m^2 f^2$. When this
is responsible for all of dark energy,
we require that
$2m^2 f^2=3M_{\rm pl}^2 H_0^2 \Omega_{\Lambda}$ and hence
\begin{equation}
m=\sqrt{\frac{3\Omega_{\Lambda}}{2}}
\frac{M_{\rm pl}}{f}H_0\,.
\label{mes}
\end{equation}
For $f=M_{\rm pl}$ and $\Omega_{\Lambda}=0.69$, it follows
that $m=1.017H_0$.
As the initial field value $\bar{\phi}_{\mathrm{init}}$ approaches
the inflection
point $\bar{\phi}=\pi f/2$,
$m$ slightly gets larger than
the value (\ref{mes}).
Moreover, if $\bar{\phi}_{\mathrm{init}}$
is close to $\pi f/2$, the observational
bound $w_{\phi}<-0.95$ tends to be
violated due to the large variation
of the ALP field.

In Fig.~\ref{fig: axionic}, the red dashed line corresponds to the case in which
the ALP field acts as all of dark energy.
We also show several values of
$w_{\phi}+1$ as black dots.
The corresponding line and the values
of $w_{\phi}+1$
are also plotted on the $(m, \pi - \bar{\phi}_{\mathrm{init}}/f)$ plane
in the right panel of Fig.~\ref{fig: axion m-phi contour}.
As $\bar{\phi}_{\mathrm{init}}$ approaches
$\pi f/2$, $w_{\phi}$ continues to
increase. The observational upper limit
$w_{\phi,{\rm max}}=0.95$ is reached around
$\bar{\phi}_{\mathrm{init}} \simeq 2.2f$.

For $\bar{\phi}_{\mathrm{init}}$ closer to
$\pi f$, the deviation of $w_{\phi}$ from
$-1$ decreases toward 0.
As we see in Fig.~\ref{fig: axionic},
even the tiny deviation like
$w_{\phi}+1=10^{-8}$ predicts the
coupling $g$ which is below the current
bound of Chandra.
If the future axion measurements were to detect the coupling $g$, this can provide a very interesting possibility for
probing the tiny deviation of the ALP
dark energy equation of state
$w_{\phi}$ from $-1$.
Under the slow-roll approximation where
the ALP kinetic energy is subdominant to
the potential energy, we have
\begin{align}
w_{\phi} =\frac{\dot{\bar{\phi}}^2/2-V}
{\dot{\bar{\phi}}^2/2+V}  \simeq -1+\frac{\dot{\bar{\phi}}^2}{V}\,,
\end{align}
where $V$ is related to $H_0$, as $3M_{\mathrm{pl}}^2 H_0^2 \simeq (1+\Omega_M/\Omega_\phi)V$.
Then, the ALP-photon coupling constant
can be estimated as
\begin{align}
    g = \frac{2|\bar\alpha|}
    {|\Delta\phi|}
    \simeq
    \frac{2|\bar\alpha|}{H_0^{-1}|
    \dot{\bar{\phi}}|}
    =1.6\times 10^{-20}\,\text{GeV}^{-1}
    \left(\frac{|\bar\alpha|}
    {0.35\,\text{deg}}\right)
    \left(\frac{0.05}{ 1+w_{\phi}  } \right)^{1/2},
    \label{eq_g_by_w}
\end{align}
where we assumed that the field excursion is dominated by the recent contribution, which is true for the thawing models of dark energy.
With this equation, the value
of $w_{\phi}$ can be estimated on the red dashed line in Fig.~\ref{fig: axionic}.
Applying the Chandra bound $g<1.4 \times
10^{-12}$~GeV$^{-1}$ to
Eq.~(\ref{eq_g_by_w}), it follows that
\begin{equation}
w_{\phi}+1>6.5 \times 10^{-18}
\left(\frac{|\bar\alpha|}
{0.35\,\text{deg}}\right)^2\,.
\label{wphibo}
\end{equation}
The field needs to vary at some extent
to explain the observed value
of isotropic cosmic birefringence.
It is interesting to note that the constraint (\ref{wphibo}) gives a
lower bound on $w_{\phi}$ larger
than $-1$. To our knowledge, this is
the first observational lower bound
of $w_{\phi}$ forbidding the cosmological constant value $-1$.

Finally, we should comment on the fact that one set of $m$ and $g$ does not necessarily determine a unique initial condition $\bar{\phi}_{\mathrm{init}}$. As we see in Fig.~\ref{fig: axion m-phi contour}, for the mass range
$m \lesssim H_0$, there are two values
of $\bar{\phi}_{\mathrm{init}}$
which correspond to the same $m$ and $g$.
One of those initial conditions
is in the region
$0<\bar{\phi}_{\rm init}/f<\pi/2$,
whereas the other is in the regime
$\pi/2<\bar{\phi}_{\rm init}/f<\pi$.
In Fig.~\ref{fig: axionic},
each set of $m$ and $g$ has two corresponding initial conditions $\bar{\phi}_{\mathrm{init}}$
in the left-side region of the red dashed line and above the green line.
On the green line, the two initial conditions are degenerate at the inflection point, $\bar{\phi}_{\mathrm{init}}/f=\pi/2$.

As the coupling $g$ increases along the red dashed line in Fig.~\ref{fig: axionic}, one of $\bar{\phi}_{\mathrm{init}}$ approaches the bottom of potential and the other does the top of potential.
As we already mentioned, the initial
condition $\bar{\phi}_{\rm init}/f \ll 1$
requires a fine tuning.
To obtain the values of $g$ whose orders
are the same in the two regimes $\bar{\phi}_{\rm init}/f \ll 1$
and $|\bar{\phi}_{\rm init}/f-\pi| \ll 1$,
we need the similar level of fine tuning for the ALP initial conditions.
The initial condition with
$|\bar{\phi}_{\rm init}/f-\pi| \ll 1$
corresponds to the case in which the
ALP field can be the source of all
of dark energy.
In the region above the blue line in
Fig.~\ref{fig: axionic}, we require the fine tuning
of $\bar{\phi}_{\mathrm{init}}$ in
both the two
regimes mentioned above.

\section{Early dark energy}
\label{sec: EDE}

Recently, some scalar-field models were
proposed to resolve or alleviate the  problem of $H_0$ tension between CMB and low-redshift measurements.
In these early dark energy (EDE) models, the scalar field is nearly frozen due to
the Hubble friction prior to a
critical scale factor $a_c$ of
order $a_{\rm eq}$, and it plays a role
of the cosmological constant with
$\tilde{w}_{\phi}$ close to $-1$.
The additional scalar-field energy density
increases the Hubble expansion rate
at early times, so the sound horizon
around the LSS is reduced by the presence
of EDE. Then, the models can be compatible
with the Planck data of CMB temperature
anisotropies with larger values of $H_0$.
For $a \geq a_c$, the scalar field
exhibits damped oscillations with
the energy density decaying faster than
those of standard matter components.
This is possible for the field potential
behaving like $V(\phi) \propto \phi^{2n}$
with $n \geq 2$ around its potential
minimum.
In this case, the contribution of EDE
to the late-time cosmic expansion
is negligible.

Previous works identified the
parameter space in which the existence of EDE remedies the Hubble tension.
In this section, we assume that the scalar field in EDE models is coupled to photon through the coupling $g\phi F_{\mu \nu}
\tilde{F}^{\mu \nu}/4$
and compute how much cosmic birefringence
is generated.
We will explore two EDE models,
(A) higher-order periodic potentials~\cite{Poulin:2018dzj,Poulin:2018cxd} and
(B) power-law potentials~\cite{Agrawal:2019lmo}, and
obtain the coupling constant for which the observed isotropic birefringence
is produced.
Therefore, the ALP not only ameliorates the Hubble tension but also explains the observed value of $\bar\alpha$
inside the parameter space derived below.

\subsection{Higher-order periodic potentials}
\label{sec: Kamion potential}

First, we consider the higher-order
periodic potentials studied in Refs.~\cite{Poulin:2018dzj,Poulin:2018cxd}:
\begin{equation}
V_{\cos}^{(n)}(\phi) = m^2f^2
\left[ 1 - \cos \left(\frac{\phi}{f}\right) \right]^n\,,
\label{earlyV1}
\end{equation}
where $f$ is the decay constant and we fix $f = M_{\mathrm{pl}}$ in the following. For $\phi\ll f$, this potential is well approximated by a power-law function $V_\mathrm{cos}^{(n)}(\phi) \simeq
(m^2 f^2/2^n)(\phi/f)^{2n}$.
In Ref.~\cite{Poulin:2018cxd}, the critical redshift $a_c$ is defined by
\begin{equation}
\rho_{\phi}(a_c) =\frac{ V_{\cos}^{(n)}
(\bar{\phi}_{\mathrm{init}})}{2}\,,
\end{equation}
which qualitatively indicates that the ALP field begins to oscillate at $a = a_c$.
The onset of oscillations can be also
roughly estimated as
$|V_{\cos_,\bar{\phi}\bar{\phi}}^{(n)}(\bar{\phi}_c)| \simeq H_c^2$,
where $\bar{\phi}_c$ and $H_c$ are the ALP background and the Hubble parameter at $a=a_c$, respectively.
In the regime $\bar{\phi} \ll f$, this
estimation
approximately translates to
$m^2 (\bar{\phi}_c/f)^{2(n-1)} \simeq H_c^2$.
For $n \geq 2$, $H_c$ is smaller than $m$.
The EDE field starts to oscillate
prior to the last scattering epoch,
so the mass $m$
should be in the range
\begin{equation}
m \gg H_{\rm LSS}
\simeq 3.3 \times 10^{-29}~{\rm eV}\,.
\label{massrange}
\end{equation}
Before and after the transition at $a=a_c$,
the field equation of state changes
from $\tilde{w}_{\phi} \simeq -1$ to
$\langle \tilde{w}_\phi \rangle \simeq(n-1)/(n+1)$,
where the latter is averaged over
oscillations.
For $a>a_c$ the energy density of $\phi$ decreases as $\rho_{\phi} \propto a^{-6n/(n+1)}$.
This means that, for $n \ge 2$, the contribution of $\rho_{\phi}$ to $H$ becomes negligible
compared to nonrelativistic matter.

In Ref.~\cite{Poulin:2018cxd}, the authors introduced the relative ratio between
the density parameters of EDE and
total matter at the transition, as
$f_{\mathrm{EDE}} \equiv \tilde{\Omega}_{\phi}(a_c)/
\tilde{\Omega}_{\mathrm{tot}}(a_c)$.
Running a Markov chain Monte Carlo (MCMC) simulation with flat priors on $\log_{10}(a_c), \Omega_{\phi}=\tilde{\Omega}_\phi(t_0), \bar\phi_\mathrm{init}$, and
six $\Lambda$CDM parameters,
they obtained the posterior
distributions of them.
For the likelihood analysis, they used
the observational data of SH0ES, Planck, 6dFGS, SDSS, BOSS DR12, and Pantheon.
The best-fit value of the Hubble constant
was found to be $H_0= 71.1\,\mathrm{km/s/Mpc}$ for $n=2$ and $H_0=71.6\,\mathrm{km/s/Mpc}$ for $n=3$,
so this EDE model can ease
the Hubble tension.
Here, we use their marginalized 2D posterior distributions of $\log_{10}(a_c)$ and $f_{\mathrm{EDE}}(a_c)$ with $n=2$ and $3$.

\begin{figure}[t]
\centering
\includegraphics[width=.45\textwidth ]{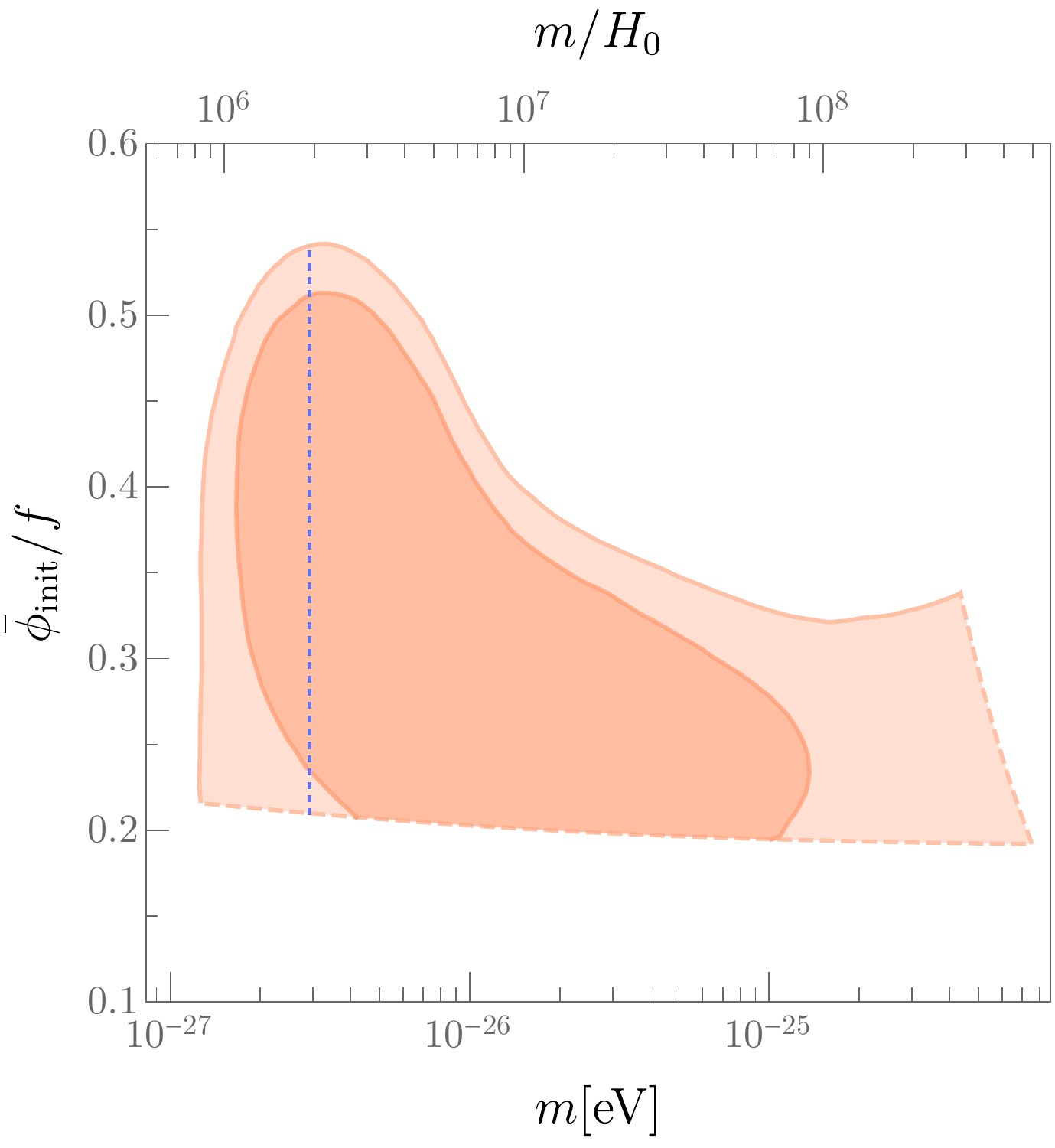}
	    \hspace{8mm}
   	    \includegraphics[width=.45\textwidth ]{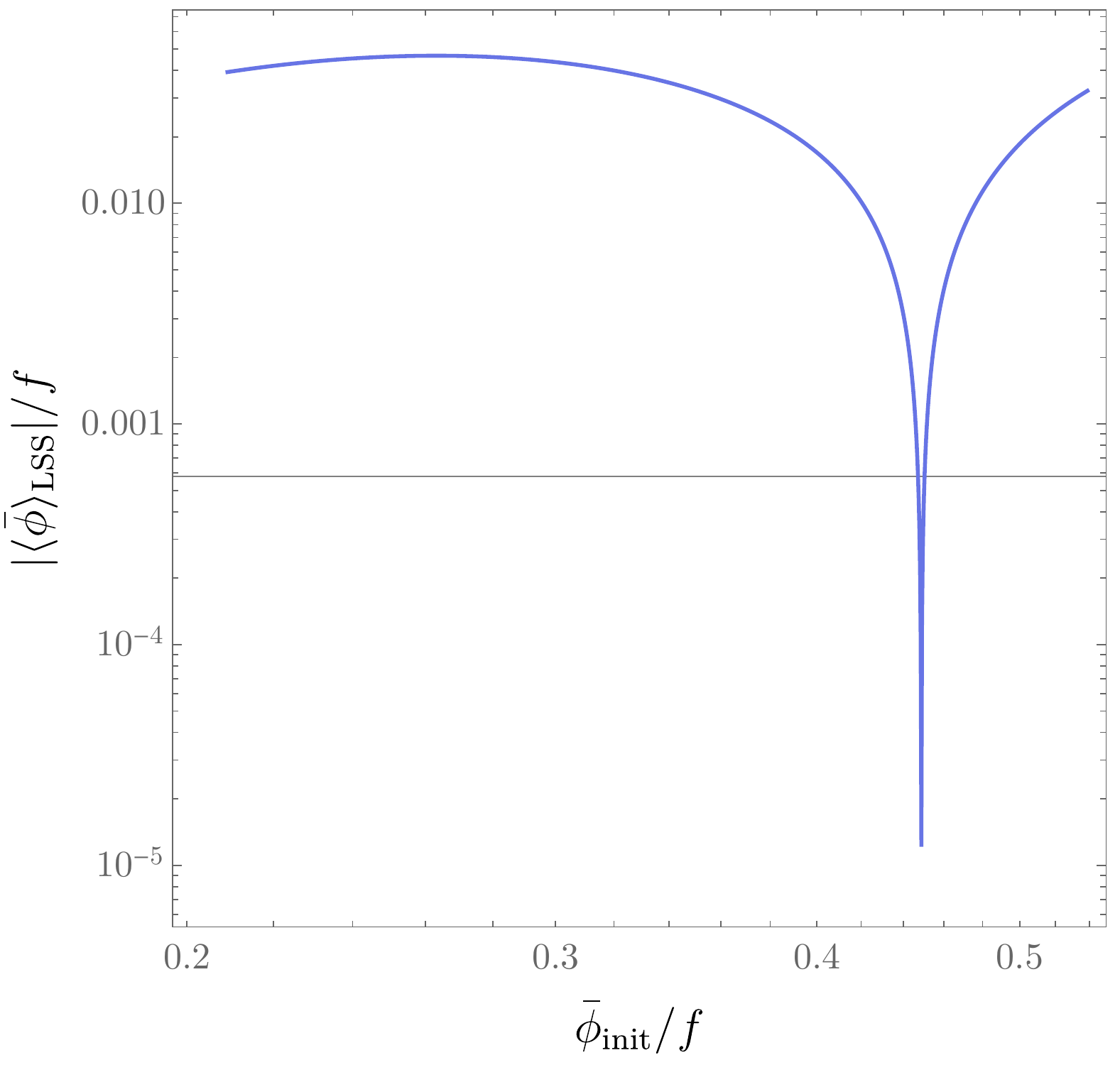}
	    \caption{
	    (Left panel) The $1\sigma$ and $2\sigma$ contours of the mass parameter $m$ and the initial EDE field value $\bar{\phi}_{\mathrm{init}}/f$
	    with the potential $V_\mathrm{cos}^{(2)}(\phi)$.
	    We convert the posterior distributions of $a_c$ and $f_{\mathrm{EDE}}(a_c)$ in Ref.~\cite{Poulin:2018cxd} by (inversely) solving the dynamics of $\bar{\phi}$.
        A sharp cut-off at the bottom and right edges corresponds to the boundary of the figure in Ref.~\cite{Poulin:2018cxd}.
        The blue dotted line denotes $m$
        and the domain of $\bar\phi_\mathrm{init}/f$ with which the right panel is depicted. (Right panel) $|\langle \bar{\phi} \rangle_{\mathrm{LSS}} |/f$ as a function of $\bar{\phi}_{\mathrm{init}}/f$ within the 2$\sigma$ contour for $m=2.9\times 10^{-27}\,\mathrm{eV}$.
	    This value of $m$ corresponds to the best-fit values of $a_c$ and $f_{\mathrm{EDE}}(a_c)$.
	    The horizontal gray line denotes the threshold defined in the main text.
	    The region below this threshold includes only $1\,\%$ of the $\bar{\phi}_{\mathrm{init}}/f$ domain and it is less likely to have such an initial value, which accidentally suppresses $|\langle \bar{\phi} \rangle_{\mathrm{LSS}} |/f$.
	    }
\label{fig: period m-phi}
\end{figure}

To calculate the field excursion $\Delta\bar\phi$ and derive the produced isotropic birefringence
in this model, one needs to know the
mass parameter $m$ and the initial field value $\bar\phi_\mathrm{init}$.
Numerically solving the dynamics of $\bar\phi$ for varying parameters, we convert the posterior distributions
of $a_c$ and $f_{\mathrm{EDE}}(a_c)$ in Ref.~\cite{Poulin:2018cxd} to
the distribution of $m$ and $\bar{\phi}_{\mathrm{init}}$.
The result is shown in the left panel of Fig.~\ref{fig: period m-phi}.
In Ref.~\cite{Poulin:2018cxd} the
posterior distribution was presented only for $f_{\mathrm{EDE}}(a_c) > 0.01$, presumably because $\phi$ is not effective to reduce
the Hubble tension
for a smaller energy fraction.
Following Ref.~\cite{Poulin:2018cxd}, we disregard the distribution for $f_{\mathrm{EDE}}(a_c) < 0.01$, and hence the contour in the left panel of Fig.~\ref{fig: period m-phi} has a sharp cut-off at its bottom edge.

For the derivation of $\bar{\alpha}$ in Eq.~\eqref{master eq}, we also compute $\langle \bar{\phi} \rangle_{\mathrm{LSS}}$ by integrating Eq.~(\ref{fieldeq2}).
As we already discussed in Sec.~\ref{sec: CB}, the suppression of $\langle \bar{\phi} \rangle_{\mathrm{LSS}}$ by the fast oscillation of $\bar{\phi}$ around the LSS
also occurs for the present potential
with the mass scale (\ref{massrange}).
To illustrate this effect, we show
the averaged value $\langle \bar{\phi} \rangle_{\mathrm{LSS}}$ as a function of the initial value $\bar\phi_\mathrm{init}$ for $m=2.9\times 10^{-27}$~eV
in the right panel of
Fig.~\ref{fig: period m-phi}.
Since $\langle \bar{\phi} \rangle_{\mathrm{LSS}}$ is obtained by convoluting the oscillating field $\bar{\phi}$ with the positive visibility function, its sign can be positive or negative depending on the phase of  $\bar{\phi}$.
When we change $\bar{\phi}_{\mathrm{init}}/f$ continuously, the oscillation phase slides and the sign of $\langle \bar{\phi} \rangle_{\mathrm{LSS}}$ flips at certain values of $\bar{\phi}_{\mathrm{init}}/f$.
Therefore $|\langle \bar{\phi} \rangle_{\mathrm{LSS}}|$
passes through zero for these initial conditions, which appear as a sharp
dip in the right panel of Fig.~\ref{fig: period m-phi}.
For such $\bar{\phi}_{\mathrm{init}}$, $\Delta \bar{\phi}$ is also subject to suppression and one apparently needs
a large $g$ to account for the
observed $\bar{\alpha}$.
Nevertheless, it is less likely to have such specifically small values of $|\langle \bar{\phi} \rangle_{\mathrm{LSS}}|$ with high precision by chance.
To quantify this fine tuning,
we introduce a threshold of $|\langle \bar{\phi} \rangle_{\mathrm{LSS}}|$ above which $99\,\%$ of the interval of $\bar{\phi}_{\mathrm{init}}/f$
within the $1\sigma$ or $2\sigma$ contour is included for a given $m$.
This threshold is shown as a solid horizontal line in the right panel of Fig.~\ref{fig: period m-phi} and the probability to realize
$|\langle \bar{\phi} \rangle_{\mathrm{LSS}}|$ less than this threshold is smaller than 1\,\%.

\begin{figure}[ht]
	    \centering
	    \includegraphics[height=3.0in,width=3.5in]{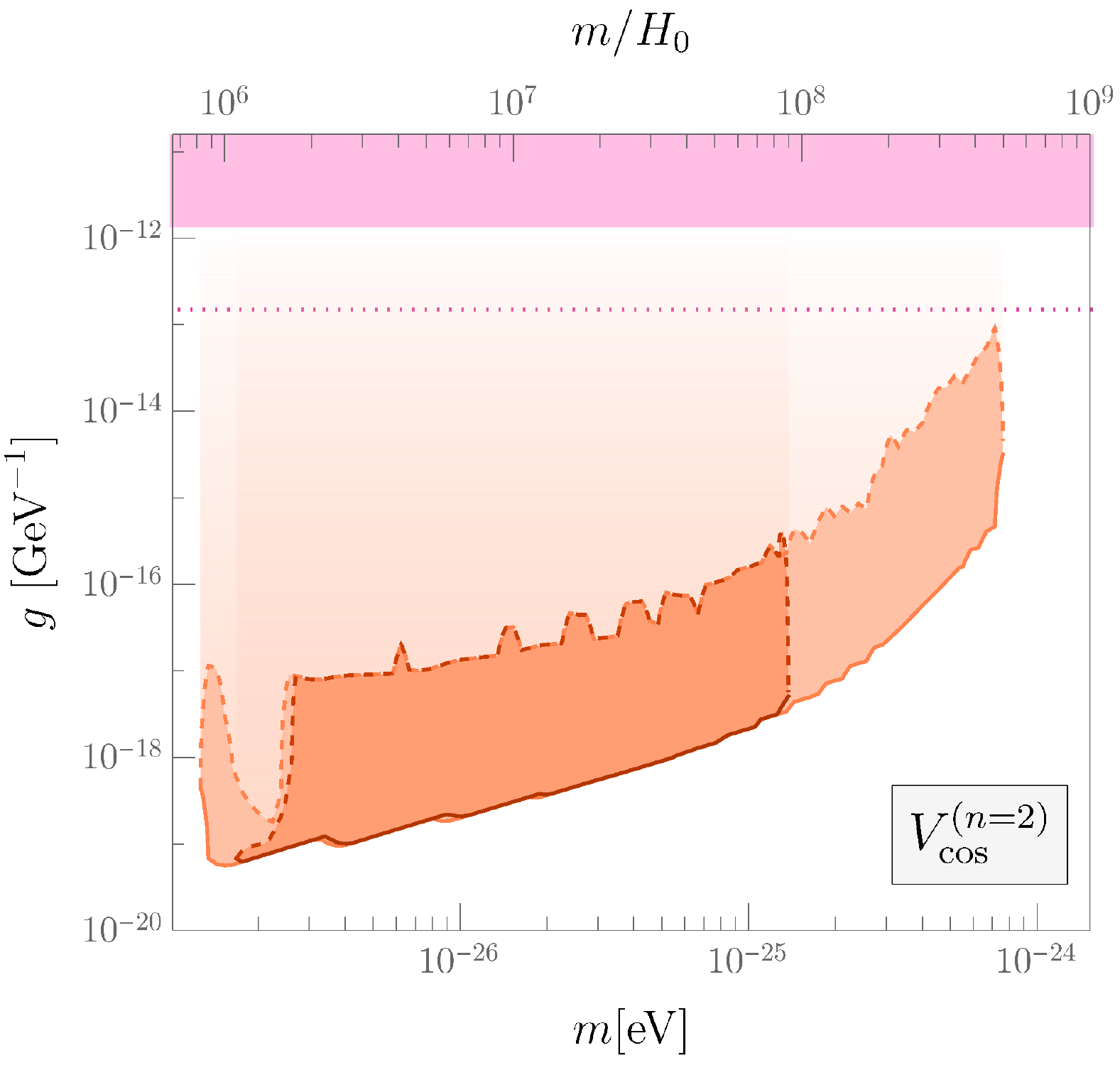}
	    \includegraphics[height=3.0in,width=3.5in]{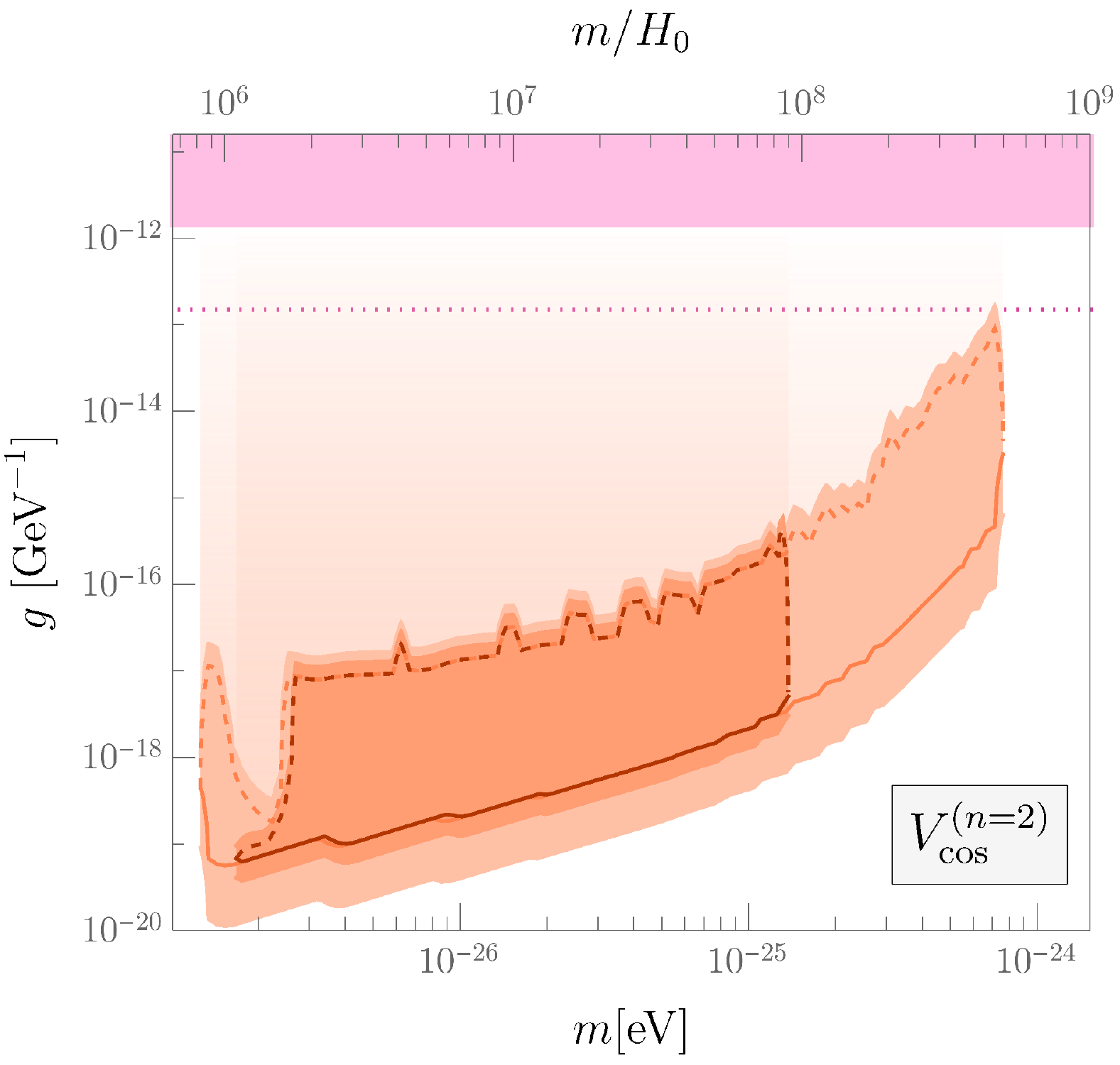}
	    \caption{
	    (Left panel) The orange shaded regions denote the $1\sigma$ (dark) and $2\sigma$ (light) contours
	    in the ($m$, $g$) plane for the EDE model with the higher-order periodic potential $V_{\cos}^{(n=2)}(\phi)$.
	    We fix the cosmic birefringence
	    to be the observed best-fit value $\bar{\alpha} = 0.35\,\mathrm{deg}$. The dashed lines denote the fine-tuning threshold below which $99\,\%$ of the initial value $\bar{\phi}_{\mathrm{init}}$ is included. As $g$ goes higher than the threshold line, it becomes less likely to realize, although it is not rigorously excluded. This feature is expressed by the gradation of the light orange color above the contour. The pink shaded region is constrained by Chandra. The pink dotted line is the projected sensitivity of Athena.
	    (Right panel) The dark and light orange regions are the $1\sigma$ and $2\sigma$ EDE model contours extended by the $1\sigma$ and $2\sigma$ uncertainties of
	    the observed $\bar\alpha$
	    (i.e., $\delta\bar\alpha=0.14$ deg and $\delta\bar\alpha=0.28$ deg), respectively.
	    The solid, dashed, and dotted lines are all the same as those in
	    the left panel.
    	}
    	\label{fig: period n2 g}
\end{figure}
\begin{figure}[ht]
	    \centering
	    \includegraphics[height=3.0in,width=3.5in]{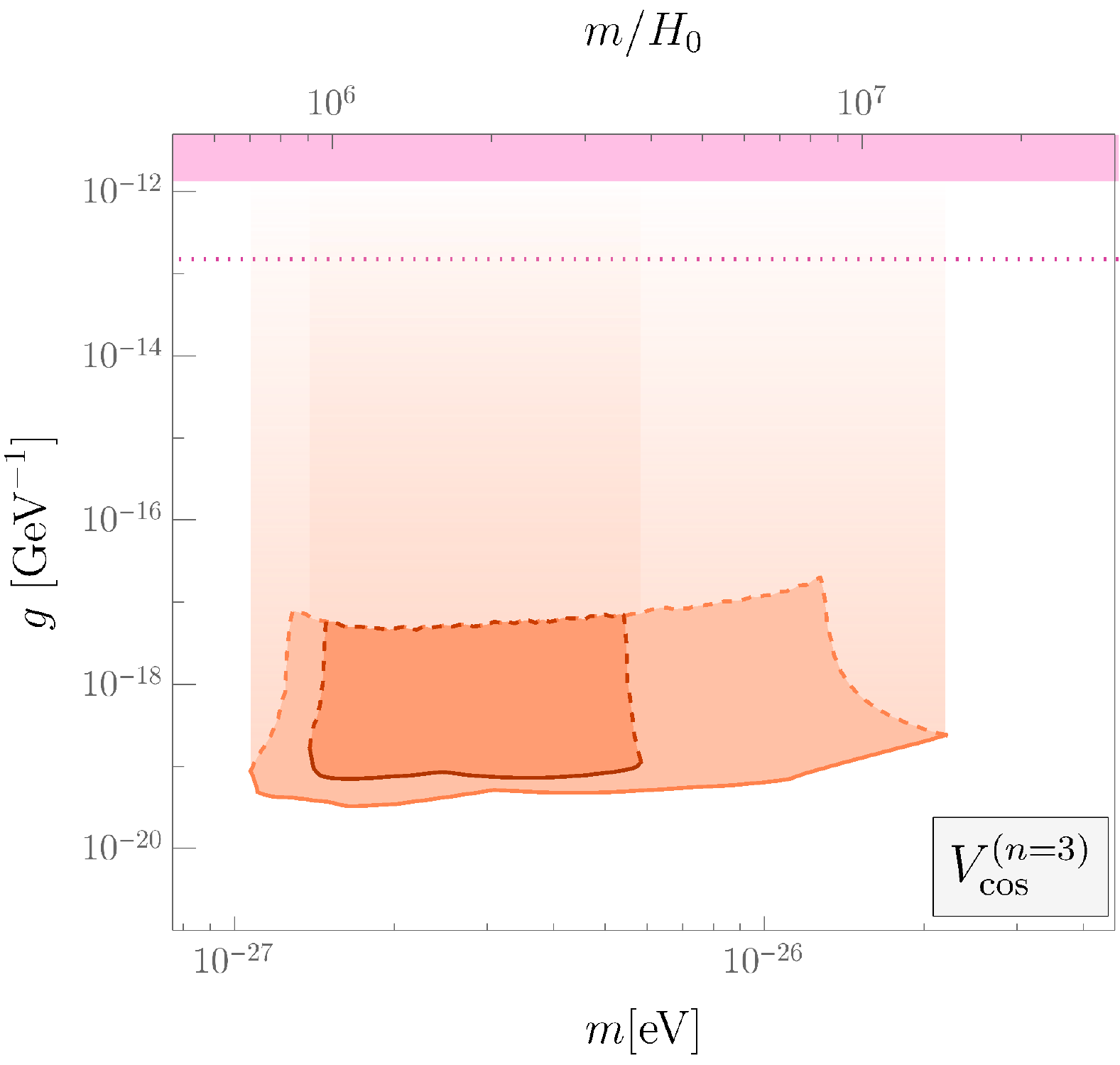}
	    \includegraphics[height=3.0in,width=3.5in]{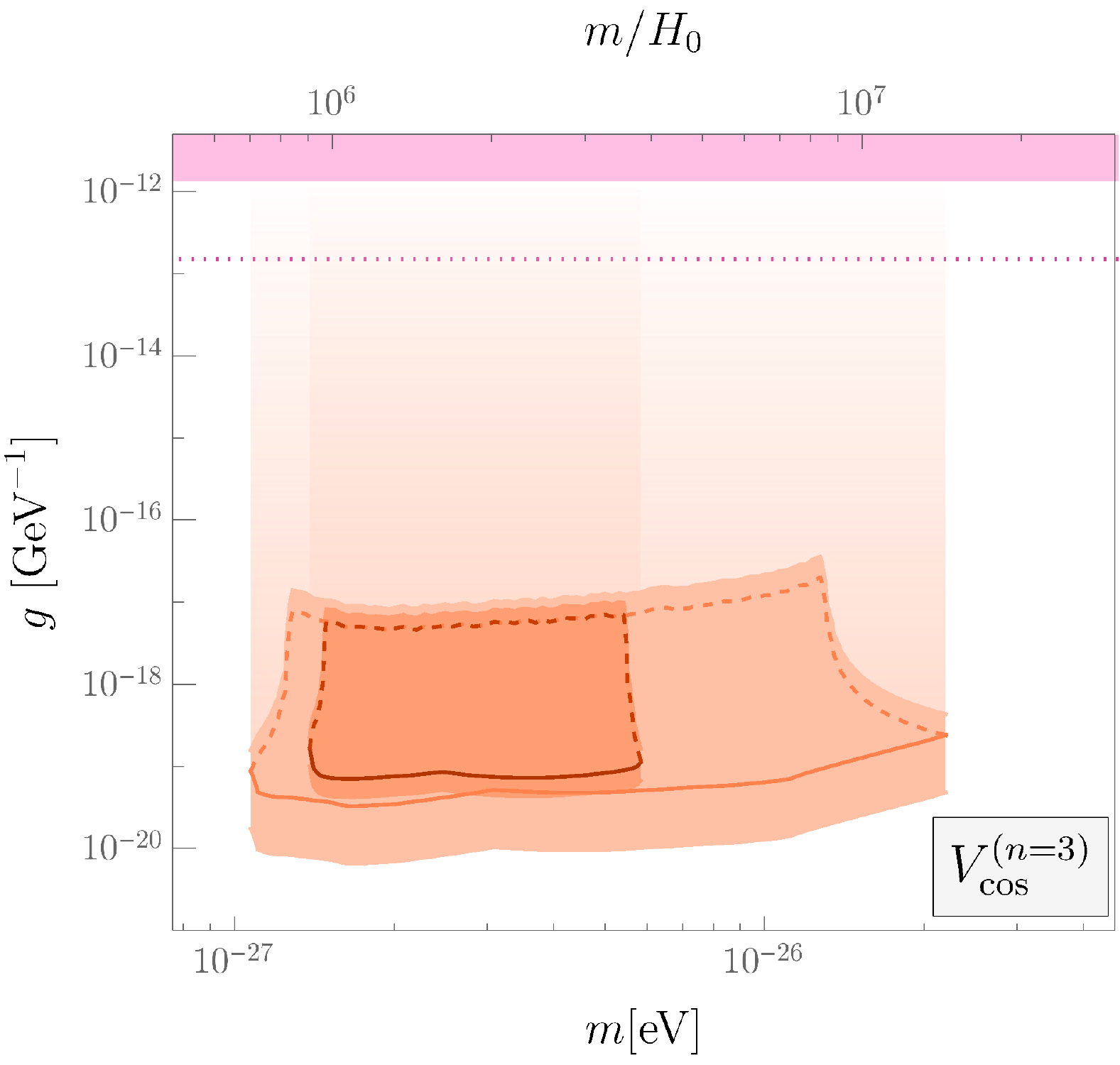}
	    \caption{
	    The plot scheme is the same as Fig.~\ref{fig: period n2 g}, while the EDE model potential is replaced by $V_{\cos}^{(n=3)}(\phi)$.
    	}
    	\label{fig: period n3 g}
\end{figure}

In Figs.~\ref{fig: period n2 g}
and~\ref{fig: period n3 g}, we plot the
$1\sigma$ and $2\sigma$ contours in the
($m$, $g$) plane which explain
the observed $\bar\alpha$ and also reduce the Hubble tension for the potential
(\ref{earlyV1}) with $n=2$ and $3$.
Since $|\langle \bar{\phi} \rangle_{\mathrm{LSS}}|$ can vanish for some special initial conditions of $\bar{\phi}$, the contour of $m$ and $\bar\phi_\mathrm{init}/f$ in the left panel
of Fig.~\ref{fig: period m-phi} does not put
an upper bound on $g$ in a rigorous sense.
Thus we show the $1\sigma$ and $2\sigma$ regions enclosed by the lower bound from the EDE contour (solid line) and the sketchy upper bound inferred by the fine-tuning threshold discussed above (dashed line). Although there is a small chance to have $g$ larger than the threshold, the probability rapidly decreases as $g$ increases.
This feature is expressed by the gradation of the light orange color above the contour.

The left panels of Figs.~\ref{fig: period n2 g} and~\ref{fig: period n3 g} do not take into account the uncertainty of
$\bar\alpha$, but we merely use its best-fit value, $\bar\alpha=0.35$ deg.
It is tricky to combine the uncertainties of the EDE model parameters and that of $\bar\alpha$, because we do not have the complete information of their posterior distributions.
To present conservative contour plots, we sweep from $\bar\alpha=(0.35-0.14)\,\mathrm{deg}$ to $(0.35+0.14)\,\mathrm{deg}$ for the $1\sigma$ EDE contour and from $\bar\alpha=(0.35-0.28)\,\mathrm{deg}$ to $(0.35+0.28)\,\mathrm{deg}$ for the $2\sigma$ EDE contour
in the right panels of Figs.~\ref{fig: period n2 g} and~\ref{fig: period n3 g}.
This treatment does not follow the general rule of error propagation, so it can  overestimate the uncertainty
to some extent.
Thus one should consider the contours in these right panels as crude but conservative constraints which do not exactly correspond to $1\sigma$ or $2\sigma$.

\subsection{Rock `n' roll model}
\label{sec: RnR}

As another interesting proposal of the EDE potential, we consider the rock `n' roll model studied in Ref.~\cite{Agrawal:2019lmo}:
\begin{equation}
    V_{\mathrm{RnR}}^{(n)}(\phi) = V_0 \left( \frac{\phi}{M_{\mathrm{Pl}}} \right)^{2n}
    =
    \frac{m^2 M_{\mathrm{Pl}}^2}{2^n} \left( \frac{\phi}{M_{\mathrm{Pl}}} \right)^{2n}\,,
    \label{earlyV2}
\end{equation}
where $V_0$, $n$, and $m$ are constants.
Here, we defined the mass $m$ such that
the higher-order periodic potential
(\ref{earlyV1}) with $f = M_{\mathrm{pl}}$ asymptotes to (\ref{earlyV2})
in the vicinity of the origin, $\phi = 0$.
For $n \ge 2$, a scalar field with this potential can work as the source of EDE.

In Ref.~\cite{Agrawal:2019lmo}, the authors introduced two parameters, the critical redshift $a_c$ defined by
$V_{\mathrm{RnR},\bar{\phi}\bar{\phi}}^{(n)}
(\bar{\phi}_c)=9H^2(a_c)$, and the energy fraction of the EDE potential to the total energy at $a_c$, $f_{\phi} \equiv V_{\mathrm{RnR}}^{(n)}(a_c)/\rho_{\mathrm{tot}}(a_c)$.
These parameters are defined in a slightly
different way in comparison to those
for the potential (\ref{earlyV1}).
Using the datasets of SH0ES, Planck, 6dFGS, SDSS, BOSS DR12, and Pantheon and
running a MCMC simulation with flat priors on $\log_{10}(a_c), f_{\phi}$, and six $\Lambda$CDM parameters,
they obtained the posterior
distributions of $a_c$ and $f_{\phi}$.
For $n=2$ the best-fit value of the
Hubble constant was found to be $H_0= 70.5\,\mathrm{km/s/Mpc}$, so
this EDE model also reduces the
Hubble tension.
Here, we use their marginalized 2D posterior distributions of $\log_{10}(a_c)$ and $f_{\phi}$ with $n=2$.

\begin{figure}[t]
	    \centering
	    \includegraphics[width=.45\textwidth ]{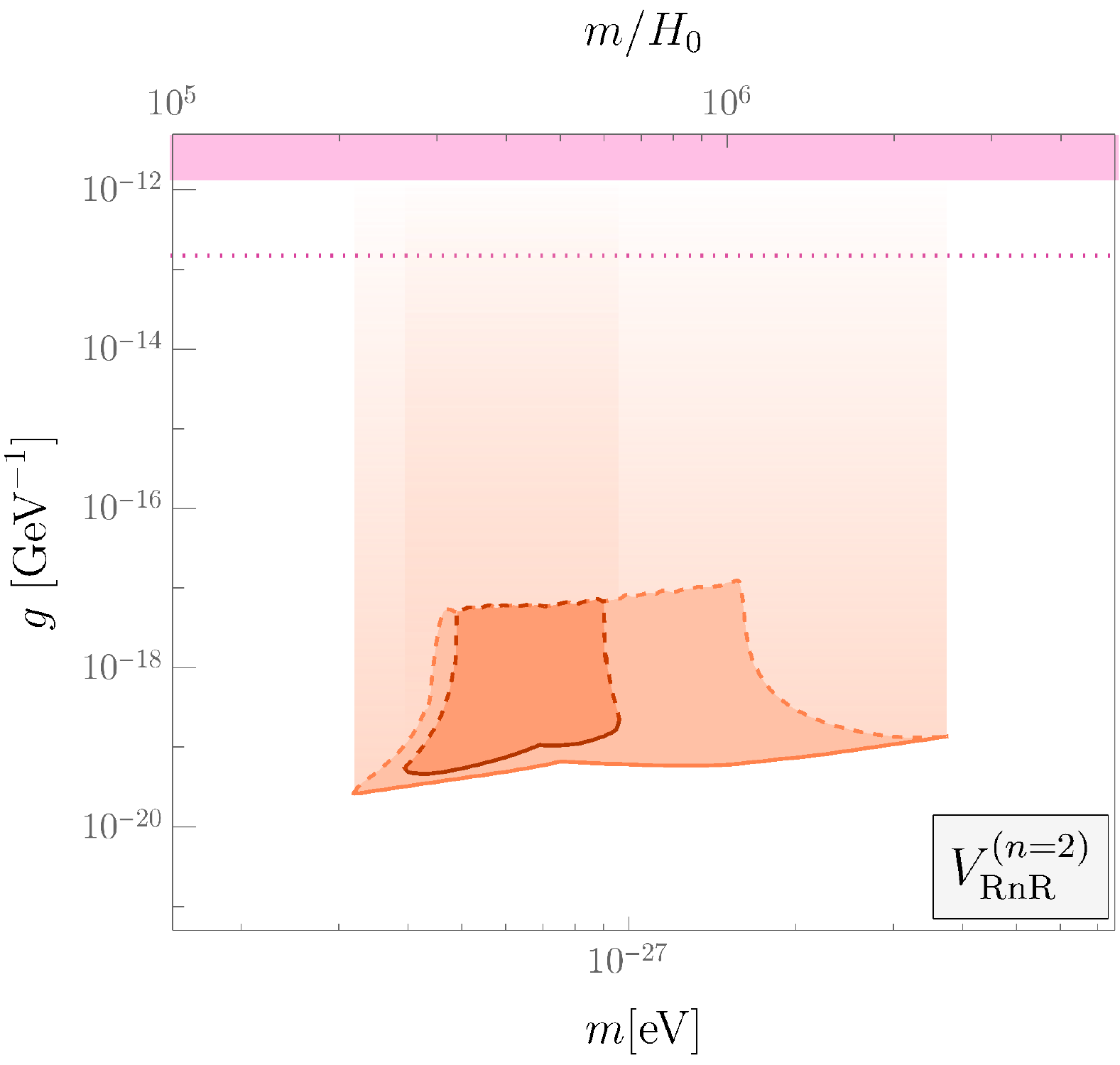}
   	    \includegraphics[width=.45\textwidth ]{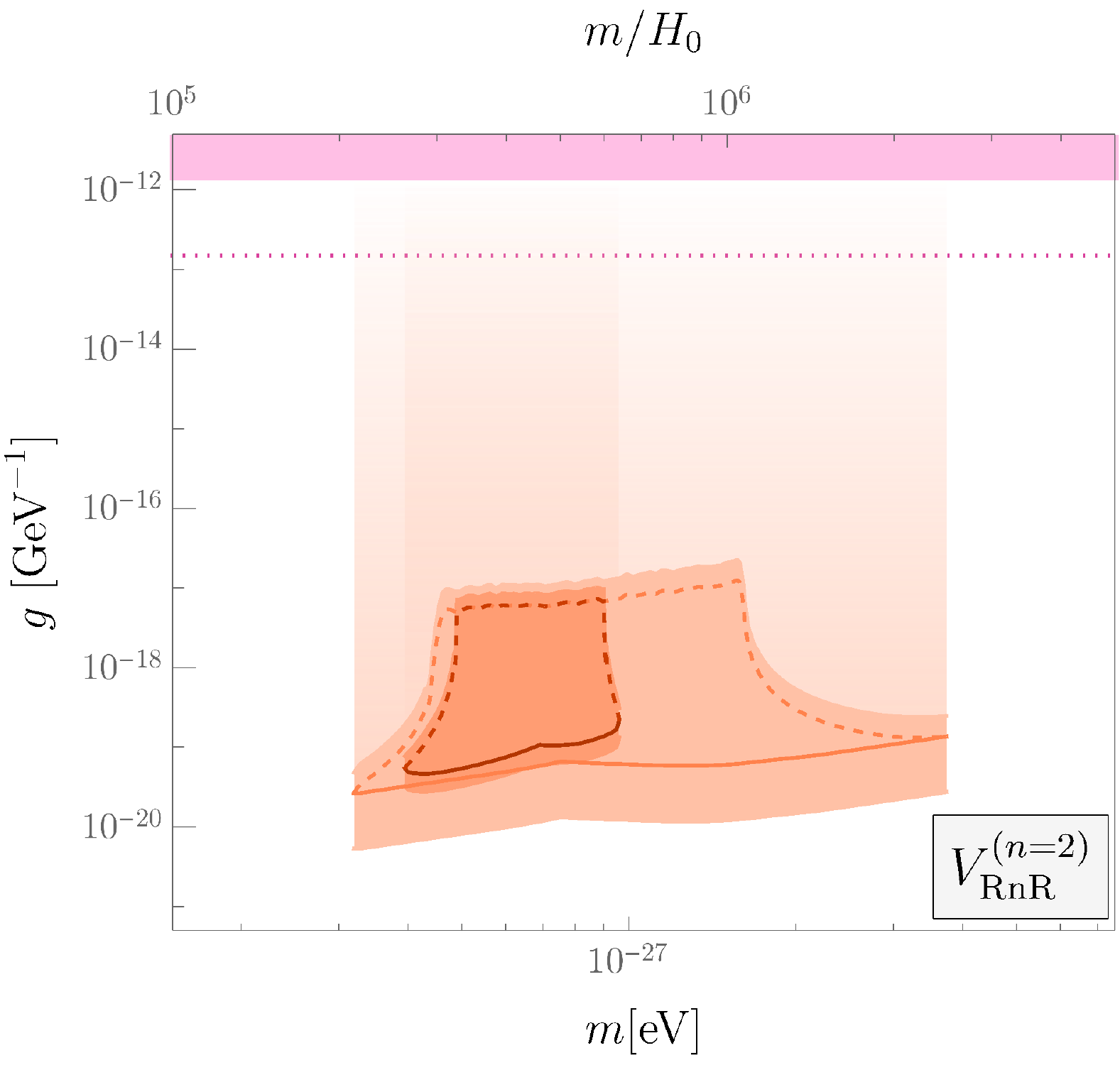}
	    \caption{
	    The plot scheme is the same as Figs.~\ref{fig: period n2 g} and \ref{fig: period n3 g}, while the rock `n' roll model with $n=2$ is employed.
    	}
    	\label{fig: RnR g}
\end{figure}

Analogous to the discussion in Sec.~\ref{sec: Kamion potential}, we translate the $(a_c, f_{\phi})$ contour
derived in Ref.~\cite{Agrawal:2019lmo} to
the distribution of $m$ and $\bar{\phi}_{\mathrm{init}}$ by solving the dynamics of $\bar{\phi}$.
In this step, we disregard the region of $(a_c, f_{\phi})$ with $f_{\phi}<0.01$ for the same reason explained in Sec.~\ref{sec: Kamion potential}.
Our results are shown in Fig.~\ref{fig: RnR g}. Compared to Fig.~\ref{fig: period n2 g} whose potential $V_{\cos}^{(2)}(\phi)$ asymptotes $V_{\mathrm{RnR}}^{(2)}(\phi)$ around the origin, the contour in Fig.~\ref{fig: RnR g} favors a lower mass $m$, with
the shrink of an allowed region of $m$.
Note that the treatments of the ALP dynamics are not identical to each other between Ref.~\cite{Poulin:2018cxd} and
\cite{Agrawal:2019lmo}, which might lead to the deviation of the results in addition to the intrinsic model difference.

\subsection{Common features in two EDE models}

We discuss some common results in the two EDE models presented in Secs.~\ref{sec: Kamion potential} and \ref{sec: RnR}.
The lower bounds on $g$ consistent with the
observed value of $\bar{\alpha}$ can be
seen in Figs.~\ref{fig: period n2 g}-\ref{fig: RnR g}. A careful reader may notice that these lower limits are even smaller than those of
the simple models obtained
in Sec.~\ref{sec: DE ALP} for similar values of $m$
(see the green lines in Figs.~\ref{fig: mass} and \ref{fig: axionic}).
This is because the suppression by the time averaging in Eq.~\eqref{eq_LSSwashout} is less significant in the EDE models.
The effective mass squared of the EDE
field around $\phi=0$ is given by
$V_{,\phi \phi}(\phi) \sim m^2(\phi/f)^{2(n-1)}$, so it decreases as
the amplitude of $\bar{\phi}$ decays
during the oscillation.
Hence the EDE field oscillates more slowly
in comparison to the scalar field with
a constant mass $m$.
Unlike the models studied
in Sec.~\ref{sec: DE ALP}, the values of $|\langle \bar{\phi} \rangle_{\mathrm{LSS}}|$
are typically larger than
$|\bar{\phi}_{\rm obs}|$
for the constrained model parameters
shown in
Figs.~\ref{fig: period n2 g}-\ref{fig: RnR g}
(apart from the 1\,\% of the dip of
$|\langle \bar{\phi} \rangle_{\mathrm{LSS}}|$
seen in the right panel of Fig.~\ref{fig: period m-phi}).

One also observes in Figs.~\ref{fig: period n2 g}-\ref{fig: RnR g} that the coupling constant $g$ accounting for the observed $\bar{\alpha}$
and reducing the Hubble tension is much smaller than the current Chandra bound
$g < 1.4\times 10^{-12}$\,GeV$^{-1}$.
Unfortunately, it looks difficult for the future observation by Athena to detect a signal of
the axion-photon coupling.
It is worth considering how to confirm our scenario where the scalar field in the EDE models with the photon coupling produces the isotropic birefringence. We will discuss
this issue in Sec.~\ref{sec: discussion}.

Finally, it is interesting to note that the contours in Figs.~\ref{fig: period n2 g}-\ref{fig: RnR g} are centered at
\begin{equation}
g\sim 10^{-18}~{\rm GeV}^{-1}\,.
\label{eq_typical_g_EDE}
\end{equation}
This means that $gf$ is a dimensionless number of order unity. We can rewrite
Eq.~\eqref{master eq}, as
\begin{equation}
    gf=\frac{2\bar\alpha}{\Delta\bar\phi/f}\simeq \frac{2\bar\alpha}{\langle \bar{\phi} \rangle_{\mathrm{LSS}}/f}\,,
\end{equation}
where we ignored the sub-leading contribution from $\bar\phi_\mathrm{obs}$.
The observed value of isotropic birefringence is a small number in radians,
$\bar\alpha \simeq 6\times 10^{-3}$,
and $\langle \bar{\phi} \rangle_{\mathrm{LSS}}$ is subject to suppression by the time averaging.
Therefore the fact that $gf=\mathcal{O}(1)$ appears to be quite non-trivial and it may imply something about Planck scale physics.
Although we fix $f=M_\mathrm{pl}$ in this paper, the same exercises can be done for other
values of $f$.

\section{Discussion}
\label{sec: discussion}

In this section, we discuss the contributions to the CMB birefringence from the ALP fluctuation, namely $\delta\phi_\mathrm{LSS}$ and $\delta\phi_\mathrm{obs}$ in Eq.~\eqref{alpha full expression},
which have been ignored so far.
The ALP fluctuation has two possible origins: One is the
quantum perturbation produced during inflation $\delta\phi^\mathrm{(inf)}$, and the other is the sourced perturbation induced by the adiabatic mode during the background ALP evolution $\delta\phi^\mathrm{(src)}$.

Let us mainly consider the massive ALP potential given by Eq.~(\ref{eq_V_quadratic}).
Since we are interested in the mass scale $m$ much smaller than the inflationary Hubble parameter $H_\mathrm{inf}$, the amplitude of $\delta\phi^\mathrm{(inf)}$
is of the order $\sim H_\mathrm{inf}/(2\pi)$, when the
primordial perturbation is produced
around the Hubble radius crossing.
However, the evolution of  $\delta\phi^\mathrm{(src)}$ as well as
$\delta\phi^\mathrm{(inf)}$ highly depends on the models and their parameters.
It requires a dedicated investigation to calculate their contributions to the birefringence in the full range
of $m$, which is beyond the scope
of this paper.

Instead, we split the parameter range into three parts,
(i) the light region: $m\lesssim H_0$,
(ii) the heavy region:
$m\gtrsim H_\mathrm{LSS}$, and
(iii) the intermediate region:
$H_0 \lesssim m \lesssim H_\mathrm{LSS}$, and briefly explore them in order.
We give simple estimations of the fluctuations and
make comments on their potential effects on the cosmic birefringence.

In the mass region $m\lesssim H_0$,
the sourced perturbation
is not relevant to the observed $\bar\alpha$
because the source term for $\delta\phi$ is proportional to $\dot{\bar\phi}\propto m^2$ and thus negligible.
Then it is straightforward to track the evolution of $\delta\phi$ originating from $\delta\phi^\mathrm{(inf)}$.
In doing so, we introduce the tensor-to-scalar ratio
$r=2H_{\rm inf}^2/(\pi M_{\rm pl}^2 P_{\zeta})$, where $P_{\zeta}$ is
the scalar power spectrum generated
during inflation.
The contributions of perturbations at
present and at the LSS
to the birefringence can be computed,
respectively, as~\cite{Fujita:2020aqt}
\begin{align}
        \frac{ |\bar\alpha_{\delta\phi}| }{0.35\,\mathrm{deg} }
        &=
        \left(
            \frac{g }{ 3.0 \times 10^{-15}\,\si{\GeV}^{-1}}
        \right)
        \left( \frac{r}{0.06} \right)^{1/2},
        \label{gdeltaphiobs}
    \\
        \frac{ A_{\alpha }}{0.033 \,\mathrm{deg}^2}
        &=
        \left( \frac{g}{1.1 \times 10^{-15}\,\si{\GeV}^{-1}} \right)^{2}
        \left( \frac{r}{0.06} \right),
        \label{gdeltaphiLSS}
\end{align}
where $A_\alpha = L(L+1)C_L^{\alpha\alpha}/(2\pi)$ characterizes the anisotropic birefringence with $C_L^{\alpha\alpha}$ being the angular power spectrum of $\alpha(\hat{\bm n })$, and we used its current
upper bound $A_{\alpha} \leq 0.033\,\mathrm{deg}^2$~\cite{Namikawa:2020ffr,Bianchini:2020osu} as a reference value.
Here $\bar\alpha_{\delta\phi}$ denotes $\bar\alpha$ contributed by $\delta\phi_\mathrm{obs}$.
Eliminating $g$ and $r$ from the above equations, we obtain the relation between $\bar{\alpha}_{\delta\phi}$
and $A_{\alpha}$, as
\begin{equation}
    |\bar\alpha_{\delta\phi}|
    = 0.13 \,{\rm deg} \times\left( \frac{ A_{\alpha }}{0.033\,\mathrm{deg}^2} \right)^{1/2}.
    \label{alphadeltaphi}
\end{equation}
It implies that, since $\delta\phi_\mathrm{LSS}$ and $\delta\phi_\mathrm{obs}$ are connected to
each other, the upper bound on the anisotropic birefringence also puts a constraint on the isotropic birefringence which is generated by the ALP fluctuation.
Therefore, we expect that $\delta\phi_\mathrm{obs}$ gives only a
subdominant contribution to the observed $\bar\alpha$ for $m \lesssim H_0$.
Nonetheless, $\delta\phi_\mathrm{obs}$ has a stochastic nature, in that Eq.~\eqref{gdeltaphiobs} is evaluated by
its root mean square and hence
there is a chance that $\delta\phi_\mathrm{obs}$ gives a larger contribution. An interested reader may refer to Ref.~\cite{Fujita:2020aqt} for more details on $\delta\phi_\mathrm{obs}$.

In the mass region $m\gtrsim H_\mathrm{LSS}$, the ALP begins to oscillate before the last scattering epoch.
In the simple ALP models, $\langle \bar\phi \rangle_\mathrm{LSS}$ is exponentially suppressed by the averaging (\ref{eq_LSSwashout}), and
the amplitude of $\bar\phi$ has been
damped by today.
In the EDE models, on the other hand,
the suppression effect is less significant
due to the field-dependent effective mass
decaying in time.
This is expected to be true for $\delta \phi_{\mathrm{LSS}}$ as well.
In fact, Ref.~\cite{Capparelli:2019rtn} computed the effective sourced fluctuation $\langle \delta\phi^\mathrm{(src)}_\mathrm{LSS} \rangle$ for
the potential (\ref{earlyV1}) with $n=2,3$ and $f=M_{\rm pl}$, and found that idealized future CMB experiments could detect the anisotropic birefringence down to $g\sim10^{-17}\,\mathrm{GeV}^{-1}$.
This value roughly corresponds
to the upper part of the contours in Figs.~\ref{fig: period n2 g} and \ref{fig: period n3 g}.
Therefore, future observations may confirm the prediction of the EDE models and further investigation is awaiting to be done.

For the higher mass region $m \gtrsim 10^{-26}$~eV,
the ALP behavior becomes closer to the major
dark matter component and its clustering may be relevant in the simple ALP models.
Considering the local dark matter density much higher than the averaged one, it may be possible that the ALP clustering boosts its fluctuation and makes $\delta\phi_\mathrm{obs}$ significantly contribute to the isotropic birefringence.
To give a crude estimate, we assume that
the ALP follows the normal dark matter distribution. Then, by scaling the dark
matter density, one finds the local ALP amplitude, as
\begin{align}
        \phi_\text{local} \sim \sqrt{\frac{2\rho_\text{local}
        \Omega_\phi}{m^2 \Omega_c}}
        = 2.5\times 10^{11}~\text{GeV}
        \left(\frac{\Omega_\phi}{10^{-2}\Omega_c} \right)^{1/2}
        \left(\frac{m}{10^{-24}\,\text{eV} } \right)^{-1}\,,
\end{align}
where $\Omega_c$ denotes the energy fraction of all dark matter, and the local dark matter density $\rho_\text{local}$ is set to be $0.4\,{\rm GeV/cm^3}$.
$\phi_\text{local}$ contributes to
the isotropic birefringence through the $\delta\phi_\mathrm{obs}$ term in Eq.~\eqref{alpha full expression}
and the corresponding coupling constant
is given by
\begin{align}
        g_{\delta\phi_\mathrm{obs}} = \frac{2\bar\alpha}{\phi_\text{local}}
        \sim
        5\times 10^{-14}\,\text{GeV}^{-1}
        \left(\frac{\bar\alpha }{0.35\,\text{deg}} \right)
        \left(\frac{\Omega_\phi}{10^{-2}\Omega_c} \right)^{-1/2}
        \left(\frac{m}{10^{-24}\,\text{eV} } \right).
    \label{eq_g_localALPhalo}
\end{align}
This value is much smaller than the naive extrapolation of Figs.~\ref{fig: mass}
and \ref{fig: axionic} to $m=10^{-24}\,$eV as well as the current upper bound.
Thus, such a heavy ALP at the observer might be able to dominate the observed isotropic birefringence in the simple models.
However, it should be noted that, in the above crude estimate, we ignore the
de Broglie wavelength of the ALP reaching or exceeding the galactic scale, the quantum pressure preventing its clustering on a smaller scale, and the oscillation of the ALP field whose period is
$\sim 100\,(m/10^{-24}\,{\rm eV})^{-1}$\,yr.
We leave the evaluation of the contribution from the local ALP density to the birefringence for future work.

Finally, the intermediate region
$H_0 \lesssim m \lesssim H_{\rm LSS}$ has not been studied well in the literature.
The ALP oscillation starts after the
decoupling of CMB photons and the source effect is negligible before that.
Eq.~\eqref{gdeltaphiLSS} is applicable
for $\delta\phi_\mathrm{LSS}$, while $\delta\phi^\mathrm{(inf)}_\mathrm{obs}$
is more suppressed than that estimated by  Eq.~\eqref{gdeltaphiobs} due to
the damped oscillation.
The ALP clustering may be negligible due to its extremely large de Broglie wavelength.
It would be interesting to investigate the sourced fluctuation $\delta\phi^\mathrm{(src)}_\mathrm{obs}$ in this mass region
in the simple ALP models.

\section{Conclusion}
\label{sec: conclusion}

Cosmic birefringence is a powerful tool to investigate the properties of
ultra-light ALPs.
The recent analysis based on the Planck 2018 polarization data reported the rotation angle of CMB polarization $\bar {\alpha} = 0.35 \pm 0.14$ deg, excluding the null hypothesis
at 99.2\,\% CL.
The isotropic birefringence indicates the difference of ALP field values between the last scattering and the detection of CMB photon caused by the dynamics of the background ALP field.
Thus, the detected signal opened up a new
window for the study of dark energy
and early dark energy.

In this paper, we studied the possible
origins of isotropic birefringence signal
by solving the ALP background dynamics
for various potentials $V(\phi)$.
The ALP effective mass, which is associated
with the second derivative
$V_{,\phi \phi}(\phi)$, is a crucial
quantity to characterize the epoch at which
the initially slow-rolling field starts
to oscillate around the potential minimum.
The field dynamics translates to the rotation angle through the relation
$\bar{\alpha} = g \Delta \bar{\phi}/2$ as Eq.~\eqref{master eq}.
We included the effect of finite thickness of the LSS, which suppresses the net rotation angle if the ALP field begins to
oscillate before the decoupling of
CMB photons.
Finally, we determined the required value of the ALP-photon coupling constant, $g$, on each parameter to explain the observed isotropic birefringence.

In Sec.~\ref{sec: DE ALP}, we investigated the simple ALP models with two potentials: the quadratic potential,
$V_{\rm mass}(\phi)
=m^2 \phi^2/2$ of Eq.~\eqref{eq_V_quadratic}, and the cosine-type potential,
$V_{\rm cos}(\phi)=
m^2 f^2 [1-\cos(\phi/f)]$ of Eq.~\eqref{eq_V_cos}.
Figs.~\ref{fig: mass} and \ref{fig: axionic} show the axion-photon coupling inferred from the observed isotropic rotation against the ALP mass.
We found that the inferred values of $g$ can be smallest for $H_0 \lesssim m \lesssim H_{\mathrm{LSS}}$, while
in other mass ranges
the smaller field variation
$\Delta \bar{\phi}$ leads to
the larger $g$.
Using the dependence of $g$ on $m$ and $\Omega_\phi$, we put lower bounds on $g$, $m$, and $\Omega_\phi$ for both potentials.
Moreover, we studied the possibility of ALP as dark energy and derived the relation between the ALP-photon coupling and
the field equation of state of  $w_\phi$, in Eq.~\eqref{eq_g_by_w}.
Using this relation, we put the lower bound of $w_\phi$ as $w_{\phi}+1 \gtrsim 6.5 \times 10^{-18}$ in Eq.~\eqref{wphibo}, which is a quite surprising result since such a small deviation of $w_\phi$ from $-1$
is far out of reach of the previous constraints derived by standard distance measurements.

In Sec.~\ref{sec: EDE}, we studied the signature of ALP-photon coupling in cosmic birefringence for two typical models of EDE: the higher-order periodic potential,
$V_{\rm \cos}^{(n)}(\phi)
=m^2f^2[1-\cos(\phi/f)]^n$ of Eq.\eqref{earlyV1}, and the rock `n' roll model,
$V_{\rm RnR}^{(n)}(\phi)=m^2 M_{\mathrm{pl}}^2 (\phi/M_\text{pl})^{2n}/2^n$ of Eq.~\eqref{earlyV2}.
The EDE is motivated to alleviate the observational tension of today's Hubble constant between
the CMB \cite{Aghanim:2018eyx} and local astrophysical observations at low redshifts  \cite{Riess:2011yx,Riess:2016jrr,Bonvin:2016crt,Riess:2018byc,Birrer:2018vtm,Riess:2019cxk}.
Since the successful EDE scenarios require the very limited ranges of ALP mass and initial conditions, we can rigorously estimate the ALP-photon coupling
based on the EDE models.
In Figs.~\ref{fig: period n2 g}, \ref{fig: period n3 g}, and \ref{fig: RnR g}, we
showed the axion-photon coupling inferred from the observed isotropic rotation with $1\sigma$ and $2\sigma$ parameter regions for the successful EDE scenarios.
The upper bound of $g$ is plotted by requiring the 1\,\% fine tuning of the ALP initial condition, and the more fine-tuned the initial condition is, the larger $g$ is.
We found that the observed isotropic birefringence typically requires $g\sim 10^{-18}$ GeV for the EDE models in Eq.~\eqref{eq_typical_g_EDE}, and it results in $gf= \mathcal O(1)$ for $f=M_\text{pl}$, which is a non-trivial coincidence.
In this paper, we focused on the higher-order periodic potential with $f=M_\text{pl}$ and leave the dependence on $f$ for future work.

In Sec.~\ref{sec: discussion},
we commented on other possible sources of isotropic birefringence by ALP.
The isotropic birefringence can be induced not only by the background dynamics but also by the fluctuation at the observer's position, $\delta\phi_{\mathrm{obs}}$.
To estimate $\delta\phi_{\mathrm{obs}}$,
we divided the ALP mass range into three different regions,
(i) the light region: $m\lesssim H_0$,
(ii) the heavy region:
$m\gtrsim H_\mathrm{LSS}$, and
(iii) the intermediate region:
$H_0 \lesssim m \lesssim H_\mathrm{LSS}$, and briefly explore each of them.
In the region (i), the ALP fluctuation is
mostly given by the primordial perturbation during inflation.
We found that $\delta\phi_{\mathrm{obs}}$ is unlikely to explain the observed isotropic birefringence since the contribution of such ALP fluctuations is constrained by the observation of anisotropic birefringence.
In the region (ii), the ALP fluctuation may be mainly produced by the gravitational growth around galaxies.
We roughly estimated the ALP field value as the sub-component of the local dark matter density, and found that the observed signal might be explained by the ALP with $g\sim 10^{-14}\,\text{GeV}^{-1}$, $m\sim 10^{-24}$ eV, and $\Omega_\phi\sim 10^{-2}\Omega_c$ in Eq.~\eqref{eq_g_localALPhalo}.
We leave the region (iii) for future work since the de Broglie wavelength of ALP is larger than the size of galaxies and the structure formation of ALP is unclear
in such a mass region.

\section*{Acknowledgements}

We would like to thank Yuto Minami and Masahiro Kawasaki for fruitful discussions and productive comments.
This work is supported by the Grant-in-Aid for Scientific Research Fund of the JSPS
Nos.~18K13537 (T.\,F.),
20J20248 (K.\,M.),
19J21974 (H.\,N.), and
19K03854 (S.\,T.).
K.\,M. is supported by World Premier International Research Center Initiative (WPI Initiative), MEXT, Japan and the Program of Excellence in Photon Science.
H.\,N. is supported by Advanced Leading Graduate Course for Photon Science.

\small
\bibliographystyle{apsrev4-1}
\bibliography{Ref}

\begin{thebibliography}{86}%
\makeatletter
\providecommand \@ifxundefined [1]{%
 \@ifx{#1\undefined}
}%
\providecommand \@ifnum [1]{%
 \ifnum #1\expandafter \@firstoftwo
 \else \expandafter \@secondoftwo
 \fi
}%
\providecommand \@ifx [1]{%
 \ifx #1\expandafter \@firstoftwo
 \else \expandafter \@secondoftwo
 \fi
}%
\providecommand \natexlab [1]{#1}%
\providecommand \enquote  [1]{``#1''}%
\providecommand \bibnamefont  [1]{#1}%
\providecommand \bibfnamefont [1]{#1}%
\providecommand \citenamefont [1]{#1}%
\providecommand \href@noop [0]{\@secondoftwo}%
\providecommand \href [0]{\begingroup \@sanitize@url \@href}%
\providecommand \@href[1]{\@@startlink{#1}\@@href}%
\providecommand \@@href[1]{\endgroup#1\@@endlink}%
\providecommand \@sanitize@url [0]{\catcode `\\12\catcode `\$12\catcode
  `\&12\catcode `\#12\catcode `\^12\catcode `\_12\catcode `\%12\relax}%
\providecommand \@@startlink[1]{}%
\providecommand \@@endlink[0]{}%
\providecommand \url  [0]{\begingroup\@sanitize@url \@url }%
\providecommand \@url [1]{\endgroup\@href {#1}{\urlprefix }}%
\providecommand \urlprefix  [0]{URL }%
\providecommand \Eprint [0]{\href }%
\providecommand \doibase [0]{http://dx.doi.org/}%
\providecommand \selectlanguage [0]{\@gobble}%
\providecommand \bibinfo  [0]{\@secondoftwo}%
\providecommand \bibfield  [0]{\@secondoftwo}%
\providecommand \translation [1]{[#1]}%
\providecommand \BibitemOpen [0]{}%
\providecommand \bibitemStop [0]{}%
\providecommand \bibitemNoStop [0]{.\EOS\space}%
\providecommand \EOS [0]{\spacefactor3000\relax}%
\providecommand \BibitemShut  [1]{\csname bibitem#1\endcsname}%
\let\auto@bib@innerbib\@empty
\bibitem [{\citenamefont {Spergel}\ \emph {et~al.}(2003)\citenamefont {Spergel}
  \emph {et~al.}}]{Spergel:2003cb}%
  \BibitemOpen
  \bibfield  {author} {\bibinfo {author} {\bibfnamefont {D.}~\bibnamefont
  {Spergel}} \emph {et~al.} (\bibinfo {collaboration} {WMAP}),\ }\href
  {\doibase 10.1086/377226} {\bibfield  {journal} {\bibinfo  {journal}
  {Astrophys. J. Suppl.}\ }\textbf {\bibinfo {volume} {148}},\ \bibinfo {pages}
  {175} (\bibinfo {year} {2003})},\ \Eprint
  {http://arxiv.org/abs/astro-ph/0302209} {arXiv:astro-ph/0302209} \BibitemShut
  {NoStop}%
\bibitem [{\citenamefont {Weiland}\ \emph {et~al.}(2011)\citenamefont
  {Weiland}, \citenamefont {Odegard}, \citenamefont {Hill}, \citenamefont
  {Wollack}, \citenamefont {Hinshaw}, \citenamefont {Greason}, \citenamefont
  {Jarosik}, \citenamefont {Page}, \citenamefont {Bennett}, \citenamefont
  {Dunkley} \emph {et~al.}}]{weiland2011seven}%
  \BibitemOpen
  \bibfield  {author} {\bibinfo {author} {\bibfnamefont {J.}~\bibnamefont
  {Weiland}}, \bibinfo {author} {\bibfnamefont {N.}~\bibnamefont {Odegard}},
  \bibinfo {author} {\bibfnamefont {R.}~\bibnamefont {Hill}}, \bibinfo {author}
  {\bibfnamefont {E.}~\bibnamefont {Wollack}}, \bibinfo {author} {\bibfnamefont
  {G.}~\bibnamefont {Hinshaw}}, \bibinfo {author} {\bibfnamefont
  {M.}~\bibnamefont {Greason}}, \bibinfo {author} {\bibfnamefont
  {N.}~\bibnamefont {Jarosik}}, \bibinfo {author} {\bibfnamefont
  {L.}~\bibnamefont {Page}}, \bibinfo {author} {\bibfnamefont {C.}~\bibnamefont
  {Bennett}}, \bibinfo {author} {\bibfnamefont {J.}~\bibnamefont {Dunkley}},
  \emph {et~al.},\ }\href@noop {} {\bibfield  {journal} {\bibinfo  {journal}
  {The Astrophysical Journal Supplement Series}\ }\textbf {\bibinfo {volume}
  {192}},\ \bibinfo {pages} {19} (\bibinfo {year} {2011})}\BibitemShut
  {NoStop}%
\bibitem [{\citenamefont {Ade}\ \emph {et~al.}(2014)\citenamefont {Ade} \emph
  {et~al.}}]{Ade:2013zuv}%
  \BibitemOpen
  \bibfield  {author} {\bibinfo {author} {\bibfnamefont {P.}~\bibnamefont
  {Ade}} \emph {et~al.} (\bibinfo {collaboration} {Planck}),\ }\href {\doibase
  10.1051/0004-6361/201321591} {\bibfield  {journal} {\bibinfo  {journal}
  {Astron. Astrophys.}\ }\textbf {\bibinfo {volume} {571}},\ \bibinfo {pages}
  {A16} (\bibinfo {year} {2014})},\ \Eprint {http://arxiv.org/abs/1303.5076}
  {arXiv:1303.5076 [astro-ph.CO]} \BibitemShut {NoStop}%
\bibitem [{\citenamefont {Aghanim}\ \emph {et~al.}(2020)\citenamefont {Aghanim}
  \emph {et~al.}}]{Aghanim:2018eyx}%
  \BibitemOpen
  \bibfield  {author} {\bibinfo {author} {\bibfnamefont {N.}~\bibnamefont
  {Aghanim}} \emph {et~al.} (\bibinfo {collaboration} {Planck}),\ }\href
  {\doibase 10.1051/0004-6361/201833910} {\bibfield  {journal} {\bibinfo
  {journal} {Astron. Astrophys.}\ }\textbf {\bibinfo {volume} {641}},\ \bibinfo
  {pages} {A6} (\bibinfo {year} {2020})},\ \Eprint
  {http://arxiv.org/abs/1807.06209} {arXiv:1807.06209 [astro-ph.CO]}
  \BibitemShut {NoStop}%
\bibitem [{\citenamefont {Minami}\ and\ \citenamefont
  {Komatsu}(2020)}]{MinamiKomatsu}%
  \BibitemOpen
  \bibfield  {author} {\bibinfo {author} {\bibfnamefont {Y.}~\bibnamefont
  {Minami}}\ and\ \bibinfo {author} {\bibfnamefont {E.}~\bibnamefont
  {Komatsu}},\ }\href {\doibase 10.1103/PhysRevLett.125.221301} {\bibfield
  {journal} {\bibinfo  {journal} {Phys. Rev. Lett.}\ }\textbf {\bibinfo
  {volume} {125}},\ \bibinfo {pages} {221301} (\bibinfo {year} {2020})},\
  \Eprint {http://arxiv.org/abs/2011.11254} {arXiv:2011.11254 [astro-ph.CO]}
  \BibitemShut {NoStop}%
\bibitem [{\citenamefont {Carroll}(1998)}]{Carroll:1998zi}%
  \BibitemOpen
  \bibfield  {author} {\bibinfo {author} {\bibfnamefont {S.~M.}\ \bibnamefont
  {Carroll}},\ }\href {\doibase 10.1103/PhysRevLett.81.3067} {\bibfield
  {journal} {\bibinfo  {journal} {Phys. Rev. Lett.}\ }\textbf {\bibinfo
  {volume} {81}},\ \bibinfo {pages} {3067} (\bibinfo {year} {1998})},\ \Eprint
  {http://arxiv.org/abs/astro-ph/9806099} {arXiv:astro-ph/9806099} \BibitemShut
  {NoStop}%
\bibitem [{\citenamefont {Lue}\ \emph {et~al.}(1999)\citenamefont {Lue},
  \citenamefont {Wang},\ and\ \citenamefont {Kamionkowski}}]{Lue:1998mq}%
  \BibitemOpen
  \bibfield  {author} {\bibinfo {author} {\bibfnamefont {A.}~\bibnamefont
  {Lue}}, \bibinfo {author} {\bibfnamefont {L.-M.}\ \bibnamefont {Wang}}, \
  and\ \bibinfo {author} {\bibfnamefont {M.}~\bibnamefont {Kamionkowski}},\
  }\href {\doibase 10.1103/PhysRevLett.83.1506} {\bibfield  {journal} {\bibinfo
   {journal} {Phys. Rev. Lett.}\ }\textbf {\bibinfo {volume} {83}},\ \bibinfo
  {pages} {1506} (\bibinfo {year} {1999})},\ \Eprint
  {http://arxiv.org/abs/astro-ph/9812088} {arXiv:astro-ph/9812088} \BibitemShut
  {NoStop}%
\bibitem [{\citenamefont {Pospelov}\ \emph {et~al.}(2009)\citenamefont
  {Pospelov}, \citenamefont {Ritz}, \citenamefont {Skordis}, \citenamefont
  {Ritz},\ and\ \citenamefont {Skordis}}]{Pospelov:2008gg}%
  \BibitemOpen
  \bibfield  {author} {\bibinfo {author} {\bibfnamefont {M.}~\bibnamefont
  {Pospelov}}, \bibinfo {author} {\bibfnamefont {A.}~\bibnamefont {Ritz}},
  \bibinfo {author} {\bibfnamefont {C.}~\bibnamefont {Skordis}}, \bibinfo
  {author} {\bibfnamefont {A.}~\bibnamefont {Ritz}}, \ and\ \bibinfo {author}
  {\bibfnamefont {C.}~\bibnamefont {Skordis}},\ }\href {\doibase
  10.1103/PhysRevLett.103.051302} {\bibfield  {journal} {\bibinfo  {journal}
  {Phys. Rev. Lett.}\ }\textbf {\bibinfo {volume} {103}},\ \bibinfo {pages}
  {051302} (\bibinfo {year} {2009})},\ \Eprint {http://arxiv.org/abs/0808.0673}
  {arXiv:0808.0673 [astro-ph]} \BibitemShut {NoStop}%
\bibitem [{\citenamefont {Finelli}\ and\ \citenamefont
  {Galaverni}(2009)}]{Finelli:2008jv}%
  \BibitemOpen
  \bibfield  {author} {\bibinfo {author} {\bibfnamefont {F.}~\bibnamefont
  {Finelli}}\ and\ \bibinfo {author} {\bibfnamefont {M.}~\bibnamefont
  {Galaverni}},\ }\href {\doibase 10.1103/PhysRevD.79.063002} {\bibfield
  {journal} {\bibinfo  {journal} {Phys. Rev. D}\ }\textbf {\bibinfo {volume}
  {79}},\ \bibinfo {pages} {063002} (\bibinfo {year} {2009})},\ \Eprint
  {http://arxiv.org/abs/0802.4210} {arXiv:0802.4210 [astro-ph]} \BibitemShut
  {NoStop}%
\bibitem [{\citenamefont {Lee}\ \emph {et~al.}(2014)\citenamefont {Lee},
  \citenamefont {Liu},\ and\ \citenamefont {Ng}}]{Lee:2013mqa}%
  \BibitemOpen
  \bibfield  {author} {\bibinfo {author} {\bibfnamefont {S.}~\bibnamefont
  {Lee}}, \bibinfo {author} {\bibfnamefont {G.-C.}\ \bibnamefont {Liu}}, \ and\
  \bibinfo {author} {\bibfnamefont {K.-W.}\ \bibnamefont {Ng}},\ }\href
  {\doibase 10.1103/PhysRevD.89.063010} {\bibfield  {journal} {\bibinfo
  {journal} {Phys. Rev. D}\ }\textbf {\bibinfo {volume} {89}},\ \bibinfo
  {pages} {063010} (\bibinfo {year} {2014})},\ \Eprint
  {http://arxiv.org/abs/1307.6298} {arXiv:1307.6298 [astro-ph.CO]} \BibitemShut
  {NoStop}%
\bibitem [{\citenamefont {Zhao}\ and\ \citenamefont {Li}(2014)}]{Zhao:2014yna}%
  \BibitemOpen
  \bibfield  {author} {\bibinfo {author} {\bibfnamefont {W.}~\bibnamefont
  {Zhao}}\ and\ \bibinfo {author} {\bibfnamefont {M.}~\bibnamefont {Li}},\
  }\href {\doibase 10.1103/PhysRevD.89.103518} {\bibfield  {journal} {\bibinfo
  {journal} {Phys. Rev. D}\ }\textbf {\bibinfo {volume} {89}},\ \bibinfo
  {pages} {103518} (\bibinfo {year} {2014})},\ \Eprint
  {http://arxiv.org/abs/1403.3997} {arXiv:1403.3997 [astro-ph.CO]} \BibitemShut
  {NoStop}%
\bibitem [{\citenamefont {Lee}\ \emph {et~al.}(2016)\citenamefont {Lee},
  \citenamefont {Liu},\ and\ \citenamefont {Ng}}]{Lee:2016jym}%
  \BibitemOpen
  \bibfield  {author} {\bibinfo {author} {\bibfnamefont {S.}~\bibnamefont
  {Lee}}, \bibinfo {author} {\bibfnamefont {G.-C.}\ \bibnamefont {Liu}}, \ and\
  \bibinfo {author} {\bibfnamefont {K.-W.}\ \bibnamefont {Ng}},\ }\href@noop {}
  {\bibfield  {journal} {\bibinfo  {journal} {The Universe}\ }\textbf {\bibinfo
  {volume} {4}},\ \bibinfo {pages} {29} (\bibinfo {year} {2016})},\ \Eprint
  {http://arxiv.org/abs/1912.12903} {arXiv:1912.12903 [astro-ph.CO]}
  \BibitemShut {NoStop}%
\bibitem [{\citenamefont {Liu}\ and\ \citenamefont {Ng}(2017)}]{Liu:2016dcg}%
  \BibitemOpen
  \bibfield  {author} {\bibinfo {author} {\bibfnamefont {G.-C.}\ \bibnamefont
  {Liu}}\ and\ \bibinfo {author} {\bibfnamefont {K.-W.}\ \bibnamefont {Ng}},\
  }\href {\doibase 10.1016/j.dark.2017.02.004} {\bibfield  {journal} {\bibinfo
  {journal} {Phys. Dark Univ.}\ }\textbf {\bibinfo {volume} {16}},\ \bibinfo
  {pages} {22} (\bibinfo {year} {2017})},\ \Eprint
  {http://arxiv.org/abs/1612.02104} {arXiv:1612.02104 [astro-ph.CO]}
  \BibitemShut {NoStop}%
\bibitem [{\citenamefont {Peccei}\ and\ \citenamefont
  {Quinn}(1977)}]{Peccei:1977hh}%
  \BibitemOpen
  \bibfield  {author} {\bibinfo {author} {\bibfnamefont {R.}~\bibnamefont
  {Peccei}}\ and\ \bibinfo {author} {\bibfnamefont {H.~R.}\ \bibnamefont
  {Quinn}},\ }\href {\doibase 10.1103/PhysRevLett.38.1440} {\bibfield
  {journal} {\bibinfo  {journal} {Phys. Rev. Lett.}\ }\textbf {\bibinfo
  {volume} {38}},\ \bibinfo {pages} {1440} (\bibinfo {year}
  {1977})}\BibitemShut {NoStop}%
\bibitem [{\citenamefont {Kim}(1979)}]{Kim:1979if}%
  \BibitemOpen
  \bibfield  {author} {\bibinfo {author} {\bibfnamefont {J.~E.}\ \bibnamefont
  {Kim}},\ }\href {\doibase 10.1103/PhysRevLett.43.103} {\bibfield  {journal}
  {\bibinfo  {journal} {Phys. Rev. Lett.}\ }\textbf {\bibinfo {volume} {43}},\
  \bibinfo {pages} {103} (\bibinfo {year} {1979})}\BibitemShut {NoStop}%
\bibitem [{\citenamefont {Shifman}\ \emph {et~al.}(1980)\citenamefont
  {Shifman}, \citenamefont {Vainshtein},\ and\ \citenamefont
  {Zakharov}}]{Shifman:1979if}%
  \BibitemOpen
  \bibfield  {author} {\bibinfo {author} {\bibfnamefont {M.~A.}\ \bibnamefont
  {Shifman}}, \bibinfo {author} {\bibfnamefont {A.}~\bibnamefont {Vainshtein}},
  \ and\ \bibinfo {author} {\bibfnamefont {V.~I.}\ \bibnamefont {Zakharov}},\
  }\href {\doibase 10.1016/0550-3213(80)90209-6} {\bibfield  {journal}
  {\bibinfo  {journal} {Nucl. Phys. B}\ }\textbf {\bibinfo {volume} {166}},\
  \bibinfo {pages} {493} (\bibinfo {year} {1980})}\BibitemShut {NoStop}%
\bibitem [{\citenamefont {Marsh}(2016)}]{Marsh:2015xka}%
  \BibitemOpen
  \bibfield  {author} {\bibinfo {author} {\bibfnamefont {D.~J.~E.}\
  \bibnamefont {Marsh}},\ }\href {\doibase 10.1016/j.physrep.2016.06.005}
  {\bibfield  {journal} {\bibinfo  {journal} {Phys. Rept.}\ }\textbf {\bibinfo
  {volume} {643}},\ \bibinfo {pages} {1} (\bibinfo {year} {2016})},\ \Eprint
  {http://arxiv.org/abs/1510.07633} {arXiv:1510.07633 [astro-ph.CO]}
  \BibitemShut {NoStop}%
\bibitem [{\citenamefont {Gong}\ \emph {et~al.}(2017)\citenamefont {Gong},
  \citenamefont {Chen},\ and\ \citenamefont {Feng}}]{Gong:2016zsb}%
  \BibitemOpen
  \bibfield  {author} {\bibinfo {author} {\bibfnamefont {Y.}~\bibnamefont
  {Gong}}, \bibinfo {author} {\bibfnamefont {X.}~\bibnamefont {Chen}}, \ and\
  \bibinfo {author} {\bibfnamefont {H.}~\bibnamefont {Feng}},\ }\href {\doibase
  10.1103/PhysRevLett.118.061101} {\bibfield  {journal} {\bibinfo  {journal}
  {Phys. Rev. Lett.}\ }\textbf {\bibinfo {volume} {118}},\ \bibinfo {pages}
  {061101} (\bibinfo {year} {2017})},\ \Eprint
  {http://arxiv.org/abs/1612.05697} {arXiv:1612.05697 [astro-ph.HE]}
  \BibitemShut {NoStop}%
\bibitem [{\citenamefont {Svrcek}\ and\ \citenamefont
  {Witten}(2006)}]{Svrcek:2006yi}%
  \BibitemOpen
  \bibfield  {author} {\bibinfo {author} {\bibfnamefont {P.}~\bibnamefont
  {Svrcek}}\ and\ \bibinfo {author} {\bibfnamefont {E.}~\bibnamefont
  {Witten}},\ }\href {\doibase 10.1088/1126-6708/2006/06/051} {\bibfield
  {journal} {\bibinfo  {journal} {JHEP}\ }\textbf {\bibinfo {volume} {06}},\
  \bibinfo {pages} {051} (\bibinfo {year} {2006})},\ \Eprint
  {http://arxiv.org/abs/hep-th/0605206} {arXiv:hep-th/0605206} \BibitemShut
  {NoStop}%
\bibitem [{\citenamefont {Arvanitaki}\ \emph {et~al.}(2010)\citenamefont
  {Arvanitaki}, \citenamefont {Dimopoulos}, \citenamefont {Dubovsky},
  \citenamefont {Kaloper},\ and\ \citenamefont
  {March-Russell}}]{Arvanitaki:2009fg}%
  \BibitemOpen
  \bibfield  {author} {\bibinfo {author} {\bibfnamefont {A.}~\bibnamefont
  {Arvanitaki}}, \bibinfo {author} {\bibfnamefont {S.}~\bibnamefont
  {Dimopoulos}}, \bibinfo {author} {\bibfnamefont {S.}~\bibnamefont
  {Dubovsky}}, \bibinfo {author} {\bibfnamefont {N.}~\bibnamefont {Kaloper}}, \
  and\ \bibinfo {author} {\bibfnamefont {J.}~\bibnamefont {March-Russell}},\
  }\href {\doibase 10.1103/PhysRevD.81.123530} {\bibfield  {journal} {\bibinfo
  {journal} {Phys. Rev. D}\ }\textbf {\bibinfo {volume} {81}},\ \bibinfo
  {pages} {123530} (\bibinfo {year} {2010})},\ \Eprint
  {http://arxiv.org/abs/0905.4720} {arXiv:0905.4720 [hep-th]} \BibitemShut
  {NoStop}%
\bibitem [{\citenamefont {Feng}\ \emph {et~al.}(2005)\citenamefont {Feng},
  \citenamefont {Li}, \citenamefont {Li},\ and\ \citenamefont
  {Zhang}}]{Feng:2004mq}%
  \BibitemOpen
  \bibfield  {author} {\bibinfo {author} {\bibfnamefont {B.}~\bibnamefont
  {Feng}}, \bibinfo {author} {\bibfnamefont {H.}~\bibnamefont {Li}}, \bibinfo
  {author} {\bibfnamefont {M.}~\bibnamefont {Li}}, \ and\ \bibinfo {author}
  {\bibfnamefont {X.}~\bibnamefont {Zhang}},\ }\href {\doibase
  10.1016/j.physletb.2005.06.009} {\bibfield  {journal} {\bibinfo  {journal}
  {Phys. Lett.}\ }\textbf {\bibinfo {volume} {B620}},\ \bibinfo {pages} {27}
  (\bibinfo {year} {2005})},\ \Eprint {http://arxiv.org/abs/hep-ph/0406269}
  {arXiv:hep-ph/0406269 [hep-ph]} \BibitemShut {NoStop}%
\bibitem [{\citenamefont {Feng}\ \emph {et~al.}(2006)\citenamefont {Feng},
  \citenamefont {Li}, \citenamefont {Xia}, \citenamefont {Chen},\ and\
  \citenamefont {Zhang}}]{Feng:2006dp}%
  \BibitemOpen
  \bibfield  {author} {\bibinfo {author} {\bibfnamefont {B.}~\bibnamefont
  {Feng}}, \bibinfo {author} {\bibfnamefont {M.}~\bibnamefont {Li}}, \bibinfo
  {author} {\bibfnamefont {J.-Q.}\ \bibnamefont {Xia}}, \bibinfo {author}
  {\bibfnamefont {X.}~\bibnamefont {Chen}}, \ and\ \bibinfo {author}
  {\bibfnamefont {X.}~\bibnamefont {Zhang}},\ }\href {\doibase
  10.1103/PhysRevLett.96.221302} {\bibfield  {journal} {\bibinfo  {journal}
  {Phys. Rev. Lett.}\ }\textbf {\bibinfo {volume} {96}},\ \bibinfo {pages}
  {221302} (\bibinfo {year} {2006})},\ \Eprint
  {http://arxiv.org/abs/astro-ph/0601095} {arXiv:astro-ph/0601095 [astro-ph]}
  \BibitemShut {NoStop}%
\bibitem [{\citenamefont {Liu}\ \emph {et~al.}(2006)\citenamefont {Liu},
  \citenamefont {Lee},\ and\ \citenamefont {Ng}}]{Liu:2006uh}%
  \BibitemOpen
  \bibfield  {author} {\bibinfo {author} {\bibfnamefont {G.-C.}\ \bibnamefont
  {Liu}}, \bibinfo {author} {\bibfnamefont {S.}~\bibnamefont {Lee}}, \ and\
  \bibinfo {author} {\bibfnamefont {K.-W.}\ \bibnamefont {Ng}},\ }\href
  {\doibase 10.1103/PhysRevLett.97.161303} {\bibfield  {journal} {\bibinfo
  {journal} {Phys. Rev. Lett.}\ }\textbf {\bibinfo {volume} {97}},\ \bibinfo
  {pages} {161303} (\bibinfo {year} {2006})},\ \Eprint
  {http://arxiv.org/abs/astro-ph/0606248} {arXiv:astro-ph/0606248 [astro-ph]}
  \BibitemShut {NoStop}%
\bibitem [{\citenamefont {Fukugita}\ and\ \citenamefont
  {Yanagida}(1994)}]{Fukugita:1994hq}%
  \BibitemOpen
  \bibfield  {author} {\bibinfo {author} {\bibfnamefont {M.}~\bibnamefont
  {Fukugita}}\ and\ \bibinfo {author} {\bibfnamefont {T.}~\bibnamefont
  {Yanagida}},\ }\href@noop {} {\  (\bibinfo {year} {1994})}\BibitemShut
  {NoStop}%
\bibitem [{\citenamefont {Frieman}\ \emph {et~al.}(1995)\citenamefont
  {Frieman}, \citenamefont {Hill}, \citenamefont {Stebbins},\ and\
  \citenamefont {Waga}}]{Frieman:1995pm}%
  \BibitemOpen
  \bibfield  {author} {\bibinfo {author} {\bibfnamefont {J.~A.}\ \bibnamefont
  {Frieman}}, \bibinfo {author} {\bibfnamefont {C.~T.}\ \bibnamefont {Hill}},
  \bibinfo {author} {\bibfnamefont {A.}~\bibnamefont {Stebbins}}, \ and\
  \bibinfo {author} {\bibfnamefont {I.}~\bibnamefont {Waga}},\ }\href {\doibase
  10.1103/PhysRevLett.75.2077} {\bibfield  {journal} {\bibinfo  {journal}
  {Phys. Rev. Lett.}\ }\textbf {\bibinfo {volume} {75}},\ \bibinfo {pages}
  {2077} (\bibinfo {year} {1995})},\ \Eprint
  {http://arxiv.org/abs/astro-ph/9505060} {arXiv:astro-ph/9505060} \BibitemShut
  {NoStop}%
\bibitem [{\citenamefont {Kim}(1999)}]{Kim:1998kx}%
  \BibitemOpen
  \bibfield  {author} {\bibinfo {author} {\bibfnamefont {J.~E.}\ \bibnamefont
  {Kim}},\ }\href {\doibase 10.1088/1126-6708/1999/05/022} {\bibfield
  {journal} {\bibinfo  {journal} {JHEP}\ }\textbf {\bibinfo {volume} {05}},\
  \bibinfo {pages} {022} (\bibinfo {year} {1999})},\ \Eprint
  {http://arxiv.org/abs/hep-ph/9811509} {arXiv:hep-ph/9811509} \BibitemShut
  {NoStop}%
\bibitem [{\citenamefont {Kim}(2000)}]{Kim:1999dc}%
  \BibitemOpen
  \bibfield  {author} {\bibinfo {author} {\bibfnamefont {J.~E.}\ \bibnamefont
  {Kim}},\ }\href {\doibase 10.1088/1126-6708/2000/06/016} {\bibfield
  {journal} {\bibinfo  {journal} {JHEP}\ }\textbf {\bibinfo {volume} {06}},\
  \bibinfo {pages} {016} (\bibinfo {year} {2000})},\ \Eprint
  {http://arxiv.org/abs/hep-ph/9907528} {arXiv:hep-ph/9907528} \BibitemShut
  {NoStop}%
\bibitem [{\citenamefont {Choi}(2000)}]{Choi:1999xn}%
  \BibitemOpen
  \bibfield  {author} {\bibinfo {author} {\bibfnamefont {K.}~\bibnamefont
  {Choi}},\ }\href {\doibase 10.1103/PhysRevD.62.043509} {\bibfield  {journal}
  {\bibinfo  {journal} {Phys. Rev. D}\ }\textbf {\bibinfo {volume} {62}},\
  \bibinfo {pages} {043509} (\bibinfo {year} {2000})},\ \Eprint
  {http://arxiv.org/abs/hep-ph/9902292} {arXiv:hep-ph/9902292} \BibitemShut
  {NoStop}%
\bibitem [{\citenamefont {Nomura}\ \emph {et~al.}(2000)\citenamefont {Nomura},
  \citenamefont {Watari},\ and\ \citenamefont {Yanagida}}]{Nomura:2000yk}%
  \BibitemOpen
  \bibfield  {author} {\bibinfo {author} {\bibfnamefont {Y.}~\bibnamefont
  {Nomura}}, \bibinfo {author} {\bibfnamefont {T.}~\bibnamefont {Watari}}, \
  and\ \bibinfo {author} {\bibfnamefont {T.}~\bibnamefont {Yanagida}},\ }\href
  {\doibase 10.1016/S0370-2693(00)00605-5} {\bibfield  {journal} {\bibinfo
  {journal} {Phys. Lett. B}\ }\textbf {\bibinfo {volume} {484}},\ \bibinfo
  {pages} {103} (\bibinfo {year} {2000})},\ \Eprint
  {http://arxiv.org/abs/hep-ph/0004182} {arXiv:hep-ph/0004182} \BibitemShut
  {NoStop}%
\bibitem [{\citenamefont {Kim}\ and\ \citenamefont
  {Nilles}(2003)}]{Kim:2002tq}%
  \BibitemOpen
  \bibfield  {author} {\bibinfo {author} {\bibfnamefont {J.~E.}\ \bibnamefont
  {Kim}}\ and\ \bibinfo {author} {\bibfnamefont {H.~P.}\ \bibnamefont
  {Nilles}},\ }\href {\doibase 10.1016/S0370-2693(02)03148-9} {\bibfield
  {journal} {\bibinfo  {journal} {Phys. Lett. B}\ }\textbf {\bibinfo {volume}
  {553}},\ \bibinfo {pages} {1} (\bibinfo {year} {2003})},\ \Eprint
  {http://arxiv.org/abs/hep-ph/0210402} {arXiv:hep-ph/0210402} \BibitemShut
  {NoStop}%
\bibitem [{\citenamefont {Hall}\ \emph {et~al.}(2005)\citenamefont {Hall},
  \citenamefont {Nomura},\ and\ \citenamefont {Oliver}}]{Hall:2005xb}%
  \BibitemOpen
  \bibfield  {author} {\bibinfo {author} {\bibfnamefont {L.~J.}\ \bibnamefont
  {Hall}}, \bibinfo {author} {\bibfnamefont {Y.}~\bibnamefont {Nomura}}, \ and\
  \bibinfo {author} {\bibfnamefont {S.~J.}\ \bibnamefont {Oliver}},\ }\href
  {\doibase 10.1103/PhysRevLett.95.141302} {\bibfield  {journal} {\bibinfo
  {journal} {Phys. Rev. Lett.}\ }\textbf {\bibinfo {volume} {95}},\ \bibinfo
  {pages} {141302} (\bibinfo {year} {2005})},\ \Eprint
  {http://arxiv.org/abs/astro-ph/0503706} {arXiv:astro-ph/0503706} \BibitemShut
  {NoStop}%
\bibitem [{\citenamefont {Kim}\ and\ \citenamefont
  {Nilles}(2009)}]{Kim:2009cp}%
  \BibitemOpen
  \bibfield  {author} {\bibinfo {author} {\bibfnamefont {J.~E.}\ \bibnamefont
  {Kim}}\ and\ \bibinfo {author} {\bibfnamefont {H.~P.}\ \bibnamefont
  {Nilles}},\ }\href {\doibase 10.1088/1475-7516/2009/05/010} {\bibfield
  {journal} {\bibinfo  {journal} {JCAP}\ }\textbf {\bibinfo {volume} {05}},\
  \bibinfo {pages} {010} (\bibinfo {year} {2009})},\ \Eprint
  {http://arxiv.org/abs/0902.3610} {arXiv:0902.3610 [hep-th]} \BibitemShut
  {NoStop}%
\bibitem [{\citenamefont {Chatzistavrakidis}\ \emph {et~al.}(2012)\citenamefont
  {Chatzistavrakidis}, \citenamefont {Erfani}, \citenamefont {Nilles},\ and\
  \citenamefont {Zavala}}]{Chatzistavrakidis:2012bb}%
  \BibitemOpen
  \bibfield  {author} {\bibinfo {author} {\bibfnamefont {A.}~\bibnamefont
  {Chatzistavrakidis}}, \bibinfo {author} {\bibfnamefont {E.}~\bibnamefont
  {Erfani}}, \bibinfo {author} {\bibfnamefont {H.~P.}\ \bibnamefont {Nilles}},
  \ and\ \bibinfo {author} {\bibfnamefont {I.}~\bibnamefont {Zavala}},\ }\href
  {\doibase 10.1088/1475-7516/2012/09/006} {\bibfield  {journal} {\bibinfo
  {journal} {JCAP}\ }\textbf {\bibinfo {volume} {09}},\ \bibinfo {pages} {006}
  (\bibinfo {year} {2012})},\ \Eprint {http://arxiv.org/abs/1207.1128}
  {arXiv:1207.1128 [hep-ph]} \BibitemShut {NoStop}%
\bibitem [{\citenamefont {Kim}\ \emph {et~al.}(2014)\citenamefont {Kim},
  \citenamefont {Semertzidis},\ and\ \citenamefont {Tsujikawa}}]{Kim:2014tfa}%
  \BibitemOpen
  \bibfield  {author} {\bibinfo {author} {\bibfnamefont {J.~E.}\ \bibnamefont
  {Kim}}, \bibinfo {author} {\bibfnamefont {Y.}~\bibnamefont {Semertzidis}}, \
  and\ \bibinfo {author} {\bibfnamefont {S.}~\bibnamefont {Tsujikawa}},\ }\href
  {\doibase 10.3389/fphy.2014.00060} {\bibfield  {journal} {\bibinfo  {journal}
  {Front. in Phys.}\ }\textbf {\bibinfo {volume} {2}},\ \bibinfo {pages} {60}
  (\bibinfo {year} {2014})},\ \Eprint {http://arxiv.org/abs/1409.2497}
  {arXiv:1409.2497 [hep-ph]} \BibitemShut {NoStop}%
\bibitem [{\citenamefont {Kang}\ \emph {et~al.}(2019)\citenamefont {Kang},
  \citenamefont {Gong}, \citenamefont {Cheng},\ and\ \citenamefont
  {Chen}}]{Kang:2019vsk}%
  \BibitemOpen
  \bibfield  {author} {\bibinfo {author} {\bibfnamefont {J.}~\bibnamefont
  {Kang}}, \bibinfo {author} {\bibfnamefont {Y.}~\bibnamefont {Gong}}, \bibinfo
  {author} {\bibfnamefont {G.}~\bibnamefont {Cheng}}, \ and\ \bibinfo {author}
  {\bibfnamefont {X.}~\bibnamefont {Chen}},\ }\href {\doibase
  10.1088/1674-4527/20/4/55} {\  (\bibinfo {year} {2019}),\
  10.1088/1674-4527/20/4/55},\ \Eprint {http://arxiv.org/abs/1912.05926}
  {arXiv:1912.05926 [astro-ph.CO]} \BibitemShut {NoStop}%
\bibitem [{\citenamefont {Caldwell}\ and\ \citenamefont
  {Linder}(2005)}]{Caldwell:2005tm}%
  \BibitemOpen
  \bibfield  {author} {\bibinfo {author} {\bibfnamefont {R.}~\bibnamefont
  {Caldwell}}\ and\ \bibinfo {author} {\bibfnamefont {E.~V.}\ \bibnamefont
  {Linder}},\ }\href {\doibase 10.1103/PhysRevLett.95.141301} {\bibfield
  {journal} {\bibinfo  {journal} {Phys. Rev. Lett.}\ }\textbf {\bibinfo
  {volume} {95}},\ \bibinfo {pages} {141301} (\bibinfo {year} {2005})},\
  \Eprint {http://arxiv.org/abs/astro-ph/0505494} {arXiv:astro-ph/0505494}
  \BibitemShut {NoStop}%
\bibitem [{\citenamefont {Chiba}\ \emph {et~al.}(2013)\citenamefont {Chiba},
  \citenamefont {De~Felice},\ and\ \citenamefont {Tsujikawa}}]{Chiba:2012cb}%
  \BibitemOpen
  \bibfield  {author} {\bibinfo {author} {\bibfnamefont {T.}~\bibnamefont
  {Chiba}}, \bibinfo {author} {\bibfnamefont {A.}~\bibnamefont {De~Felice}}, \
  and\ \bibinfo {author} {\bibfnamefont {S.}~\bibnamefont {Tsujikawa}},\ }\href
  {\doibase 10.1103/PhysRevD.87.083505} {\bibfield  {journal} {\bibinfo
  {journal} {Phys. Rev. D}\ }\textbf {\bibinfo {volume} {87}},\ \bibinfo
  {pages} {083505} (\bibinfo {year} {2013})},\ \Eprint
  {http://arxiv.org/abs/1210.3859} {arXiv:1210.3859 [astro-ph.CO]} \BibitemShut
  {NoStop}%
\bibitem [{\citenamefont {Tsujikawa}(2013)}]{Tsujikawa:2013fta}%
  \BibitemOpen
  \bibfield  {author} {\bibinfo {author} {\bibfnamefont {S.}~\bibnamefont
  {Tsujikawa}},\ }\href {\doibase 10.1088/0264-9381/30/21/214003} {\bibfield
  {journal} {\bibinfo  {journal} {Class. Quant. Grav.}\ }\textbf {\bibinfo
  {volume} {30}},\ \bibinfo {pages} {214003} (\bibinfo {year} {2013})},\
  \Eprint {http://arxiv.org/abs/1304.1961} {arXiv:1304.1961 [gr-qc]}
  \BibitemShut {NoStop}%
\bibitem [{\citenamefont {Durrive}\ \emph {et~al.}(2018)\citenamefont
  {Durrive}, \citenamefont {Ooba}, \citenamefont {Ichiki},\ and\ \citenamefont
  {Sugiyama}}]{Durrive:2018quo}%
  \BibitemOpen
  \bibfield  {author} {\bibinfo {author} {\bibfnamefont {J.-B.}\ \bibnamefont
  {Durrive}}, \bibinfo {author} {\bibfnamefont {J.}~\bibnamefont {Ooba}},
  \bibinfo {author} {\bibfnamefont {K.}~\bibnamefont {Ichiki}}, \ and\ \bibinfo
  {author} {\bibfnamefont {N.}~\bibnamefont {Sugiyama}},\ }\href {\doibase
  10.1103/PhysRevD.97.043503} {\bibfield  {journal} {\bibinfo  {journal} {Phys.
  Rev. D}\ }\textbf {\bibinfo {volume} {97}},\ \bibinfo {pages} {043503}
  (\bibinfo {year} {2018})},\ \Eprint {http://arxiv.org/abs/1801.09446}
  {arXiv:1801.09446 [astro-ph.CO]} \BibitemShut {NoStop}%
\bibitem [{\citenamefont {Riess}\ \emph {et~al.}(2011)\citenamefont {Riess},
  \citenamefont {Macri}, \citenamefont {Casertano}, \citenamefont {Lampeitl},
  \citenamefont {Ferguson}, \citenamefont {Filippenko}, \citenamefont {Jha},
  \citenamefont {Li},\ and\ \citenamefont {Chornock}}]{Riess:2011yx}%
  \BibitemOpen
  \bibfield  {author} {\bibinfo {author} {\bibfnamefont {A.~G.}\ \bibnamefont
  {Riess}}, \bibinfo {author} {\bibfnamefont {L.}~\bibnamefont {Macri}},
  \bibinfo {author} {\bibfnamefont {S.}~\bibnamefont {Casertano}}, \bibinfo
  {author} {\bibfnamefont {H.}~\bibnamefont {Lampeitl}}, \bibinfo {author}
  {\bibfnamefont {H.~C.}\ \bibnamefont {Ferguson}}, \bibinfo {author}
  {\bibfnamefont {A.~V.}\ \bibnamefont {Filippenko}}, \bibinfo {author}
  {\bibfnamefont {S.~W.}\ \bibnamefont {Jha}}, \bibinfo {author} {\bibfnamefont
  {W.}~\bibnamefont {Li}}, \ and\ \bibinfo {author} {\bibfnamefont
  {R.}~\bibnamefont {Chornock}},\ }\href {\doibase 10.1088/0004-637X/732/2/129}
  {\bibfield  {journal} {\bibinfo  {journal} {Astrophys. J.}\ }\textbf
  {\bibinfo {volume} {730}},\ \bibinfo {pages} {119} (\bibinfo {year}
  {2011})},\ \bibinfo {note} {[Erratum: Astrophys.J. 732, 129 (2011)]},\
  \Eprint {http://arxiv.org/abs/1103.2976} {arXiv:1103.2976 [astro-ph.CO]}
  \BibitemShut {NoStop}%
\bibitem [{\citenamefont {Riess}\ \emph {et~al.}(2016)\citenamefont {Riess}
  \emph {et~al.}}]{Riess:2016jrr}%
  \BibitemOpen
  \bibfield  {author} {\bibinfo {author} {\bibfnamefont {A.~G.}\ \bibnamefont
  {Riess}} \emph {et~al.},\ }\href {\doibase 10.3847/0004-637X/826/1/56}
  {\bibfield  {journal} {\bibinfo  {journal} {Astrophys. J.}\ }\textbf
  {\bibinfo {volume} {826}},\ \bibinfo {pages} {56} (\bibinfo {year} {2016})},\
  \Eprint {http://arxiv.org/abs/1604.01424} {arXiv:1604.01424 [astro-ph.CO]}
  \BibitemShut {NoStop}%
\bibitem [{\citenamefont {Bonvin}\ \emph {et~al.}(2017)\citenamefont {Bonvin}
  \emph {et~al.}}]{Bonvin:2016crt}%
  \BibitemOpen
  \bibfield  {author} {\bibinfo {author} {\bibfnamefont {V.}~\bibnamefont
  {Bonvin}} \emph {et~al.},\ }\href {\doibase 10.1093/mnras/stw3006} {\bibfield
   {journal} {\bibinfo  {journal} {Mon. Not. Roy. Astron. Soc.}\ }\textbf
  {\bibinfo {volume} {465}},\ \bibinfo {pages} {4914} (\bibinfo {year}
  {2017})},\ \Eprint {http://arxiv.org/abs/1607.01790} {arXiv:1607.01790
  [astro-ph.CO]} \BibitemShut {NoStop}%
\bibitem [{\citenamefont {Riess}\ \emph {et~al.}(2018)\citenamefont {Riess}
  \emph {et~al.}}]{Riess:2018byc}%
  \BibitemOpen
  \bibfield  {author} {\bibinfo {author} {\bibfnamefont {A.~G.}\ \bibnamefont
  {Riess}} \emph {et~al.},\ }\href {\doibase 10.3847/1538-4357/aac82e}
  {\bibfield  {journal} {\bibinfo  {journal} {Astrophys. J.}\ }\textbf
  {\bibinfo {volume} {861}},\ \bibinfo {pages} {126} (\bibinfo {year}
  {2018})},\ \Eprint {http://arxiv.org/abs/1804.10655} {arXiv:1804.10655
  [astro-ph.CO]} \BibitemShut {NoStop}%
\bibitem [{\citenamefont {Birrer}\ \emph {et~al.}(2019)\citenamefont {Birrer}
  \emph {et~al.}}]{Birrer:2018vtm}%
  \BibitemOpen
  \bibfield  {author} {\bibinfo {author} {\bibfnamefont {S.}~\bibnamefont
  {Birrer}} \emph {et~al.},\ }\href {\doibase 10.1093/mnras/stz200} {\bibfield
  {journal} {\bibinfo  {journal} {Mon. Not. Roy. Astron. Soc.}\ }\textbf
  {\bibinfo {volume} {484}},\ \bibinfo {pages} {4726} (\bibinfo {year}
  {2019})},\ \Eprint {http://arxiv.org/abs/1809.01274} {arXiv:1809.01274
  [astro-ph.CO]} \BibitemShut {NoStop}%
\bibitem [{\citenamefont {Riess}\ \emph {et~al.}(2019)\citenamefont {Riess},
  \citenamefont {Casertano}, \citenamefont {Yuan}, \citenamefont {Macri},\ and\
  \citenamefont {Scolnic}}]{Riess:2019cxk}%
  \BibitemOpen
  \bibfield  {author} {\bibinfo {author} {\bibfnamefont {A.~G.}\ \bibnamefont
  {Riess}}, \bibinfo {author} {\bibfnamefont {S.}~\bibnamefont {Casertano}},
  \bibinfo {author} {\bibfnamefont {W.}~\bibnamefont {Yuan}}, \bibinfo {author}
  {\bibfnamefont {L.~M.}\ \bibnamefont {Macri}}, \ and\ \bibinfo {author}
  {\bibfnamefont {D.}~\bibnamefont {Scolnic}},\ }\href {\doibase
  10.3847/1538-4357/ab1422} {\bibfield  {journal} {\bibinfo  {journal}
  {Astrophys. J.}\ }\textbf {\bibinfo {volume} {876}},\ \bibinfo {pages} {85}
  (\bibinfo {year} {2019})},\ \Eprint {http://arxiv.org/abs/1903.07603}
  {arXiv:1903.07603 [astro-ph.CO]} \BibitemShut {NoStop}%
\bibitem [{\citenamefont {Wyman}\ \emph {et~al.}(2014)\citenamefont {Wyman},
  \citenamefont {Rudd}, \citenamefont {Vanderveld},\ and\ \citenamefont
  {Hu}}]{Wyman:2013lza}%
  \BibitemOpen
  \bibfield  {author} {\bibinfo {author} {\bibfnamefont {M.}~\bibnamefont
  {Wyman}}, \bibinfo {author} {\bibfnamefont {D.~H.}\ \bibnamefont {Rudd}},
  \bibinfo {author} {\bibfnamefont {R.}~\bibnamefont {Vanderveld}}, \ and\
  \bibinfo {author} {\bibfnamefont {W.}~\bibnamefont {Hu}},\ }\href {\doibase
  10.1103/PhysRevLett.112.051302} {\bibfield  {journal} {\bibinfo  {journal}
  {Phys. Rev. Lett.}\ }\textbf {\bibinfo {volume} {112}},\ \bibinfo {pages}
  {051302} (\bibinfo {year} {2014})},\ \Eprint {http://arxiv.org/abs/1307.7715}
  {arXiv:1307.7715 [astro-ph.CO]} \BibitemShut {NoStop}%
\bibitem [{\citenamefont {Di~Valentino}\ \emph {et~al.}(2016)\citenamefont
  {Di~Valentino}, \citenamefont {Melchiorri},\ and\ \citenamefont
  {Silk}}]{DiValentino:2016hlg}%
  \BibitemOpen
  \bibfield  {author} {\bibinfo {author} {\bibfnamefont {E.}~\bibnamefont
  {Di~Valentino}}, \bibinfo {author} {\bibfnamefont {A.}~\bibnamefont
  {Melchiorri}}, \ and\ \bibinfo {author} {\bibfnamefont {J.}~\bibnamefont
  {Silk}},\ }\href {\doibase 10.1016/j.physletb.2016.08.043} {\bibfield
  {journal} {\bibinfo  {journal} {Phys. Lett. B}\ }\textbf {\bibinfo {volume}
  {761}},\ \bibinfo {pages} {242} (\bibinfo {year} {2016})},\ \Eprint
  {http://arxiv.org/abs/1606.00634} {arXiv:1606.00634 [astro-ph.CO]}
  \BibitemShut {NoStop}%
\bibitem [{\citenamefont {Zhao}\ \emph {et~al.}(2017)\citenamefont {Zhao} \emph
  {et~al.}}]{Zhao:2017cud}%
  \BibitemOpen
  \bibfield  {author} {\bibinfo {author} {\bibfnamefont {G.-B.}\ \bibnamefont
  {Zhao}} \emph {et~al.},\ }\href {\doibase 10.1038/s41550-017-0216-z}
  {\bibfield  {journal} {\bibinfo  {journal} {Nature Astron.}\ }\textbf
  {\bibinfo {volume} {1}},\ \bibinfo {pages} {627} (\bibinfo {year} {2017})},\
  \Eprint {http://arxiv.org/abs/1701.08165} {arXiv:1701.08165 [astro-ph.CO]}
  \BibitemShut {NoStop}%
\bibitem [{\citenamefont {Di~Valentino}\ \emph
  {et~al.}(2017{\natexlab{a}})\citenamefont {Di~Valentino}, \citenamefont
  {Melchiorri}, \citenamefont {Linder},\ and\ \citenamefont
  {Silk}}]{DiValentino:2017zyq}%
  \BibitemOpen
  \bibfield  {author} {\bibinfo {author} {\bibfnamefont {E.}~\bibnamefont
  {Di~Valentino}}, \bibinfo {author} {\bibfnamefont {A.}~\bibnamefont
  {Melchiorri}}, \bibinfo {author} {\bibfnamefont {E.~V.}\ \bibnamefont
  {Linder}}, \ and\ \bibinfo {author} {\bibfnamefont {J.}~\bibnamefont
  {Silk}},\ }\href {\doibase 10.1103/PhysRevD.96.023523} {\bibfield  {journal}
  {\bibinfo  {journal} {Phys. Rev. D}\ }\textbf {\bibinfo {volume} {96}},\
  \bibinfo {pages} {023523} (\bibinfo {year} {2017}{\natexlab{a}})},\ \Eprint
  {http://arxiv.org/abs/1704.00762} {arXiv:1704.00762 [astro-ph.CO]}
  \BibitemShut {NoStop}%
\bibitem [{\citenamefont {Di~Valentino}\ \emph
  {et~al.}(2017{\natexlab{b}})\citenamefont {Di~Valentino}, \citenamefont
  {Melchiorri},\ and\ \citenamefont {Mena}}]{DiValentino:2017iww}%
  \BibitemOpen
  \bibfield  {author} {\bibinfo {author} {\bibfnamefont {E.}~\bibnamefont
  {Di~Valentino}}, \bibinfo {author} {\bibfnamefont {A.}~\bibnamefont
  {Melchiorri}}, \ and\ \bibinfo {author} {\bibfnamefont {O.}~\bibnamefont
  {Mena}},\ }\href {\doibase 10.1103/PhysRevD.96.043503} {\bibfield  {journal}
  {\bibinfo  {journal} {Phys. Rev. D}\ }\textbf {\bibinfo {volume} {96}},\
  \bibinfo {pages} {043503} (\bibinfo {year} {2017}{\natexlab{b}})},\ \Eprint
  {http://arxiv.org/abs/1704.08342} {arXiv:1704.08342 [astro-ph.CO]}
  \BibitemShut {NoStop}%
\bibitem [{\citenamefont {Di~Valentino}\ \emph {et~al.}(2018)\citenamefont
  {Di~Valentino}, \citenamefont {Linder},\ and\ \citenamefont
  {Melchiorri}}]{DiValentino:2017rcr}%
  \BibitemOpen
  \bibfield  {author} {\bibinfo {author} {\bibfnamefont {E.}~\bibnamefont
  {Di~Valentino}}, \bibinfo {author} {\bibfnamefont {E.~V.}\ \bibnamefont
  {Linder}}, \ and\ \bibinfo {author} {\bibfnamefont {A.}~\bibnamefont
  {Melchiorri}},\ }\href {\doibase 10.1103/PhysRevD.97.043528} {\bibfield
  {journal} {\bibinfo  {journal} {Phys. Rev. D}\ }\textbf {\bibinfo {volume}
  {97}},\ \bibinfo {pages} {043528} (\bibinfo {year} {2018})},\ \Eprint
  {http://arxiv.org/abs/1710.02153} {arXiv:1710.02153 [astro-ph.CO]}
  \BibitemShut {NoStop}%
\bibitem [{\citenamefont {Khosravi}\ \emph {et~al.}(2019)\citenamefont
  {Khosravi}, \citenamefont {Baghram}, \citenamefont {Afshordi},\ and\
  \citenamefont {Altamirano}}]{Khosravi:2017hfi}%
  \BibitemOpen
  \bibfield  {author} {\bibinfo {author} {\bibfnamefont {N.}~\bibnamefont
  {Khosravi}}, \bibinfo {author} {\bibfnamefont {S.}~\bibnamefont {Baghram}},
  \bibinfo {author} {\bibfnamefont {N.}~\bibnamefont {Afshordi}}, \ and\
  \bibinfo {author} {\bibfnamefont {N.}~\bibnamefont {Altamirano}},\ }\href
  {\doibase 10.1103/PhysRevD.99.103526} {\bibfield  {journal} {\bibinfo
  {journal} {Phys. Rev. D}\ }\textbf {\bibinfo {volume} {99}},\ \bibinfo
  {pages} {103526} (\bibinfo {year} {2019})},\ \Eprint
  {http://arxiv.org/abs/1710.09366} {arXiv:1710.09366 [astro-ph.CO]}
  \BibitemShut {NoStop}%
\bibitem [{\citenamefont {M\"ortsell}\ and\ \citenamefont
  {Dhawan}(2018)}]{Mortsell:2018mfj}%
  \BibitemOpen
  \bibfield  {author} {\bibinfo {author} {\bibfnamefont {E.}~\bibnamefont
  {M\"ortsell}}\ and\ \bibinfo {author} {\bibfnamefont {S.}~\bibnamefont
  {Dhawan}},\ }\href {\doibase 10.1088/1475-7516/2018/09/025} {\bibfield
  {journal} {\bibinfo  {journal} {JCAP}\ }\textbf {\bibinfo {volume} {09}},\
  \bibinfo {pages} {025} (\bibinfo {year} {2018})},\ \Eprint
  {http://arxiv.org/abs/1801.07260} {arXiv:1801.07260 [astro-ph.CO]}
  \BibitemShut {NoStop}%
\bibitem [{\citenamefont {Poulin}\ \emph
  {et~al.}(2018{\natexlab{a}})\citenamefont {Poulin}, \citenamefont {Boddy},
  \citenamefont {Bird},\ and\ \citenamefont {Kamionkowski}}]{Poulin:2018zxs}%
  \BibitemOpen
  \bibfield  {author} {\bibinfo {author} {\bibfnamefont {V.}~\bibnamefont
  {Poulin}}, \bibinfo {author} {\bibfnamefont {K.~K.}\ \bibnamefont {Boddy}},
  \bibinfo {author} {\bibfnamefont {S.}~\bibnamefont {Bird}}, \ and\ \bibinfo
  {author} {\bibfnamefont {M.}~\bibnamefont {Kamionkowski}},\ }\href {\doibase
  10.1103/PhysRevD.97.123504} {\bibfield  {journal} {\bibinfo  {journal} {Phys.
  Rev. D}\ }\textbf {\bibinfo {volume} {97}},\ \bibinfo {pages} {123504}
  (\bibinfo {year} {2018}{\natexlab{a}})},\ \Eprint
  {http://arxiv.org/abs/1803.02474} {arXiv:1803.02474 [astro-ph.CO]}
  \BibitemShut {NoStop}%
\bibitem [{\citenamefont {Pandey}\ \emph {et~al.}(2020)\citenamefont {Pandey},
  \citenamefont {Karwal},\ and\ \citenamefont {Das}}]{Pandey:2019plg}%
  \BibitemOpen
  \bibfield  {author} {\bibinfo {author} {\bibfnamefont {K.~L.}\ \bibnamefont
  {Pandey}}, \bibinfo {author} {\bibfnamefont {T.}~\bibnamefont {Karwal}}, \
  and\ \bibinfo {author} {\bibfnamefont {S.}~\bibnamefont {Das}},\ }\href
  {\doibase 10.1088/1475-7516/2020/07/026} {\bibfield  {journal} {\bibinfo
  {journal} {JCAP}\ }\textbf {\bibinfo {volume} {07}},\ \bibinfo {pages} {026}
  (\bibinfo {year} {2020})},\ \Eprint {http://arxiv.org/abs/1902.10636}
  {arXiv:1902.10636 [astro-ph.CO]} \BibitemShut {NoStop}%
\bibitem [{\citenamefont {Vattis}\ \emph {et~al.}(2019)\citenamefont {Vattis},
  \citenamefont {Koushiappas},\ and\ \citenamefont {Loeb}}]{Vattis:2019efj}%
  \BibitemOpen
  \bibfield  {author} {\bibinfo {author} {\bibfnamefont {K.}~\bibnamefont
  {Vattis}}, \bibinfo {author} {\bibfnamefont {S.~M.}\ \bibnamefont
  {Koushiappas}}, \ and\ \bibinfo {author} {\bibfnamefont {A.}~\bibnamefont
  {Loeb}},\ }\href {\doibase 10.1103/PhysRevD.99.121302} {\bibfield  {journal}
  {\bibinfo  {journal} {Phys. Rev. D}\ }\textbf {\bibinfo {volume} {99}},\
  \bibinfo {pages} {121302} (\bibinfo {year} {2019})},\ \Eprint
  {http://arxiv.org/abs/1903.06220} {arXiv:1903.06220 [astro-ph.CO]}
  \BibitemShut {NoStop}%
\bibitem [{\citenamefont {Alexander}\ and\ \citenamefont
  {McDonough}(2019)}]{Alexander:2019rsc}%
  \BibitemOpen
  \bibfield  {author} {\bibinfo {author} {\bibfnamefont {S.}~\bibnamefont
  {Alexander}}\ and\ \bibinfo {author} {\bibfnamefont {E.}~\bibnamefont
  {McDonough}},\ }\href {\doibase 10.1016/j.physletb.2019.134830} {\bibfield
  {journal} {\bibinfo  {journal} {Phys. Lett. B}\ }\textbf {\bibinfo {volume}
  {797}},\ \bibinfo {pages} {134830} (\bibinfo {year} {2019})},\ \Eprint
  {http://arxiv.org/abs/1904.08912} {arXiv:1904.08912 [astro-ph.CO]}
  \BibitemShut {NoStop}%
\bibitem [{\citenamefont {Vagnozzi}(2020)}]{Vagnozzi:2019ezj}%
  \BibitemOpen
  \bibfield  {author} {\bibinfo {author} {\bibfnamefont {S.}~\bibnamefont
  {Vagnozzi}},\ }\href {\doibase 10.1103/PhysRevD.102.023518} {\bibfield
  {journal} {\bibinfo  {journal} {Phys. Rev. D}\ }\textbf {\bibinfo {volume}
  {102}},\ \bibinfo {pages} {023518} (\bibinfo {year} {2020})},\ \Eprint
  {http://arxiv.org/abs/1907.07569} {arXiv:1907.07569 [astro-ph.CO]}
  \BibitemShut {NoStop}%
\bibitem [{\citenamefont {Knox}\ and\ \citenamefont
  {Millea}(2020)}]{Knox:2019rjx}%
  \BibitemOpen
  \bibfield  {author} {\bibinfo {author} {\bibfnamefont {L.}~\bibnamefont
  {Knox}}\ and\ \bibinfo {author} {\bibfnamefont {M.}~\bibnamefont {Millea}},\
  }\href {\doibase 10.1103/PhysRevD.101.043533} {\bibfield  {journal} {\bibinfo
   {journal} {Phys. Rev. D}\ }\textbf {\bibinfo {volume} {101}},\ \bibinfo
  {pages} {043533} (\bibinfo {year} {2020})},\ \Eprint
  {http://arxiv.org/abs/1908.03663} {arXiv:1908.03663 [astro-ph.CO]}
  \BibitemShut {NoStop}%
\bibitem [{\citenamefont {Sekiguchi}\ and\ \citenamefont
  {Takahashi}(2020)}]{Sekiguchi:2020teg}%
  \BibitemOpen
  \bibfield  {author} {\bibinfo {author} {\bibfnamefont {T.}~\bibnamefont
  {Sekiguchi}}\ and\ \bibinfo {author} {\bibfnamefont {T.}~\bibnamefont
  {Takahashi}},\ }\href@noop {} {\  (\bibinfo {year} {2020})},\ \Eprint
  {http://arxiv.org/abs/2007.03381} {arXiv:2007.03381 [astro-ph.CO]}
  \BibitemShut {NoStop}%
\bibitem [{\citenamefont {Karwal}\ and\ \citenamefont
  {Kamionkowski}(2016)}]{Karwal:2016vyq}%
  \BibitemOpen
  \bibfield  {author} {\bibinfo {author} {\bibfnamefont {T.}~\bibnamefont
  {Karwal}}\ and\ \bibinfo {author} {\bibfnamefont {M.}~\bibnamefont
  {Kamionkowski}},\ }\href {\doibase 10.1103/PhysRevD.94.103523} {\bibfield
  {journal} {\bibinfo  {journal} {Phys. Rev. D}\ }\textbf {\bibinfo {volume}
  {94}},\ \bibinfo {pages} {103523} (\bibinfo {year} {2016})},\ \Eprint
  {http://arxiv.org/abs/1608.01309} {arXiv:1608.01309 [astro-ph.CO]}
  \BibitemShut {NoStop}%
\bibitem [{\citenamefont {Poulin}\ \emph {et~al.}(2019)\citenamefont {Poulin},
  \citenamefont {Smith}, \citenamefont {Karwal},\ and\ \citenamefont
  {Kamionkowski}}]{Poulin:2018cxd}%
  \BibitemOpen
  \bibfield  {author} {\bibinfo {author} {\bibfnamefont {V.}~\bibnamefont
  {Poulin}}, \bibinfo {author} {\bibfnamefont {T.~L.}\ \bibnamefont {Smith}},
  \bibinfo {author} {\bibfnamefont {T.}~\bibnamefont {Karwal}}, \ and\ \bibinfo
  {author} {\bibfnamefont {M.}~\bibnamefont {Kamionkowski}},\ }\href {\doibase
  10.1103/PhysRevLett.122.221301} {\bibfield  {journal} {\bibinfo  {journal}
  {Phys. Rev. Lett.}\ }\textbf {\bibinfo {volume} {122}},\ \bibinfo {pages}
  {221301} (\bibinfo {year} {2019})},\ \Eprint
  {http://arxiv.org/abs/1811.04083} {arXiv:1811.04083 [astro-ph.CO]}
  \BibitemShut {NoStop}%
\bibitem [{\citenamefont {Agrawal}\ \emph {et~al.}(2019)\citenamefont
  {Agrawal}, \citenamefont {Cyr-Racine}, \citenamefont {Pinner},\ and\
  \citenamefont {Randall}}]{Agrawal:2019lmo}%
  \BibitemOpen
  \bibfield  {author} {\bibinfo {author} {\bibfnamefont {P.}~\bibnamefont
  {Agrawal}}, \bibinfo {author} {\bibfnamefont {F.-Y.}\ \bibnamefont
  {Cyr-Racine}}, \bibinfo {author} {\bibfnamefont {D.}~\bibnamefont {Pinner}},
  \ and\ \bibinfo {author} {\bibfnamefont {L.}~\bibnamefont {Randall}},\
  }\href@noop {} {\  (\bibinfo {year} {2019})},\ \Eprint
  {http://arxiv.org/abs/1904.01016} {arXiv:1904.01016 [astro-ph.CO]}
  \BibitemShut {NoStop}%
\bibitem [{\citenamefont {Lin}\ \emph {et~al.}(2019)\citenamefont {Lin},
  \citenamefont {Benevento}, \citenamefont {Hu},\ and\ \citenamefont
  {Raveri}}]{Lin:2019qug}%
  \BibitemOpen
  \bibfield  {author} {\bibinfo {author} {\bibfnamefont {M.-X.}\ \bibnamefont
  {Lin}}, \bibinfo {author} {\bibfnamefont {G.}~\bibnamefont {Benevento}},
  \bibinfo {author} {\bibfnamefont {W.}~\bibnamefont {Hu}}, \ and\ \bibinfo
  {author} {\bibfnamefont {M.}~\bibnamefont {Raveri}},\ }\href {\doibase
  10.1103/PhysRevD.100.063542} {\bibfield  {journal} {\bibinfo  {journal}
  {Phys. Rev. D}\ }\textbf {\bibinfo {volume} {100}},\ \bibinfo {pages}
  {063542} (\bibinfo {year} {2019})},\ \Eprint
  {http://arxiv.org/abs/1905.12618} {arXiv:1905.12618 [astro-ph.CO]}
  \BibitemShut {NoStop}%
\bibitem [{\citenamefont {Smith}\ \emph {et~al.}(2020)\citenamefont {Smith},
  \citenamefont {Poulin},\ and\ \citenamefont {Amin}}]{Smith:2019ihp}%
  \BibitemOpen
  \bibfield  {author} {\bibinfo {author} {\bibfnamefont {T.~L.}\ \bibnamefont
  {Smith}}, \bibinfo {author} {\bibfnamefont {V.}~\bibnamefont {Poulin}}, \
  and\ \bibinfo {author} {\bibfnamefont {M.~A.}\ \bibnamefont {Amin}},\ }\href
  {\doibase 10.1103/PhysRevD.101.063523} {\bibfield  {journal} {\bibinfo
  {journal} {Phys. Rev. D}\ }\textbf {\bibinfo {volume} {101}},\ \bibinfo
  {pages} {063523} (\bibinfo {year} {2020})},\ \Eprint
  {http://arxiv.org/abs/1908.06995} {arXiv:1908.06995 [astro-ph.CO]}
  \BibitemShut {NoStop}%
\bibitem [{\citenamefont {Niedermann}\ and\ \citenamefont
  {Sloth}(2019)}]{Niedermann:2019olb}%
  \BibitemOpen
  \bibfield  {author} {\bibinfo {author} {\bibfnamefont {F.}~\bibnamefont
  {Niedermann}}\ and\ \bibinfo {author} {\bibfnamefont {M.~S.}\ \bibnamefont
  {Sloth}},\ }\href@noop {} {\  (\bibinfo {year} {2019})},\ \Eprint
  {http://arxiv.org/abs/1910.10739} {arXiv:1910.10739 [astro-ph.CO]}
  \BibitemShut {NoStop}%
\bibitem [{\citenamefont {Berghaus}\ and\ \citenamefont
  {Karwal}(2020)}]{Berghaus:2019cls}%
  \BibitemOpen
  \bibfield  {author} {\bibinfo {author} {\bibfnamefont {K.~V.}\ \bibnamefont
  {Berghaus}}\ and\ \bibinfo {author} {\bibfnamefont {T.}~\bibnamefont
  {Karwal}},\ }\href {\doibase 10.1103/PhysRevD.101.083537} {\bibfield
  {journal} {\bibinfo  {journal} {Phys. Rev. D}\ }\textbf {\bibinfo {volume}
  {101}},\ \bibinfo {pages} {083537} (\bibinfo {year} {2020})},\ \Eprint
  {http://arxiv.org/abs/1911.06281} {arXiv:1911.06281 [astro-ph.CO]}
  \BibitemShut {NoStop}%
\bibitem [{\citenamefont {Sakstein}\ and\ \citenamefont
  {Trodden}(2020)}]{Sakstein:2019fmf}%
  \BibitemOpen
  \bibfield  {author} {\bibinfo {author} {\bibfnamefont {J.}~\bibnamefont
  {Sakstein}}\ and\ \bibinfo {author} {\bibfnamefont {M.}~\bibnamefont
  {Trodden}},\ }\href {\doibase 10.1103/PhysRevLett.124.161301} {\bibfield
  {journal} {\bibinfo  {journal} {Phys. Rev. Lett.}\ }\textbf {\bibinfo
  {volume} {124}},\ \bibinfo {pages} {161301} (\bibinfo {year} {2020})},\
  \Eprint {http://arxiv.org/abs/1911.11760} {arXiv:1911.11760 [astro-ph.CO]}
  \BibitemShut {NoStop}%
\bibitem [{\citenamefont {Ye}\ and\ \citenamefont
  {Piao}(2020{\natexlab{a}})}]{Ye:2020btb}%
  \BibitemOpen
  \bibfield  {author} {\bibinfo {author} {\bibfnamefont {G.}~\bibnamefont
  {Ye}}\ and\ \bibinfo {author} {\bibfnamefont {Y.-S.}\ \bibnamefont {Piao}},\
  }\href {\doibase 10.1103/PhysRevD.101.083507} {\bibfield  {journal} {\bibinfo
   {journal} {Phys. Rev. D}\ }\textbf {\bibinfo {volume} {101}},\ \bibinfo
  {pages} {083507} (\bibinfo {year} {2020}{\natexlab{a}})},\ \Eprint
  {http://arxiv.org/abs/2001.02451} {arXiv:2001.02451 [astro-ph.CO]}
  \BibitemShut {NoStop}%
\bibitem [{\citenamefont {Niedermann}\ and\ \citenamefont
  {Sloth}(2020)}]{Niedermann:2020dwg}%
  \BibitemOpen
  \bibfield  {author} {\bibinfo {author} {\bibfnamefont {F.}~\bibnamefont
  {Niedermann}}\ and\ \bibinfo {author} {\bibfnamefont {M.~S.}\ \bibnamefont
  {Sloth}},\ }\href {\doibase 10.1103/PhysRevD.102.063527} {\bibfield
  {journal} {\bibinfo  {journal} {Phys. Rev. D}\ }\textbf {\bibinfo {volume}
  {102}},\ \bibinfo {pages} {063527} (\bibinfo {year} {2020})},\ \Eprint
  {http://arxiv.org/abs/2006.06686} {arXiv:2006.06686 [astro-ph.CO]}
  \BibitemShut {NoStop}%
\bibitem [{\citenamefont {Ye}\ and\ \citenamefont
  {Piao}(2020{\natexlab{b}})}]{Ye:2020oix}%
  \BibitemOpen
  \bibfield  {author} {\bibinfo {author} {\bibfnamefont {G.}~\bibnamefont
  {Ye}}\ and\ \bibinfo {author} {\bibfnamefont {Y.-S.}\ \bibnamefont {Piao}},\
  }\href {\doibase 10.1103/PhysRevD.102.083523} {\bibfield  {journal} {\bibinfo
   {journal} {Phys. Rev. D}\ }\textbf {\bibinfo {volume} {102}},\ \bibinfo
  {pages} {083523} (\bibinfo {year} {2020}{\natexlab{b}})},\ \Eprint
  {http://arxiv.org/abs/2008.10832} {arXiv:2008.10832 [astro-ph.CO]}
  \BibitemShut {NoStop}%
\bibitem [{\citenamefont {Fujita}\ \emph {et~al.}(2020)\citenamefont {Fujita},
  \citenamefont {Minami}, \citenamefont {Murai},\ and\ \citenamefont
  {Nakatsuka}}]{Fujita:2020aqt}%
  \BibitemOpen
  \bibfield  {author} {\bibinfo {author} {\bibfnamefont {T.}~\bibnamefont
  {Fujita}}, \bibinfo {author} {\bibfnamefont {Y.}~\bibnamefont {Minami}},
  \bibinfo {author} {\bibfnamefont {K.}~\bibnamefont {Murai}}, \ and\ \bibinfo
  {author} {\bibfnamefont {H.}~\bibnamefont {Nakatsuka}},\ }\href@noop {} {\
  (\bibinfo {year} {2020})},\ \Eprint {http://arxiv.org/abs/2008.02473}
  {arXiv:2008.02473 [astro-ph.CO]} \BibitemShut {NoStop}%
\bibitem [{\citenamefont {Capparelli}\ \emph {et~al.}(2020)\citenamefont
  {Capparelli}, \citenamefont {Caldwell},\ and\ \citenamefont
  {Melchiorri}}]{Capparelli:2019rtn}%
  \BibitemOpen
  \bibfield  {author} {\bibinfo {author} {\bibfnamefont {L.~M.}\ \bibnamefont
  {Capparelli}}, \bibinfo {author} {\bibfnamefont {R.~R.}\ \bibnamefont
  {Caldwell}}, \ and\ \bibinfo {author} {\bibfnamefont {A.}~\bibnamefont
  {Melchiorri}},\ }\href {\doibase 10.1103/PhysRevD.101.123529} {\bibfield
  {journal} {\bibinfo  {journal} {Phys. Rev. D}\ }\textbf {\bibinfo {volume}
  {101}},\ \bibinfo {pages} {123529} (\bibinfo {year} {2020})},\ \Eprint
  {http://arxiv.org/abs/1909.04621} {arXiv:1909.04621 [astro-ph.CO]}
  \BibitemShut {NoStop}%
\bibitem [{\citenamefont {Harari}\ and\ \citenamefont
  {Sikivie}(1992)}]{Harari:1992ea}%
  \BibitemOpen
  \bibfield  {author} {\bibinfo {author} {\bibfnamefont {D.}~\bibnamefont
  {Harari}}\ and\ \bibinfo {author} {\bibfnamefont {P.}~\bibnamefont
  {Sikivie}},\ }\href {\doibase 10.1016/0370-2693(92)91363-E} {\bibfield
  {journal} {\bibinfo  {journal} {Phys. Lett. B}\ }\textbf {\bibinfo {volume}
  {289}},\ \bibinfo {pages} {67} (\bibinfo {year} {1992})}\BibitemShut
  {NoStop}%
\bibitem [{\citenamefont {Bianchini}\ \emph {et~al.}(2020)\citenamefont
  {Bianchini} \emph {et~al.}}]{Bianchini:2020osu}%
  \BibitemOpen
  \bibfield  {author} {\bibinfo {author} {\bibfnamefont {F.}~\bibnamefont
  {Bianchini}} \emph {et~al.} (\bibinfo {collaboration} {SPT}),\ }\href
  {\doibase 10.1103/PhysRevD.102.083504} {\bibfield  {journal} {\bibinfo
  {journal} {Phys. Rev. D}\ }\textbf {\bibinfo {volume} {102}},\ \bibinfo
  {pages} {083504} (\bibinfo {year} {2020})},\ \Eprint
  {http://arxiv.org/abs/2006.08061} {arXiv:2006.08061 [astro-ph.CO]}
  \BibitemShut {NoStop}%
\bibitem [{\citenamefont {Namikawa}\ \emph {et~al.}(2020)\citenamefont
  {Namikawa} \emph {et~al.}}]{Namikawa:2020ffr}%
  \BibitemOpen
  \bibfield  {author} {\bibinfo {author} {\bibfnamefont {T.}~\bibnamefont
  {Namikawa}} \emph {et~al.},\ }\href {\doibase 10.1103/PhysRevD.101.083527}
  {\bibfield  {journal} {\bibinfo  {journal} {Phys. Rev. D}\ }\textbf {\bibinfo
  {volume} {101}},\ \bibinfo {pages} {083527} (\bibinfo {year} {2020})},\
  \Eprint {http://arxiv.org/abs/2001.10465} {arXiv:2001.10465 [astro-ph.CO]}
  \BibitemShut {NoStop}%
\bibitem [{\citenamefont {Weinberg}(2008)}]{Weinberg:2008zzc}%
  \BibitemOpen
  \bibfield  {author} {\bibinfo {author} {\bibfnamefont {S.}~\bibnamefont
  {Weinberg}},\ }\href@noop {} {\emph {\bibinfo {title} {{Cosmology}}}}\
  (\bibinfo  {publisher} {{Oxford University Press}},\ \bibinfo {year}
  {2008})\BibitemShut {NoStop}%
\bibitem [{\citenamefont {Hlozek}\ \emph {et~al.}(2015)\citenamefont {Hlozek},
  \citenamefont {Grin}, \citenamefont {Marsh},\ and\ \citenamefont
  {Ferreira}}]{Hlozek:2014lca}%
  \BibitemOpen
  \bibfield  {author} {\bibinfo {author} {\bibfnamefont {R.}~\bibnamefont
  {Hlozek}}, \bibinfo {author} {\bibfnamefont {D.}~\bibnamefont {Grin}},
  \bibinfo {author} {\bibfnamefont {D.~J.~E.}\ \bibnamefont {Marsh}}, \ and\
  \bibinfo {author} {\bibfnamefont {P.~G.}\ \bibnamefont {Ferreira}},\ }\href
  {\doibase 10.1103/PhysRevD.91.103512} {\bibfield  {journal} {\bibinfo
  {journal} {Phys. Rev. D}\ }\textbf {\bibinfo {volume} {91}},\ \bibinfo
  {pages} {103512} (\bibinfo {year} {2015})},\ \Eprint
  {http://arxiv.org/abs/1410.2896} {arXiv:1410.2896 [astro-ph.CO]} \BibitemShut
  {NoStop}%
\bibitem [{\citenamefont {Anastassopoulos}\ \emph {et~al.}(2017)\citenamefont
  {Anastassopoulos} \emph {et~al.}}]{Anastassopoulos:2017ftl}%
  \BibitemOpen
  \bibfield  {author} {\bibinfo {author} {\bibfnamefont {V.}~\bibnamefont
  {Anastassopoulos}} \emph {et~al.} (\bibinfo {collaboration} {CAST}),\ }\href
  {\doibase 10.1038/nphys4109} {\bibfield  {journal} {\bibinfo  {journal}
  {Nature Phys.}\ }\textbf {\bibinfo {volume} {13}},\ \bibinfo {pages} {584}
  (\bibinfo {year} {2017})},\ \Eprint {http://arxiv.org/abs/1705.02290}
  {arXiv:1705.02290 [hep-ex]} \BibitemShut {NoStop}%
\bibitem [{\citenamefont {Payez}\ \emph {et~al.}(2015)\citenamefont {Payez},
  \citenamefont {Evoli}, \citenamefont {Fischer}, \citenamefont {Giannotti},
  \citenamefont {Mirizzi},\ and\ \citenamefont {Ringwald}}]{Payez:2014xsa}%
  \BibitemOpen
  \bibfield  {author} {\bibinfo {author} {\bibfnamefont {A.}~\bibnamefont
  {Payez}}, \bibinfo {author} {\bibfnamefont {C.}~\bibnamefont {Evoli}},
  \bibinfo {author} {\bibfnamefont {T.}~\bibnamefont {Fischer}}, \bibinfo
  {author} {\bibfnamefont {M.}~\bibnamefont {Giannotti}}, \bibinfo {author}
  {\bibfnamefont {A.}~\bibnamefont {Mirizzi}}, \ and\ \bibinfo {author}
  {\bibfnamefont {A.}~\bibnamefont {Ringwald}},\ }\href {\doibase
  10.1088/1475-7516/2015/02/006} {\bibfield  {journal} {\bibinfo  {journal}
  {JCAP}\ }\textbf {\bibinfo {volume} {02}},\ \bibinfo {pages} {006} (\bibinfo
  {year} {2015})},\ \Eprint {http://arxiv.org/abs/1410.3747} {arXiv:1410.3747
  [astro-ph.HE]} \BibitemShut {NoStop}%
\bibitem [{\citenamefont {Berg}\ \emph {et~al.}(2017)\citenamefont {Berg},
  \citenamefont {Conlon}, \citenamefont {Day}, \citenamefont {Jennings},
  \citenamefont {Krippendorf}, \citenamefont {Powell},\ and\ \citenamefont
  {Rummel}}]{Berg:2016ese}%
  \BibitemOpen
  \bibfield  {author} {\bibinfo {author} {\bibfnamefont {M.}~\bibnamefont
  {Berg}}, \bibinfo {author} {\bibfnamefont {J.~P.}\ \bibnamefont {Conlon}},
  \bibinfo {author} {\bibfnamefont {F.}~\bibnamefont {Day}}, \bibinfo {author}
  {\bibfnamefont {N.}~\bibnamefont {Jennings}}, \bibinfo {author}
  {\bibfnamefont {S.}~\bibnamefont {Krippendorf}}, \bibinfo {author}
  {\bibfnamefont {A.~J.}\ \bibnamefont {Powell}}, \ and\ \bibinfo {author}
  {\bibfnamefont {M.}~\bibnamefont {Rummel}},\ }\href {\doibase
  10.3847/1538-4357/aa8b16} {\bibfield  {journal} {\bibinfo  {journal}
  {Astrophys. J.}\ }\textbf {\bibinfo {volume} {847}},\ \bibinfo {pages} {101}
  (\bibinfo {year} {2017})},\ \Eprint {http://arxiv.org/abs/1605.01043}
  {arXiv:1605.01043 [astro-ph.HE]} \BibitemShut {NoStop}%
\bibitem [{\citenamefont {B{\"a}hre}\ \emph {et~al.}(2013)\citenamefont
  {B{\"a}hre} \emph {et~al.}}]{Bahre:2013ywa}%
  \BibitemOpen
  \bibfield  {author} {\bibinfo {author} {\bibfnamefont {R.}~\bibnamefont
  {B{\"a}hre}} \emph {et~al.},\ }\href {\doibase 10.1088/1748-0221/8/09/T09001}
  {\bibfield  {journal} {\bibinfo  {journal} {JINST}\ }\textbf {\bibinfo
  {volume} {8}},\ \bibinfo {pages} {T09001} (\bibinfo {year} {2013})},\ \Eprint
  {http://arxiv.org/abs/1302.5647} {arXiv:1302.5647 [physics.ins-det]}
  \BibitemShut {NoStop}%
\bibitem [{\citenamefont {Armengaud}\ \emph {et~al.}(2014)\citenamefont
  {Armengaud} \emph {et~al.}}]{Armengaud:2014gea}%
  \BibitemOpen
  \bibfield  {author} {\bibinfo {author} {\bibfnamefont {E.}~\bibnamefont
  {Armengaud}} \emph {et~al.},\ }\href {\doibase 10.1088/1748-0221/9/05/T05002}
  {\bibfield  {journal} {\bibinfo  {journal} {JINST}\ }\textbf {\bibinfo
  {volume} {9}},\ \bibinfo {pages} {T05002} (\bibinfo {year} {2014})},\ \Eprint
  {http://arxiv.org/abs/1401.3233} {arXiv:1401.3233 [physics.ins-det]}
  \BibitemShut {NoStop}%
\bibitem [{\citenamefont {Irastorza}\ and\ \citenamefont
  {Redondo}(2018)}]{Irastorza:2018dyq}%
  \BibitemOpen
  \bibfield  {author} {\bibinfo {author} {\bibfnamefont {I.~G.}\ \bibnamefont
  {Irastorza}}\ and\ \bibinfo {author} {\bibfnamefont {J.}~\bibnamefont
  {Redondo}},\ }\href {\doibase 10.1016/j.ppnp.2018.05.003} {\bibfield
  {journal} {\bibinfo  {journal} {Prog. Part. Nucl. Phys.}\ }\textbf {\bibinfo
  {volume} {102}},\ \bibinfo {pages} {89} (\bibinfo {year} {2018})},\ \Eprint
  {http://arxiv.org/abs/1801.08127} {arXiv:1801.08127 [hep-ph]} \BibitemShut
  {NoStop}%
\bibitem [{\citenamefont {Conlon}\ \emph {et~al.}(2018)\citenamefont {Conlon},
  \citenamefont {Day}, \citenamefont {Jennings}, \citenamefont {Krippendorf},\
  and\ \citenamefont {Muia}}]{Conlon:2017ofb}%
  \BibitemOpen
  \bibfield  {author} {\bibinfo {author} {\bibfnamefont {J.~P.}\ \bibnamefont
  {Conlon}}, \bibinfo {author} {\bibfnamefont {F.}~\bibnamefont {Day}},
  \bibinfo {author} {\bibfnamefont {N.}~\bibnamefont {Jennings}}, \bibinfo
  {author} {\bibfnamefont {S.}~\bibnamefont {Krippendorf}}, \ and\ \bibinfo
  {author} {\bibfnamefont {F.}~\bibnamefont {Muia}},\ }\href {\doibase
  10.1093/mnras/stx2652} {\bibfield  {journal} {\bibinfo  {journal} {Mon. Not.
  Roy. Astron. Soc.}\ }\textbf {\bibinfo {volume} {473}},\ \bibinfo {pages}
  {4932} (\bibinfo {year} {2018})},\ \Eprint {http://arxiv.org/abs/1707.00176}
  {arXiv:1707.00176 [astro-ph.HE]} \BibitemShut {NoStop}%
\bibitem [{\citenamefont {Poulin}\ \emph
  {et~al.}(2018{\natexlab{b}})\citenamefont {Poulin}, \citenamefont {Smith},
  \citenamefont {Grin}, \citenamefont {Karwal},\ and\ \citenamefont
  {Kamionkowski}}]{Poulin:2018dzj}%
  \BibitemOpen
  \bibfield  {author} {\bibinfo {author} {\bibfnamefont {V.}~\bibnamefont
  {Poulin}}, \bibinfo {author} {\bibfnamefont {T.~L.}\ \bibnamefont {Smith}},
  \bibinfo {author} {\bibfnamefont {D.}~\bibnamefont {Grin}}, \bibinfo {author}
  {\bibfnamefont {T.}~\bibnamefont {Karwal}}, \ and\ \bibinfo {author}
  {\bibfnamefont {M.}~\bibnamefont {Kamionkowski}},\ }\href {\doibase
  10.1103/PhysRevD.98.083525} {\bibfield  {journal} {\bibinfo  {journal} {Phys.
  Rev. D}\ }\textbf {\bibinfo {volume} {98}},\ \bibinfo {pages} {083525}
  (\bibinfo {year} {2018}{\natexlab{b}})},\ \Eprint
  {http://arxiv.org/abs/1806.10608} {arXiv:1806.10608 [astro-ph.CO]}
  \BibitemShut {NoStop}%
\end{thebibliography}%

\end{document}